\documentclass[a4paper,superscriptaddress,amsmath,amssymb,11pt]{article}
\pdfoutput=1 % if your are submitting a pdflatex (i.e. if you have
\usepackage{EPJCpub} % for details on the use of the package, please
\usepackage[T1]{fontenc} % if needed
\usepackage[utf8]{inputenc}
\usepackage{graphicx}% Include figure files
\usepackage{dcolumn}% Align table columns on decimal point
\usepackage{bm}% bold math
\usepackage{float}
\usepackage{multirow}
\usepackage{verbatim}
\usepackage{amsmath}
\usepackage{graphicx,subfigure,epsfig}
\usepackage{hyperref}
\hypersetup{
    colorlinks=true,
    citecolor=blue,
    linkcolor=blue,
    filecolor=magenta,
    urlcolor=blue,}%cyan
\usepackage{color}

\newcommand{\be}{\begin{equation}}
\newcommand{\ee}{\end{equation}}

\title{Gravitational ringing and superradiant instabilities of the Kerr-like black holes in a dark matter halo}

\author[a]{Dong Liu,}
\author[a]{Yi Yang,}
\author[b]{Ali \"Ovg\"un,}
\author[a,1]{Zheng-Wen Long\note{Corresponding author.}}
\author[a]{and Zhaoyi Xu}

\affiliation[a]{College of Physics, Guizhou University, Guiyang, 550025, China}
\affiliation[b]{Physics Department, Eastern Mediterranean University, Famagusta, 99628 North Cyprus via Mersin 10, Turkey}

\emailAdd{dongliuvv@yeah.net}%{gs.dongliu19@gzu.edu.cn}dongliuvv@yeah.net
\emailAdd{yangyigz@yeah.net}%{gs.yangyi17@gzu.edu.cn}
\emailAdd{ali.ovgun@emu.edu.tr}
\emailAdd{zwlong@gzu.edu.cn}
\emailAdd{zyxu@gzu.edu.cn}

\abstract{%Supermassive black holes from the center of galaxies may be immersed in a dark matter halo. This dark matter halo may form a ``cusp” structure around the black hole and disappear at a certain distance from the black hole. Based on this interesting physical background, we use the continued fraction method to study quasinormal modes (QNM) and quasibound states (QBS) of the Kerr-like black holes immersed in a dark matter halo (by considering cold dark matter (CDM) model and scalar field dark matter (SFDM) model in the LSB galaxy ESO$1200211$), and compare them with Kerr black hole. Besides, we also study the impacts of dark matter parameters on the QNM/QBS of black holes at the specific circumstances. Our main conclusions are as follows: (1) In the massless scalar field, for the same state, QNM frequency of kerr black hole is greater than that of SFDM model, and SFDM model is greater than that of CDM model. (2) In the massive scalar field, by testing the states of QBS frequencies with different $l,m,a$, we confirm the existence of the superradiant instabilities when the black holes both in CDM model and SFDM model. We prove that the maximum instabilities of black hole both in CDM model and SFDM model occur at the state $l=1,m=1, a=0.99$. (3) In black hole units, for CDM model in ESO$1200211$, the maximum instability of black hole occurs for $M\mu \approx 0.42$ and the maximum instability growth rate is approximately $\tau^{-1} \approx 1.50601311 \times 10^{-7} (GM/c^3)^{-1}$. For SFDM model, the maximum instability occurs for $M\mu \approx 0.42 $ and the maximum instability growth rate is approximately $\tau^{-1} \approx 1.50601515 \times 10^{-7} (GM/c^3)^{-1}$. (4) Compared the maximum instabilities of black holes in CDM/SFDM models with Kerr black hole, the maximum instability of Kerr-like black holes in a dark matter halo are all smaller than that of Kerr black hole. The maximum instability difference of black hole between CDM model and Kerr black hole is approximately $2.38\times10^{-13}$. For SFDM model, the difference is approximately $3.50\times10^{-14}$. (5) The impacts of dark matter parameters on the QNM/QBS of black holes are studied. At the state $l=m=1,a=0.99$, the real part and imaginary part of QNM/QBS decrease with the increasing of the dark matter parameter. In the future, these results may be used for gravitational wave detection of supermassive black holes, and may provide an effective method for detecting the existence of dark matter.
Supermassive black holes from the center of galaxy may be immersed in a dark matter halo. This dark matter halo may form a ``cusp” structure around the black hole and disappear at a certain distance from the black hole. Based on this interesting physical background, we use the continued fraction method to study gravitational ringring of the Kerr-like black holes immersed in a dark matter halo, i.e., quasinormal modes (QNM) and quasibound states (QBS). We consider these gravitational ringring of black holes both in cold dark matter (CDM) model and scalar field dark matter (SFDM) model at the LSB galaxy, and compare them with Kerr black hole. By testing the states of QNM/QBS frequencies with different parameters $l,m,a$, we confirm the existence of the superradiant instabilities when the black holes both in CDM model and SFDM model. Besides, we also study the impacts of dark matter parameters on the QNM/QBS of black holes at the specific circumstances. In the future, these results may be used for gravitational wave detection of supermassive black holes, and may provide an effective method for detecting the existence of dark matter.}

\keywords{Dark Matter, Superradiant Instability, Quasibound State, Quasinormal Mode}

\begin{document}
\maketitle
\flushbottom

\section{Introduction}
\label{s1}
Now, there is a large number of observational data indicating the existence of dark matter (DM), such as the cosmic microwave background radiation (CMBR), the rotation curves (RC) of the galaxies and the large-scale structure of the universe. Based on these observational data, astronomers have proposed lots of dark matter models. Among all these dark matter models, cold dark matter (CDM) model \cite{Navarro:1995iw,Navarro:1996gj} and scalar field dark matter (SFDM) model have received much attention. In these dark matter models, the most important part is to obtain the spatial density distribution function of dark matter. At present, the distribution of dark matter in the large-scale structure of galaxies is clear \cite{Planck:2015mvg}. However, the distribution of dark matter around supermassive black holes (BHs) is unclear \cite{Bullock:2017xww}. Therefore, studying the distribution of dark matter around black holes is a very significant project. The dark matter around the black hole is usually characterized by its density parameter and characteristic radius. It is generally believed that the distribution of dark matter around black holes is a big ``spike” \cite{Gondolo:1999ef,Sadeghian:2013laa,Fields:2014pia}. Fortunately, with our efforts, we have derived black hole (BH) spacetime metrics both in a dark matter halo \cite{Xu_2018} and a dark matter spike \cite{Xu:2021dkv}, respectively, and generalized them to the case of rotation. In addition, through the black hole photos released by the Event Horizon Telescope (EHT) \cite{EventHorizonTelescope:2019dse} and the observation of gravitational waves by the Laser Interferometer Gravitational Wave Observatory Scientific Collaboration and Virgo Collaboration (LIGO Scientific/Virgo) \cite{LIGO1,LIGO2,LIGO3,LIGO4}, the existence of black holes can be basically determined.\\
\indent On the other hand, the term ``black hole” was originally coined by Wheeler \cite{Ruffini:1971bza}. For a black hole, he proposed the famous ``No-hair theorem” \cite{1973grav}. The object described by the No-hair theorem is a static isolated black hole. However, in our universe, it is hard to find an isolated black hole, suggesting that there may be various complex matter fields around the black hole, causing the black hole to be in a perturbed state. When a black hole is perturbed by the matter field, its initial perturbation can be represented by a complex frequency of excited oscillation mode, which is so-called quasinormal modes (QNM) \cite{Moderski:2005hf}. This perturbation is usually divided into three stages, the second stage is the QNM. C. V. Vishveshwara show that when a black hole is in a perturbed state, it can emit gravitational waves \cite{Vishveshwara:1970zz}. And this kind of gravitational wave is just dominated by the QNM with a complex frequency. The real part of this frequency represents the oscillation frequency of the black hole when perturbed, while the imaginary part represents the rate of oscillation, also known as damping \cite{Konoplya:2011qq}. This QNM mode is directly related to the oscillating properties of black holes, so the study of QNM is helpful to understand the information about black holes. Besides, this QNM is an important source of gravitational waves from the supernova collapse. Kostas D. Kokkotas et al. show that the gravitational collapse of massive rotating stars and the merger of the binary star system can generate gravitational waves which are directly related to the QNM \cite{Kokkotas:1999bd}. Now, there are many methods to calculate the QNM, such as WKB method \cite{Schutz:1985km,PhysRevD.35.3621}, P{\"o}schl-Teller potential approximation \cite{Poschl:1933zz,Ferrari:1984zz,Churilova:2021nnc} and continued fraction method \cite{Leaver:1985ax,Dolan:2007mj}. The QNM is a characteristic ``sound” of a black hole which can provide us with a new method to verify a black hole in our universe \cite{Konoplya:2011qq}.

Besides, according to the fact that no matter can escape from a black hole, the solution at the event horizon of the black hole is a pure incoming wave. However, for the behavior of the solution of the equation at infinity, it can be divided into two types. One of types is QNM we introduced before and another is the quasibound states (QBS). The solution of QNM is a pure outgoing wave when it is far away from the black hole, while QBS is only shown as an exponential decay when it is far away from the black hole \cite{Dolan:2007mj}. According to the superradiant phenomenon, a rotating black hole produces wave amplification, and when the perturbation field is large, the superradiant phenomenon can cause instability \cite{Siqueira:2022tbc}. This superradiant instability is closely related to the existence of QBS, which may have important astrophysical implications. Besides, if a rotating black hole is trapped inside an perfectly reflective cavity, it can cause a phenomenon called the ``black hole bomb” \cite{Press:1972zz,Cardoso:2004nk}. That's because any initial perturbations would be magnified near the black hole and reflected back to the mirror, creating instability, which grows exponentially with the time. Through this amplified scattered wave, we can extract certain rotational energy from the ergosphere of the black hole. However, it should be noted that the rotation energy of a black hole is limited, and the process of extracting energy should take into account the backreaction effects. In fact, these studies show that Kerr black holes are prone to instabilities under massive fields because the mass term effectively suppresses the field \cite{Damour:1976kh,Detweiler:1980uk,Cardoso:2005vk,Hod:2012px}. Here, what we are interested in is whether the massive field of the Kerr-like black hole in the dark matter halo will also produce the instability similar to the black hole bomb. This is one of our study points.

In recent years, the QBS of black holes \cite{Yoshino:2013ofa,Konoplya:2006br,Myung:2022krb,Mai:2021yny,Huang:2020pga,Vieira:2022pxd,Vieira:2021nha} and the QNM  \cite{Gundlach:1993tp,PhysRevD.63.084014,PhysRevD.64.044024,PhysRevD.77.124007,Fernandes:2021qvr,McManus:2019ulj,Assumpcao:2018bka,Konoplya_2019,Zhou:2021cef,Gonzalez:2017zdz,Javed:2021ymu,Ponglertsakul:2020ufm} have been extensively studied. Among the recent researches connected to our work, we list the following papers. In Ref.\cite{Dolan:2007mj}, they study the instabilities of rotating black holes in the massive scalar field. M. Richartz et al. use the scalar field perturbation to study the eigenfrequencies of the Kerr-like black holes \cite{Siqueira:2022tbc}. Besides, with the release of black hole photos, people are also increasingly concerned about the interaction of other celestial bodies (or matter) around the black hole \cite{Vicente:2022ivh,pantig2022a}. Cardoso et al. introduce an exact solution for a black hole immersed in a galactic-like distribution of matter and use gravitational perturbations to study the quasinormal modes of this black hole \cite{Cardoso:2021wlq}. Based on this, Konoplya studies the matter field perturbations, the greybody factors and the Unruh temperature \cite{Konoplya:2021ube} of the exact solution of this black hole. On the other hand, an interesting and important question is how these parameters of dark matter affect the QNM/QBS. These dark matter parameters generally come from different galaxies, and they are usually the results obtained through fitting processing on the basis of observation data \cite{Robles:2012uy,Fernandez-Hernandez:2017pgq}. In recent years, the research on black hole immersed in dark matter and Schwarzschild black has received extensive attention \cite{Daghigh:2022pcr,Liu:2022lrg,Konoplya:2022hbl,Qin:2023nog,Bamber:2022pbs}. C. Zhang et al. studied the QNM of a spherically symmetric black hole in a dark matter halo using gravitational perturbation \cite{Zhang:2021bdr,Zhang:2022roh}. However, there are relatively few studies between black hole in dark matter and Kerr black hole. Comparing the differences between them is also a meaningful and important work. As the QNM/QBS of the Kerr black hole have become more familiar, these comparisons can help us understand black holes in a dark matter halo to some extent. So, in this paper, we will also try to study these questions. In the longer term, with the rapid development of gravitational wave detectors in the future, it is possible to detect signals from such black holes immersed in dark matter, which also provides an effective method for testing dark matter models.\\
\indent In this paper, we mainly use the scalar field perturbation to study QNM/QBS of the Kerr-like black holes in a dark matter halo, and use the continued fraction method to calculate QNM/QBS frequencies. For the same time, these results are compared with the Kerr black hole. We will also analyze and verify the existence of the superradiant instabilities of black holes. In addition, we will also study the impacts of dark matter parameters on the QNM/QBS of black holes at the specific circumstances. This work is an in-depth study based on our previous work \cite{Xu:2017bpz,Xu:2017vse,Liu:2021xfb}. Although these previous works are on spherically symmetric black holes, these valuable experiences give us enough confidence to solve the case of axisymmetric black holes.\\
\indent This paper is organized as follows. In Section \ref{s2}, we briefly introduce the spacetime properties of Kerr-like black holes in a dark matter halo, and the discussion of extremal black holes of them. In particular, we unify the dark matter parameters into the black hole units. In Section \ref{s3}, we introduce scalar field perturbation of this Kerr-like black holes, and show that how to use ansatz to decompose the complex perturbation equations into radial and angular equations. In Section \ref{s4}, we mainly introduce the continued fraction method, which includes the asymptotic solution of the Kerr-like black hole oscillation behavior at the specific boundary conditions, the derivation of the $3$-term recurrence formula of the radial/angular equations and the continued fraction equation. In Section \ref{s5}, we mainly use the continued fraction method to calculate the QNM/QBS frequencies of the Kerr-like black holes both in CDM/SFDM models, and compare their results with Kerr black hole. At the same time, we will verify the existence of superradiant instabilities and give the figures of the maximum instabilities of the Kerr-like black holes in a dark matter halo. In addition, we will also study the impacts of dark matter parameters on the QNM/QBS of black holes at the specific circumstances. Finally, Section \ref{s6} is our conclusions and discussions. In this paper, we use mostly the black hole units that $G=c=M_\text{BH}=r_{\text{BH}}=1$ and the radius of black hole (BH) is given by $r_{\text{BH}}=GM_\text{BH}/c^2$.

\section{The spacetime of the Kerr-like BHs in a dark matter halo}\label{s2}
In this section, we will review the Kerr-like black hole metrics we obtained in a dark matter halo \cite{Xu_2018}. Both of them have the following form in four-dimensional coordinates,
\begin{equation}
\begin{aligned}
ds^{2}=&-\left(1-\frac{r^{2}+2Mr-r^{2}f(r)}{\Sigma^{2}}\right) d t^{2} +\frac{\Sigma^{2}}{\Delta} d r^{2}+\Sigma^{2} d \theta^{2}+\frac{A\sin ^{2} \theta}{\Sigma^{2}} d \phi^{2} \\
&-\frac{2 (r^{2}+2Mr-r^{2}f(r)) a \sin ^{2} \theta}{\Sigma^{2}} d \phi d t  ,
\label{e21}
\end{aligned}
\end{equation}
with
\begin{equation}
\Delta= r^{2}f(r)-2Mr+a^{2}, \quad  \Sigma^2=r^2+a^2\cos(\theta )^2, \quad A=\left(r^{2}+a^{2}\right)^{2}- a^{2} \Delta   \sin ^{2} \theta ,
\label{e22}
\end{equation}
where, $M$ is the mass of a black hole and $a$ is the rotation parameter. $f(r)$ represents the factor term for considering dark matter. For cold dark matter (CDM) model, $f(r)$ has the form of the following,
\begin{equation}
f_{\rm c}(r)=\left(1+\frac{r}{R_{\rm c}}\right)^{-\frac{8 \pi  \rho_{\rm c} R_{\rm c}^{3}}{ r}},
\label{e23}
\end{equation}
and for scalar field dark matter (SFDM) model, $f(r)$ has the form of the following,
\begin{eqnarray}
f_{\rm s}(r)=\exp \left(-\frac{8 \rho_{{\rm s}} R_{\rm s}^{2}}{\pi} \frac{\sin (\pi r / R_{\rm s})}{\pi r / R_{\rm s}}\right),
\label{e24}
\end{eqnarray}
here, the parameter $\rho$ is the density of the universe when a dark matter halo collapses, and $R$ means its characteristic radius in this halo.
%%%%%%%%%%%%%%%%%%%%%%%%%%%%%%%%%%%%%%%%%%%%%%%%%%%%%%%%%%%
\begin{table}[t!]
\setlength{\abovecaptionskip}{0cm}
\setlength{\belowcaptionskip}{0.3cm}
\centering
\caption{The values of density parameters $\rho$ and characteristic radius $R$ from different galaxies in CDM model and SFDM model. These data are excerpted in the galaxies from Refs. \cite{Robles:2012uy,Fernandez-Hernandez:2017pgq}.}
\scalebox{0.8}{
\begin{tabular}{clclclclclclclclclclcl}
\hline \hline
\multicolumn{2}{c}{Galaxies}                                   & \multicolumn{4}{c}{ESO1200211}                        & \multicolumn{4}{c}{ESO1870510}                         & \multicolumn{4}{c}{ESO3020120}                         & \multicolumn{4}{c}{ESO3050090}                          & \multicolumn{4}{c}{ESO4880049}                         \\ \hline
\multicolumn{2}{c}{$(\times 10^{-3} M_{\bigodot}/pc^{3}; kpc)$} & \multicolumn{2}{c}{$\rho$} & \multicolumn{2}{c}{$R$}  & \multicolumn{2}{c}{$\rho$} & \multicolumn{2}{c}{$R$}   & \multicolumn{2}{c}{$\rho$} & \multicolumn{2}{c}{$R$}   & \multicolumn{2}{c}{$\rho$} & \multicolumn{2}{c}{$R$}    & \multicolumn{2}{c}{$\rho$} & \multicolumn{2}{c}{$R$}   \\ \hline
\multicolumn{2}{c}{The CDM model}                                & \multicolumn{2}{c}{2.45}   & \multicolumn{2}{c}{5.7}  & \multicolumn{2}{c}{0.761}  & \multicolumn{2}{c}{31.82} & \multicolumn{2}{c}{2.65}   & \multicolumn{2}{c}{19.72} & \multicolumn{2}{c}{0.0328} & \multicolumn{2}{c}{705.67} & \multicolumn{2}{c}{1.42}   & \multicolumn{2}{c}{52.27} \\
\multicolumn{2}{c}{The SFDM model}                               & \multicolumn{2}{c}{13.66}  & \multicolumn{2}{c}{2.92} & \multicolumn{2}{c}{32.55}  & \multicolumn{2}{c}{2.93}  & \multicolumn{2}{c}{22.74}  & \multicolumn{2}{c}{8.86}  & \multicolumn{2}{c}{21.50}  & \multicolumn{2}{c}{4.81}   & \multicolumn{2}{c}{54.29}  & \multicolumn{2}{c}{5.36}\\ \hline \hline
\end{tabular}}
\label{t11}
\end{table}
%%%%%%%%%%%%%%%%%%%%%%%%%%%%%%%%%%%%%%
The corresponding dark matter parameters from the different galaxies have been fitted by the Authors in these Refs. \cite{Robles:2012uy,Fernandez-Hernandez:2017pgq}, and we present some of the dark matter parameters in Table \ref{t11}. As can be seen from this table, these dark matter parameters correspond to specific galaxies. They are the result of a fit to the observed data. Here, we mainly take the Low Surface Brightness (LSB) galaxy ESO$1200211$ as our research object, and compare it with Kerr black hole.\footnote{Unless otherwise specified, in the following discussion, the values we used of the density parameter $\rho_{\text{c/s}}$ and characteristic radius $R_{\text{c/s}}$ both in CDM/SFDM models are selected from the LSB galaxy ESO1200211. The subscripts c and s represent the first letters of CDM model and SFDM model, respectively.} The mass of black hole at the center of the LSB galaxy is approximately $M_{\text{BH}} = 5.62 \times 10^{6}M_{\bigodot}$ \cite{Subramanian:2015siv}. For CDM model in the galaxy ESO$1200211$, the density parameter $\rho_c=2.45 \times 10^{-3}$ $M_{\bigodot}/pc^{3}$ and the characteristic radius $R_c=5.7$ $kpc$. The values of $R_\text{c}$ and $\rho_\text{c}$ in black hole units are $R_\text{c}=5.7 kpc/(G \times M_{\text{BH}}/c^2)\approx 2.05 \times 10^{10} r_{\text{BH}} = 2.05 \times 10^{10}$ and $\rho_\text{c}=2.45 \times 10^{-3}M_{\bigodot}/pc^3/(M_\text{BH}/(4/3\times \pi\times (G \times M_{\text{BH}}/c^2)^3))=3.90 \times 10^{-29} \rho_{\text{BH}}=3.90 \times 10^{-29}\times\frac{M_\text{BH}}{4/3\times \pi \times r_{BH}^3} \approx 9.33\times10^{-30}$. For SFDM model in the galaxy ESO$1200211$, the density parameter $\rho_s=13.66 \times 10^{-3}$ $M_{\bigodot}/pc^{3}$ and the characteristic radius $R_s=2.92$ $kpc$. The values of $R_\text{s}$ and $\rho_\text{s}$ in black hole units are $R_\text{s}=2.92 kpc/(G \times M_{\text{BH}}/c^2)\approx 1.05 \times 10^{10} r_{\text{BH}} = 1.05 \times 10^{10}$ and $\rho_\text{s}=13.66\times 10^{-3}M_{\bigodot}/pc^3/(M_\text{BH}/(4/3\times \pi\times (G \times M_{\text{BH}}/c^2)^3))=2.17 \times 10^{-28} \rho_{\text{BH}}=2.17 \times 10^{-28} \times\frac{M_\text{BH}}{4/3\times \pi \times r_{BH}^3} \approx 5.20\times10^{-29}$. In this way, all the black hole parameters in Eq.(\ref{e21}) are guaranteed to be in black hole units. On the other hand, if the impacts of dark matter on black holes is not considered, that is, $\rho=0$, then the dark matter term will become $f(r)=1$. The Kerr-like black hole metric in a dark matter halo will degenerate into the Kerr metric.
\begin{figure*}[t!]
\centering
{
\includegraphics[width=0.31\columnwidth]{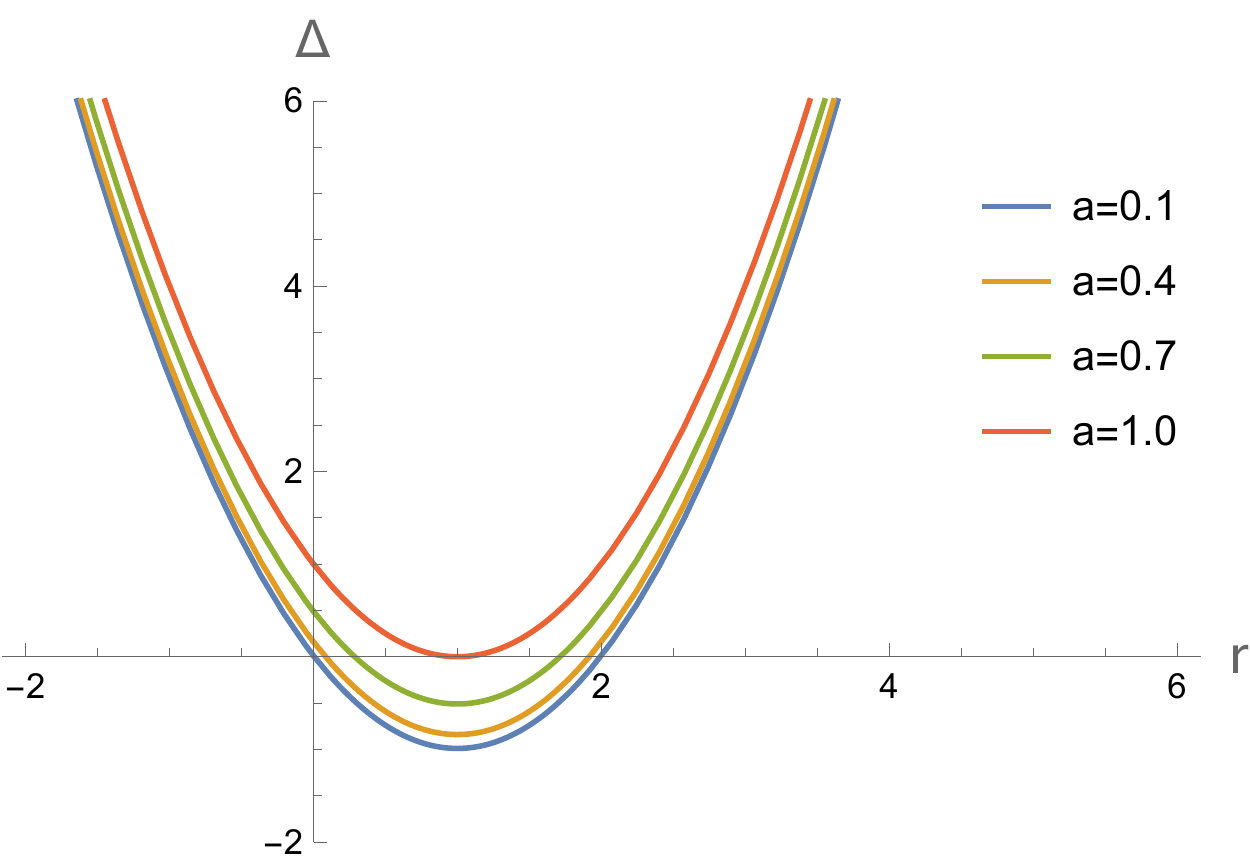}
}
{
\includegraphics[width=0.31\columnwidth]{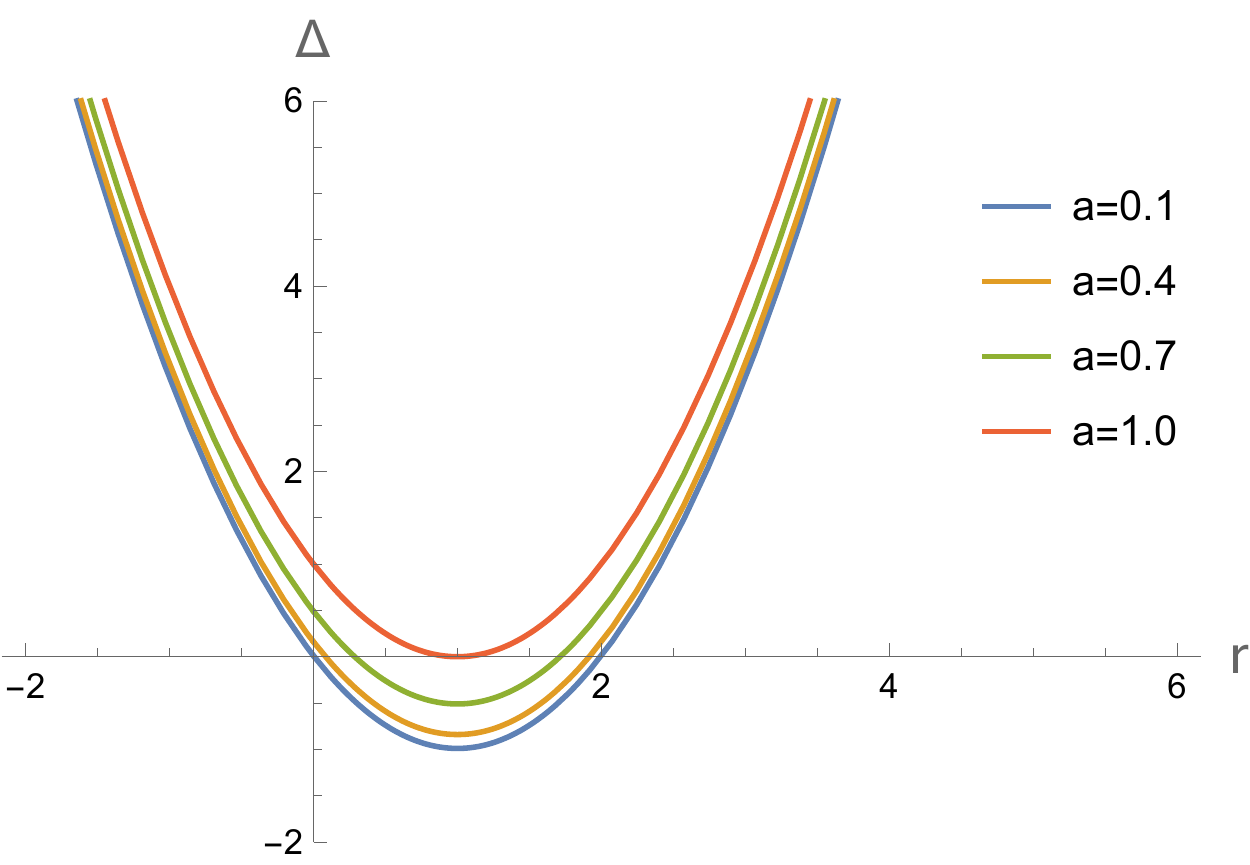}
}
{
\includegraphics[width=0.31\columnwidth]{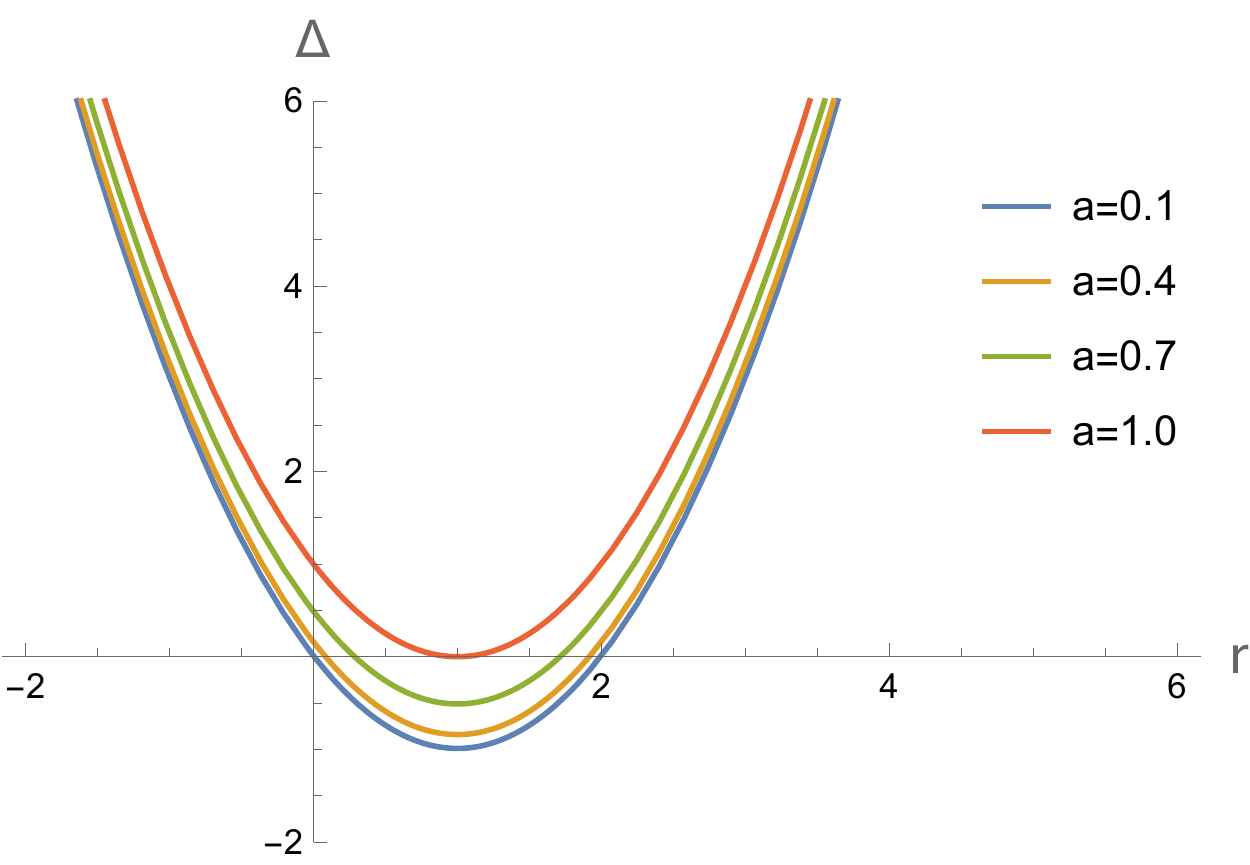}
}
\caption{The functional image of $\Delta$ as a function of independent variable $r$ in CDM model (left panel), SFDM model and the Kerr spacetime (right panel) , respectively. The function image of $\Delta$ for different rotation parameters $a$ are given in each panel. The main calculation parameters are $M=1$, $\rho_c=2.45 \times 10^{-3}$ $M_{\bigodot}/pc^{3}$, $R_c=5.7$ $kpc$, $\rho_s=13.66 \times 10^{-3}$ $M_{\bigodot}/pc^{3}$, $R_s=2.92$ $kpc$. We have converted these main calculation parameters to the black hole units before plotting.}
\label{nf1}
\end{figure*}
Besides, in these two black holes, the location of the event horizons can be obtained from Eq. (\ref{e22}),
\begin{equation}
\Delta= r^{2}f(r)-2Mr+a^{2} \approx (r-r_+)(r-r_-),
\label{e25}
\end{equation}
where, $r_+$ and $r_-$ are the outer and inner horizon of this rotating black hole, respectively. Here, we also present pictures of the root of the $\Delta$ function for these three types of event horizon in Figure \ref{nf1}. We find that with the rotation parameter $a$ increases, the roots of the $\Delta$ function gradually change from two to one, and then to zero. This process indicates the transition from a rotation black hole to an extremal black hole. For the case of extremal black hole, it appears as the coincidence of the inner and outer horizons, and thermodynamically, it appears as Hawking temperature equals to zero \cite{Biswas:2021gvq}. Next, let's calculate the extremal value of the rotation parameter $a$ of the extremal black hole. Using the following ansatz, $a^2$ can be rewritten as a function of $r$ in black hole units $(M=1)$,
\begin{equation}
a^2(r)=- r^{2}f(r)+2r.
\label{e26}
\end{equation}
According to the definition of Eq.(\ref{e26}), the coefficient of the highest order term of the independent variable $r$ is negative. Then, the functional image of Eq. (\ref{e26}) opens downwards and then this function has a maximum value. So, the extremal value of the rotation parameter $a$ can be transformed into the maximal value of Eq. (\ref{e26}). At this time, we only need to solve its first derivative, that is, $d a^2(r)/dr=0$. Then, taking the maximum value $r$ back into Eq. (\ref{e26}) to get the maximum value of $a^2$. Finally, after square root, the extremal value of the rotation parameter $a$ can be obtained. In black hole units, for CDM and SFDM models, the rotation parameters obtained by the numerical calculation are $a_c \approx 1.00000021$ and $a_s \approx 1.000000031$ with the mass $M=1$. The rotation parameter of black hole in a dark matter halo are very close to those of Kerr black hole $(a_k=1)$, but they are larger than that of Kerr black hole. Note that Eq. (\ref{e25}) approximates the transcendental function with a quadratic function. As an example, we will discuss and analyze the possibility of this approximation mainly from $f(r)$ and $\Delta(r)$ with the parameters $\rho_c=2.45 \times 10^{-3}$ $M_{\bigodot}/pc^{3}$, $R_c=5.7$ $kpc$ in CDM model and the parameters $\rho_s=13.66 \times 10^{-3}$ $M_{\bigodot}/pc^{3}$, $R_s=2.92$ $kpc$ in SFDM model.  Firstly, for $f(r)$, we plot the functional image of $f(r)$ in CDM and SFDM models in Figure \ref{nnf2}, as a function of $r$. Our results show that $f(r)$ is almost $f(r)=1$ over the whole space, but slightly less than $1$. The biggest difference $\delta f(r)$ between $f_{c/s}(r)$ and $f(r)=1$ is approximately $1 \times 10^{-7}$. In other words, it is possible to replace this transcendental function with a quadratic function. However, to ensure the rationality of the results, we set the minimum number of digits of precision of the calculation results to $20$ in the following discussion. Secondly, we set the variable $\delta \Delta (r) = r^{2}f(r) - 2Mr + a^{2} - (r-r_+)(r-r_-)$. Then, we take the numerical results of the inner and outer horizons into $\delta \Delta (r)$, and record the results in the following Tables \ref{nnt1} and \ref{nnt2}. Our results show that within the scope of the inner and outer horizons, the maximum difference $\delta \Delta(r)$ between the transcendental function and the quadratic function is approximately $\delta \Delta(r)=$ $3.73 \times 10^{-13}$, or even smaller. This difference $\delta \Delta(r)$ relative to $\Delta(r)$ itself is so small that it is negligible. Therefore, based on these two points, in the discussion of this work, there is some rationality when Eq. (\ref{e25}) is established.\footnote{Similarly, using this method we introduced above can test the rationality of other dark matter parameters within the Eq. (\ref{e25}).}
\begin{figure*}[t!]
\setlength{\abovecaptionskip}{0cm}
\setlength{\belowcaptionskip}{0cm}
\centering
{
\includegraphics[width=.45\columnwidth]{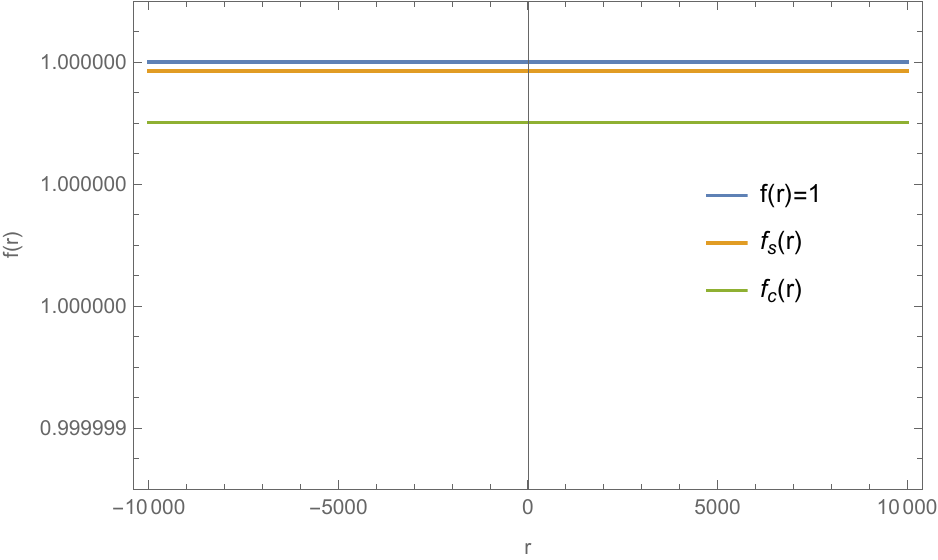}
}
{
\includegraphics[width=.45\columnwidth]{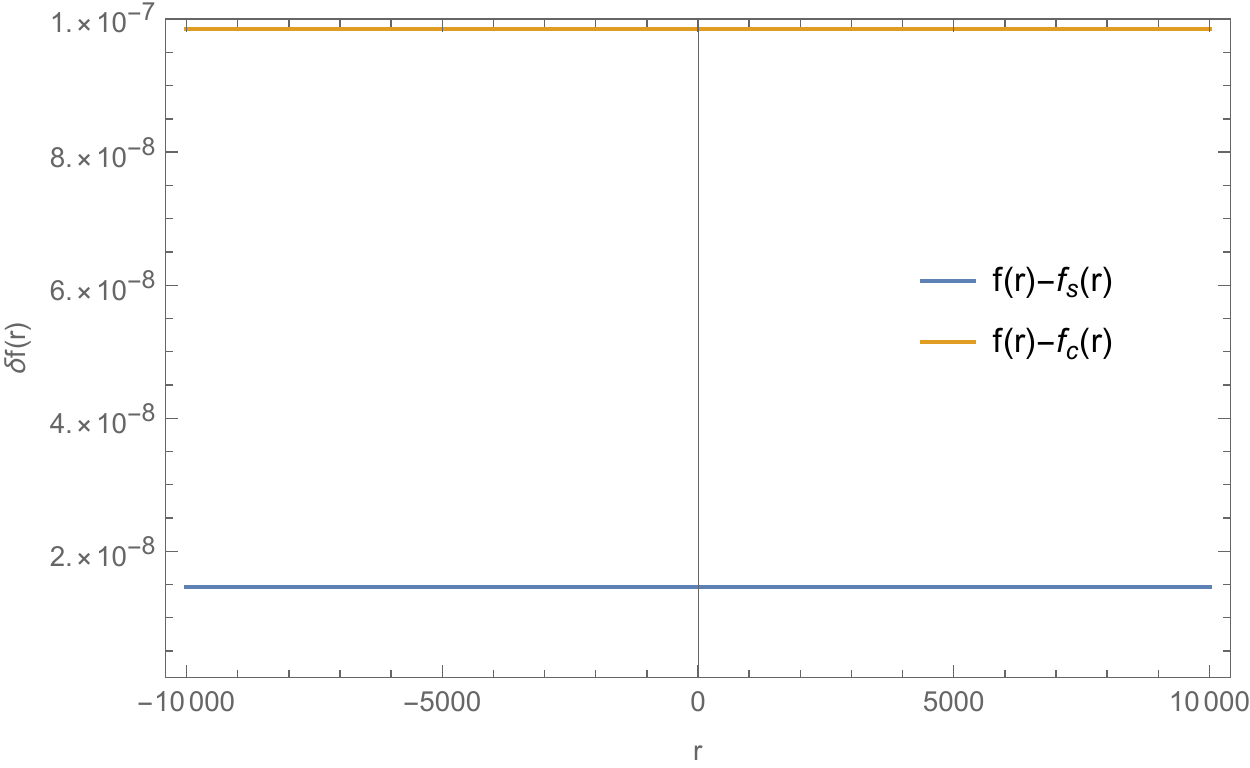}
}
\caption{The functional image of $f(r)$ and $\delta f(r)$ of the Kerr-like black hole in SFDM, CDM models. The main calculation parameters are $M=1$, $a=0.5$, $\rho_c=2.45 \times 10^{-3}$ $M_{\bigodot}/pc^{3}$, $R_c=5.7$ $kpc$, $\rho_s=13.66 \times 10^{-3}$ $M_{\bigodot}/pc^{3}$, $R_s=2.92$ $kpc$.}
\label{nnf2}
\end{figure*}
%%%%%%%%%%%%%%%%%%%%%%%%%%%%%%%%%%%%%%%%%%%%%%%%%%%%
\begin{table}[t!]
\setlength{\abovecaptionskip}{0cm}
\setlength{\belowcaptionskip}{0cm}
\centering
\caption{The variable $\delta \Delta (r)$ with different rotation parameter $a$ in SFDM model. The main calculation parameters are $M=1$, $\rho_s=13.66 \times 10^{-3}$ $M_{\bigodot}/pc^{3}$, $R_s=2.92$ $kpc$.}
\scalebox{.8}{
\begin{tabular}{ccccccccc}
\hline \hline
$a$ &  & $r_-$      &  & $r_+$   &  & $\mid\delta \Delta_{r_+}\mid$                         &  & $\mid\delta \Delta_{r_-}\mid$               \\ \hline
0.1 &  & 0.005012562893195716   &  & 1.9949874663047338 &  & $4.44089 \times 10^{-16}$                    &  & $1.89058\times10^{-18}$ \\
0.2 &  & 0.020204102883687628   &  & 1.9797952631424170  &  & $4.44089\times10^{-16}$                      &  & $4.66207\times10^{-18}$ \\
0.3 &  & 0.046060798566820020   &  & 1.9539392306311094 &  & $4.44089\times10^{-16}$                      &  & $2.86229\times10^{-17}$ \\
0.4 &  & 0.083484860953322430   &  & 1.9165151682446070 &  & 0.                                             &  & $2.77556\times10^{-17}$ \\
0.5 &  & 0.133974596064272700   &  & 1.8660254331336568 &  & $4.44089\times10^{-16}$                      &  & $1.04083\times10^{-17}$ \\
0.6 &  & 0.199999996350258800   &  & 1.8000000295629035     &  & $4.44089\times10^{-16}$                      &  & 0.                                   \\
0.7 &  & 0.285857156310486860   &  & 1.7141428728874426 &  & 0.                                             &  & $1.38778\times10^{-16}$ \\
0.8 &  & 0.399999998053471400   &  & 1.6000000311444580     &  & 0.                                             &  & $1.11022\times10^{-16}$ \\
0.9 &  & 0.564110100316966000   &  & 1.4358899288809635 &  & 0.                                             &  & $5.55112\times10^{-17}$ \\
1.0 &  & 0.999879188422415000   &  & 1.0001208407755144 &  & $2.22045\times10^{-16}$                      &  & 0.\\
\hline \hline
\end{tabular}}
\label{nnt1}
\end{table}
%%%%%%%%%%%%%%%%%%%%%%%%%%%%%%%%%%%%%%%%%%%%%%%%%%%%%%5
\begin{table}[t!]
\setlength{\abovecaptionskip}{0cm}
\setlength{\belowcaptionskip}{0cm}
\centering
\caption{The variable $\delta \Delta (r)$ with different rotation parameter $a$ in CDM model. The main calculation parameters are $M=1$, $\rho_c=2.45 \times 10^{-3}$ $M_{\bigodot}/pc^{3}$, $R_c=5.7$ $kpc$.}
\scalebox{.8}{
\begin{tabular}{ccccccccc}
\hline \hline
$a$ &  & $r_-$      &  & $r_+$   &  & $\mid\delta \Delta_{r_+}\mid$     &  & $\mid\delta \Delta_{r_-}\mid$           \\ \hline
0.1 &  & 0.0050125628921358140   &  & 1.9949876341954378 &  & $3.73923\times10^{-13}$ &  & $4.48213\times10^{-16}$  \\
0.2 &  & 0.0202041028662000946   &  & 1.9797960942213726  &  & $1.99840\times10^{-14}$    &  & $3.66699\times10^{-15}$  \\
0.3 &  & 0.0460607984734717900   &  & 1.9539393986141016 &  & $3.32179\times10^{-13}$  &  & $7.67615\times10^{-17}$  \\
0.4 &  & 0.0834848606347179000   &  & 1.9165153364534340 &  & $1.06581\times10^{-14}$   &  & $1.47920\times10^{-14}$  \\
0.5 &  & 0.1339745951943550600   &  & 1.8660256018932184 &  & $2.84217\times10^{-14}$   &  & $2.51084\times10^{-14}$   \\
0.6 &  & 0.1999999975364056300   &  & 1.8000001995511679     &  & $4.84057\times10^{-14}$   &  & $3.97529\times10^{-14}$   \\
0.7 &  & 0.2858571515078815500   &  & 1.7141430455796920 &  & $2.24709\times10^{-13}$   &  & $4.17583\times10^{-14}$  \\
0.8 &  & 0.3999999868608307000   &  & 1.6000002102267428     &  & $3.55271\times10^{-13}$  &  & $2.02061\times10^{-14}$  \\
0.9 &  & 0.5641100696751329000   &  & 1.4358901274124407 &  & $3.06422\times10^{-13}$  &  & $2.34257\times10^{-14}$  \\
1.0 &  & 0.9996861816814463000   &  & 1.0003140154061272 &  & $1.02363\times10^{-13}$  &  & $6.97220\times10^{-14}$\\
\hline \hline
\end{tabular}}
\label{nnt2}
\end{table}
Next, to make our expression more compact, we continue to use $f(r)$ to represent the dark matter term. Then, we can obtain a covariant metric tensor $g_{\mu\nu}$ from the Eq.(\ref{e21}),
\begin{equation}
g_{\mu \nu}=\left(
\begin{array}{cccc}
 -(1-\frac{r^2+2Mr-r^2 f(r)}{\Sigma^2 }) & 0 & 0 & -\frac{a \sin ^2(\theta ) \left(r^2+2Mr-r^2 f(r)\right)}{\Sigma^2 } \\
 0 & \frac{\Sigma^2 }{\Delta } & 0 & 0 \\
 0 & 0 & \Sigma^2  & 0 \\
- \frac{a \sin ^2(\theta ) \left(r^2+2Mr-r^2 f(r)\right)}{\Sigma^2 } & 0 & 0 & \frac{A \sin ^2(\theta )}{\Sigma^2 } \\
\end{array}
\right).
\label{e27}
\end{equation}
\noindent With the Eq.(\ref{e27}), we can calculate the determinant of this metric,
\begin{eqnarray}
g=det(g_{\mu\nu})=-\Sigma^4\sin(\theta )^2 .
\label{e28}
\end{eqnarray}
From Eqs.(\ref{e27}) and (\ref{e28}), we get the contravariant form of the metric,

\begin{equation}
g^{\mu \nu}=\left(
\begin{array}{cccc}
 -\frac{A}{\Delta \Sigma^2 } & 0 & 0 & \frac{a r (r f(r)-2 M-r)}{\Delta  \Sigma^2 } \\
 0 & \frac{\Delta}{\Sigma^2 } & 0 & 0 \\
 0 & 0 & \frac{1}{\Sigma^2 } & 0 \\
 \frac{a r (r f(r)-2 M-r)}{\Delta \Sigma^2 } & 0 & 0 & \frac{ \left(r^2 f(r)-2 M r-r^2+\Sigma^2 \right)}{\Delta  \Sigma^2 \sin ^2(\theta ) } \\
\end{array}
\right) .
\label{e29}
\end{equation}

\section{Scalar field perturbation of the Kerr-like BHs in a dark matter halo}\label{s3}
In this section, we will study the scalar field perturbation of Kerr-like black hole and derive the radial and angular equations of scalar particles in a dark matter halo. In a curved spacetime, the equation of the motion of scalar particles can be described by the Klein-Gordon (K-G) equation,
\begin{equation}
\frac{1}{\sqrt{-g}}{\partial_\sigma}(\sqrt{-g}g^{\sigma\nu} {\partial _\nu}\Psi)=\mu ^{2}\Psi ,
\label{e31}
\end{equation}
where $\mu$ is the mass of the scalar particle. With Eqs.(\ref{e28}) and (\ref{e29}), we can get the following form,
\begin{equation}
\begin{aligned}
&-\frac{A}{\Delta \Sigma^2 }\partial ^2_t\Psi +\frac{ar(-2M-r+rf(r))}{\Delta  \Sigma ^2  }\partial _t\partial _\phi \Psi +\frac{1}{\Sigma ^2  \sin(\theta )}\partial _\theta (\sin(\theta )\partial _\theta )\Psi +\frac{1}{\Sigma ^2 }\partial _r(\Delta  \partial _r)\Psi \\
&+\frac{ar(-2M-r+rf(r))}{\Delta  \Sigma  ^2 }\partial _\phi \partial _t \Psi + \frac{-2Mr-r^2+\Sigma ^2+r^2f(r)}{\Delta  \Sigma ^2  \sin(\theta )^2}\partial _\phi ^2\Psi =\mu^2 \Psi.
\label{e32}
\end{aligned}
\end{equation}
Eq.(\ref{e32}) is a complex second-order partial differential equation, but it can be separated by using the following ansatz,
\begin{equation}
\Psi (t,r,\theta ,\phi )=e^{-i\omega t}e^{im\phi }R(r)S(\theta ),
\label{e33}
\end{equation}
where, $m$ is the azimuthal quantum number and $\omega$ is the frequency of this system. Therefore, we can get ordinary differential equations about angular and radial parts. If we introduce the variable $x=\cos \theta$, with $x\in [-1,1]$, the angular part can be written as
\begin{equation}
\begin{aligned}
(1-x^2)\frac{d^2S(x)}{dx^2}-2x\frac{dS(x)}{dx} +\left(\Lambda _{lm} + a^2k^2x^2-\frac{m^2}{1-x^2} \right)S(x)=0,
\label{e34}
\end{aligned}
\end{equation}
where, $k^2=\omega ^2 - \mu^2$ and $\Lambda_{lm}$ is a separation constant. Eq.(\ref{e34}) is also known as the spheroidal equation, and $\Lambda_{lm}$ is the eigenvalue of this equation \cite{Berti:2005gp}. Normally, the eigenvalue $\Lambda_{lm}$ has no analytical expression in this case. A simple method is to calculate its value by calling the SpheroidalEigenvalue command in Mathematica. However, in the case of non-rotating limit, the spheroidal function can be reduced to a spherical harmonic function, that is, $S_{lm} \rightarrow  Y_{lm}$, and $\Lambda_{lm}=l(l+1)$. The eigenvalue at this time can be uniquely determined, and it is related to the angular quantum number $l$.\\
\indent Another is radial equation, and it can be written as
\begin{equation}
\begin{aligned}
\Delta ^2\frac{d^2R(r)}{dr^2}+\Delta \frac{d\Delta }{dr}\frac{dR(r)}{dr} +\left ( K^2(r)-(\lambda +\mu _0^2r^2)\Delta  \right )R(r)=0
\label{e35}
\end{aligned}
\end{equation}
where, $\Delta=(r-r_+)(r-r_-)$, $K(r)=\omega(r^2+a^2)-am$ and $\lambda=\Lambda_{lm}+a^2\omega^2-2am\omega$. There are two eigenvalues $\Lambda_{lm}, \omega$ in this equation. Among them, the eigenvalue $\omega$ is the frequency of this system, which is a complex number. The real part of the frequency represents the oscillation frequency and its imaginary part represents the decay rate. Where, the relationship between decay rate and time scale is $\tau^{-1}=M \times \text{Im}(\omega)$, that is growth rate. To solve the radial equation, we need to first determine the separation constant $\Lambda_{lm}$ in the angular equation. In the following section, instead of using the $Mathematica$ function, we will demonstrate how to use the numerical method to determine the eigenvalues both in the angular and radial equations.

\section{Continued fraction method}\label{s4}
In this section, we will introduce the numerical method, that is, the continued fraction method. The rotating black hole is different from the general spherically symmetric black hole because there are two eigenvalues $\Lambda_{lm}$, $\omega$ unclear in Eqs. (\ref{e34}) and (\ref{e35}). Up to now, the continued fraction method is considered to be one of the most direct and effective methods to solve these eigenvalues. Among these two eigenvalues, $\omega$ is the specific frequency of the black hole in Physics, and this frequency is related to the QNM/QBS \cite{Moderski:2005hf,Dolan:2007mj}. So, the continued fraction method also can be used to solve the QNM/QBS of rotating black holes. This method was first used by Leaver to study the QNM of the kerr black holes \cite{Leaver:1985ax}. Next, we will strictly follow the continued fraction method used in these Refs. \cite{Dolan:2007mj,Leaver:1985ax,Konoplya:2006br} and generalize it to apply to our Kerr-like black holes. Firstly, from the angular equation (\ref{e34}), we find that there are two regular singular points ($x=-1$ and $x=1$) and one irregular singular point $(x=\infty)$. Therefore, their asymptotic behaviors at the location of the singular points can be written as
\begin{equation}
\begin{aligned}
\lim_{x\rightarrow -1}S(x) \sim (1+x)^{\frac{\left | m \right |}{2}}, \quad
\lim_{x\rightarrow +1}S(x) \sim (1-x)^{\frac{\left | m \right |}{2}}.
\end{aligned}
\label{e41}
\end{equation}
Taking Eq.(\ref{e41}) into account, this eigenfunction $S(x)$ to Eq.(\ref{e34}) has the following series solution,
\begin{equation}
S(x)=\exp(a k x)(1-x)^{\frac{\left | m \right |}{2}}(1+x)^{\frac{\left | m \right |}{2}}\sum_{n=0}^{\infty }b_n(1+x)^n
\label{e42}
\end{equation}
Putting Eq.(\ref{e42}) into Eq.(\ref{e34}) for calculation, it can be found that $b_n$ must satisfy the following $3$-term recurrence relation,
\begin{equation}
\begin{cases}
 & \alpha _0b_1+\beta _0b_0=0, \\
 &  \alpha _nb_{n+1}+\beta _nb_n+\gamma _nb_{n-1}=0, \quad n\geq 1
\end{cases}
\label{e43}
\end{equation}
where, $b_0=1$ and the coefficients $\alpha _n$, $\beta _n$, $\gamma _n$ are as follows,
\begin{equation}
\begin{aligned}
\begin{cases}
 & \alpha _n= -2(n+1)(\left | m \right |+n+1),\\
 & \beta _n=-\Lambda _{lm}+\left | m \right | (-2ak+2n+1)-ak(ak+2)+m^2+n^2-4akn+n, \\
 & \gamma _n=  2ak(\left | m \right |+n).
\end{cases}
\end{aligned}
\label{e44}
\end{equation}
If we define the convergence series $R_{n} = b_{n}/b_{n-1}$, the recurrence relation can be rewritten as
\begin{equation}
R_n=\frac{-\gamma _n}{\beta _n+\alpha _nR_{n+1}}.
\label{e45}
\end{equation}
With Eq.(\ref{e43}), we also have $R_1=b_1/b_0=-\beta_0/\alpha_0$. So, we can get the main equation of the continued fraction method,
 \begin{equation}
\begin{aligned}
0=\beta _0-\frac{\alpha _0\gamma _1}{\beta _1-\frac{\alpha _1\gamma _2}{\beta _2-\frac{\alpha _2\gamma _3}{\beta _3-\cdots  }}}.
\end{aligned}
\label{e46}
\end{equation}
Given $a$, $k$, $m$, these coefficients only depend on the eigenvalue $\Lambda_{lm}$. So, the infinite continued fraction is an equation for $\Lambda_{lm}$, and $\Lambda_{lm}$ is a root of the continued fraction equation or any of its inversions \cite{Leaver:1985ax,Berti:2005gp}.\\% Schr\"{o}dinger-like
\indent Similarly, a solution $R(r)$ of the radial equation (\ref{e35}) can also be found since it is a spherical wave equation with similar boundary conditions as Eq. ({\ref{e34}}).
Therefore, we can also use the continued fraction method to solve the eigenfrequency $\omega$ in Eq. (\ref{e35}) under the help of previous discussion. On the other hand, for the perturbation theory of the black hole, the essence of the radial equation can be simplified as a wave equation. The solution to the wave equation is directly related to the location of the boundary conditions of this system, which at the event horizon and at infinity. For the behavior away from the black hole, it can be divided into two types, namely QNM and QBS. The solution of this equation is generally related to the oscillating mode of the matter field. QNM and QBS are the modes with complex frequencies. Its real part is the oscillation frequency of a black hole and the imaginary part is the decay rate of this oscillation. For the QNM, its solution is usually represented by pure incoming waves at the event horizon and pure outgoing waves at infinity. So, we require that
\begin{equation}
\begin{aligned}
&\lim_{r\rightarrow r_+}R(r)\sim (r-r_+)^{-i\alpha }, \quad \alpha = \frac{ r_+^2+a^2}{r_+ - r_-}(\omega -m\Omega),  \\
&\lim_{r\rightarrow \infty }R(r)\sim  \exp(-\bar{k} r) r^ {\beta-1},  \quad \beta=\frac{(r_+ + r_-)(\mu ^2-2\omega ^2)}{2 \bar{k}},
\end{aligned}
\label{e47}
\end{equation}
where, $\Omega=a/(r_+ ^2+a^2)$, $\bar{k}=\pm\sqrt{\mu^2-\omega^2}$. For QNM, its behavior is pure outgoing waves far away from the black hole $(\text{Re}(\bar{k})<0)$ and for QBS, its behavior becomes exponentially decay away from the black hole $( \text{Im}(\bar{k})>0)$ \cite{Dolan:2007mj,Siqueira:2022tbc}. \\
\indent Now, back to Eq.(\ref{e35}), there are two regular singular points ($r=r_+$, $r=r_-$) and one irregular singular point ($r=\infty$) in this equation. Meanwhile,
taking Eq.(\ref{e47}) into account, the appropriate series solution has the following form
\begin{equation}
\begin{aligned}
R(r)=\exp(-\bar{k}r)(\frac{r-r_+}{r-r_-})^{-i\alpha }(r-r_-)^{\beta-1}\sum_{n=0}^{\infty}a_n(\frac{r-r_+}{r-r_-})^n,
\end{aligned}
\label{e48}
\end{equation}
Putting Eq.(\ref{e48}) into Eq.(\ref{e35}), it can be found that $a_n$ must satisfy the following $3$-term recurrence relation,
\begin{equation}
\begin{cases}
 & \alpha _0 a_1+\beta _0 a_0=0, \\
 &  \alpha _n a_{n+1}+\beta _n a_n+\gamma _n a_{n-1}=0, \quad n\geq 1
\end{cases}
\label{e49}
\end{equation}
where, $a_0=1$ and the coefficients $\alpha _n$, $\beta _n$, $\gamma _n$ are depend on these two eigenvalues $\Lambda_{lm}$ and $\omega$. Since these coefficients are too complex, we do not show them here. The advantage of Eq.(\ref{e48}) is that it changes the position of the singular points in the complex plane so that, for the variable $\bar{x}=(r-r_+)/(r-r_-)$, the event horizon is at $\bar{x}=0$, infinity is at $\bar{x}=1$, and the other singular points are outside the unit circle. So, to find the eigenvalues $\Lambda_{lm}$ and $\omega$, we have to solve two continued fraction equations as before
\begin{equation}
\beta _0-\frac{\alpha _0\gamma_1}{\beta _1-}\frac{\alpha _1\gamma_2}{\beta _2-}\frac{\alpha _2\gamma_3}{\beta _3-} \cdot \cdot \cdot =0,
\label{e411}
\end{equation}
Finally, this infinite continued fraction equation $R_n$ needs to be truncated at some order $n\in N$ to ensure the convergence of this equation \cite{Yoshino:2013ofa}. Regarding the equation of convergence, Nollert had proposed some solutions to guarantee the convergence rate of this method \cite{Nollert:1993zz,Zhidenko:2006rs}. From the recurrence relation (\ref{e49}), it can be found that the convergence series $a_{n+1}/a_n$ satisfies the following relation at the order $n^{-1}$,

\begin{equation}
\frac{a_{n+1}}{a_n}=1-\sqrt{\frac{-2\bar{k}(r_+-r_-)}{n}}-\left [ \frac{3}{4}+\frac{(r_++r_-)\bar{k}}{2}-2r_+\bar{k}+\frac{\omega ^2(r_++r_-)}{2\bar{k}} \right ]\frac{1}{n}
\label{e412}
\end{equation}
\section{Numerical results}\label{s5}
In this section, we will use the continued fraction method introduced in the previous section to calculate the QNM/QBS of the Kerr-like black holes immersed in a dark matter halo under the scalar field, and compare them with Kerr black hole. We will also analyze and confirm the existence of the superradiant instability of black hole, that is an unstable modes $(\text{Im}(\omega) > 0)$. In addition, we will also study the impacts of dark matter parameters on the QNM/QBS of black holes at the specific circumstances. As previously analyzed, for the angular equation, given $a, k, m$, the continued fraction equation only depend on the eigenvalue $\Lambda_{lm}$. Similarly, for the radial equation, given $M,a,\mu$ and $m$, the continued fraction equation will depend on eigenfrequency $\omega$. In other words, the eigenvalues $\omega$ and $\Lambda_{lm}$ are the roots of these two continued fraction equations. Now, we successfully converted the radial/angular equations into two infinitely continued fraction equations. What needs to be done in the next step is to make a truncation at the appropriate position and ensure the convergence of the equation. So, we will truncate the continued fraction equation at the term $N$ and use a root finding algorithm to solve the equation. Before it, one necessary step still needs to do is that giving the initial guess value of frequency $\omega$. And the accuracy of this initial guess value will directly determine whether the results of QNM/QBS are reliable. It is seem to be a fact that we can usually equate certain physical behaviors of extremely slowly rotating black holes to the that of Schwarzschild black holes in their limit state. H. Witek et al. investigate this issue of the Schwarzschild backgrounds and the slow-rotating limit in their work. Their results are in agreement with the accurate values provided by the slow-rotation \cite{Witek:2012tr}. Besides, in Refs. \cite{Siqueira:2022tbc,Ponglertsakul:2020ufm}, the authors use Schwarzschild fundamental mode as their initial guess to analyze the QNM/QBS behaviors of the Kerr-like black holes. But to the behaviors of QBS, S. R. Dolan et al. have shown that, in the nonrelativistic limit, the frequency of a Schwarzschild black hole of the massive scalar field bound has the following form \cite{Lasenby:2002mc,Dolan:2007mj},
\begin{equation}
\hbar \omega _n\approx (1-\frac{M^2\mu ^2}{2\bar{n}^2})\mu c^2,
\label{e51}
\end{equation}
where, $\bar{n}=n+l+1$ and $\bar{n}$ is the principal quantum number of this system. Therefore, in this paper, we choose the Schwarzschild fundamental mode $(n=0)$ and the slowly rotating parameter $(a=0.0001)$ as our initial guess and then use a root-finding algorithm for these eigenvalues. Finally, we set the truncated term to $(N=400)$ and the minimum number of digits of precision of all the calculation results to $20$. Considering the actual need (distinguishing the difference) for precision in the following analysis, we keep $9$ significant figures for QNM frequencies, and $20$ significant figures for QBS frequencies.

\subsection{Quasinormal modes}
First of all, to ensure the feasibility of our method, it is necessary for us to calculate the frequencies of fundamental quasinormal modes (QNM) for Kerr black hole with our method, and compare them with the data recorded in other published articles. Without loss of generality, we calculate the QNM frequencies of the Kerr black hole at the states of angular quantum number $l=0$ and $l=1$ in the massless scalar field. Note that when the angle quantum number $l=1$, the value of the azimuthal quantum number $m$ should be in three values $m=-1,0,1$. Finally, we implemented the continued fraction algorithm and our results about the massless scalar field giving in Table \ref{t1} are in good agreement with the data recorded in Refs.\cite{Leaver:1985ax,Konoplya:2006br,Dolan:2007mj}. Among these references, the authors all used the continued fraction method to calculate the QNM frequencies of rotating black holes. These evidences show that our method can provide an important guarantee for the calculation of the QNM frequencies in a dark matter halo. %Besides, we also give the QNM frequencies for the states $l=0,m=0$ and $l=1,m=0$ in Table \ref{t1}.
\begin{table*}[tbp]
\setlength{\abovecaptionskip}{0cm} %%%tiaozhengjianju
\setlength{\belowcaptionskip}{0.3cm}
\centering
\caption{The fundamental quasinormal modes frequencies of the Kerr black hole for the states $l=0$ and $l=1$ in massless scalar field. The corresponding calculation parameter is $M=1$.}
\scalebox{0.7}{
\begin{tabular}{clclclclclclclclcl}
\hline \hline
\multicolumn{2}{c}{}     & \multicolumn{4}{c}{$l=0,m=0$}                                & \multicolumn{4}{c}{$l=1,m=-1$}                               & \multicolumn{4}{c}{$l=1,m=0$}                                & \multicolumn{4}{c}{$l=1,m=1$}                                \\ \hline
\multicolumn{2}{c}{$a$}  & \multicolumn{2}{c}{Re}       & \multicolumn{2}{c}{-Im}       & \multicolumn{2}{c}{Re}       & \multicolumn{2}{c}{-Im}       & \multicolumn{2}{c}{Re}       & \multicolumn{2}{c}{-Im}       & \multicolumn{2}{c}{Re}       & \multicolumn{2}{c}{-Im}       \\ \hline
\multicolumn{2}{c}{0.00} & \multicolumn{2}{c}{0.110454939} & \multicolumn{2}{c}{0.104895717}  & \multicolumn{2}{c}{0.292928418} & \multicolumn{2}{c}{0.0976600224} & \multicolumn{2}{c}{0.292936133} & \multicolumn{2}{c}{0.0976599888} & \multicolumn{2}{c}{0.292943849} & \multicolumn{2}{c}{0.0976599553} \\
\multicolumn{2}{c}{0.10} & \multicolumn{2}{c}{0.110533136} & \multicolumn{2}{c}{0.104801456}  & \multicolumn{2}{c}{0.285570275} & \multicolumn{2}{c}{0.0976259573} & \multicolumn{2}{c}{0.293127002} & \multicolumn{2}{c}{0.0975792571} & \multicolumn{2}{c}{0.301044531} & \multicolumn{2}{c}{0.0975471619} \\
\multicolumn{2}{c}{0.30} & \multicolumn{2}{c}{0.111157488} & \multicolumn{2}{c}{0.104008564}  & \multicolumn{2}{c}{0.272634571} & \multicolumn{2}{c}{0.0972279292} & \multicolumn{2}{c}{0.294679905} & \multicolumn{2}{c}{0.0969029596} & \multicolumn{2}{c}{0.320126458} & \multicolumn{2}{c}{0.0966913169} \\
\multicolumn{2}{c}{0.50} & \multicolumn{2}{c}{0.112380936} & \multicolumn{2}{c}{0.102183171}  & \multicolumn{2}{c}{0.261572212} & \multicolumn{2}{c}{0.0965054631} & \multicolumn{2}{c}{0.297930448} & \multicolumn{2}{c}{0.0953641953} & \multicolumn{2}{c}{0.344753181} & \multicolumn{2}{c}{0.0943945190} \\
\multicolumn{2}{c}{0.70} & \multicolumn{2}{c}{0.113979195} & \multicolumn{2}{c}{0.0986318974} & \multicolumn{2}{c}{0.251928312} & \multicolumn{2}{c}{0.0955465444} & \multicolumn{2}{c}{0.303188351} & \multicolumn{2}{c}{0.0924361444} & \multicolumn{2}{c}{0.379158531} & \multicolumn{2}{c}{0.0888481922} \\
\multicolumn{2}{c}{0.90} & \multicolumn{2}{c}{0.113847816} & \multicolumn{2}{c}{0.0915692550} & \multicolumn{2}{c}{0.243370881} & \multicolumn{2}{c}{0.0944221601} & \multicolumn{2}{c}{0.310761820} & \multicolumn{2}{c}{0.0866543540} & \multicolumn{2}{c}{0.437233808} & \multicolumn{2}{c}{0.0718481320} \\
\multicolumn{2}{c}{0.99} & \multicolumn{2}{c}{0.110440986} & \multicolumn{2}{c}{0.0894882986} & \multicolumn{2}{c}{0.239809603} & \multicolumn{2}{c}{0.0938822253} & \multicolumn{2}{c}{0.314579189} & \multicolumn{2}{c}{0.0822824858} & \multicolumn{2}{c}{0.493423284} & \multicolumn{2}{c}{0.0367119849}
\\ \hline \hline
\end{tabular}}
\label{t1}
\end{table*}
%%%%%\mu=0 SFDM
\begin{table}[tbp]
\setlength{\abovecaptionskip}{0cm} %%%tiaozhengjianju
\setlength{\belowcaptionskip}{0.3cm}
\centering
\caption{The fundamental quasinormal modes frequencies of the Kerr-like black hole for the states $l=0$ and $l=1$ in the massless scalar field with the SFDM model. The corresponding calculation parameter is $M=1$, $\rho_s=13.66 \times 10^{-3}$ $M_{\bigodot}/pc^{3}$ and $R_s=2.92$ $kpc$. We have converted these main calculation parameters to the black hole units before calculating.}
\scalebox{0.7}{
\begin{tabular}{clclclclclclclclcl}
\hline\hline
\multicolumn{2}{c}{}       & \multicolumn{4}{c}{$l=0,m=0$}                                 & \multicolumn{4}{c}{$l=1,m=-1$}                               & \multicolumn{4}{c}{$l=1,m=0$}                                & \multicolumn{4}{c}{$l=1,m=1$}                                \\ \hline
\multicolumn{2}{c}{$a$}    & \multicolumn{2}{c}{Re}        & \multicolumn{2}{c}{-Im}       & \multicolumn{2}{c}{Re}       & \multicolumn{2}{c}{-Im}       & \multicolumn{2}{c}{Re}       & \multicolumn{2}{c}{-Im}       & \multicolumn{2}{c}{Re}       & \multicolumn{2}{c}{-Im}       \\ \hline
\multicolumn{2}{c}{0.00} & \multicolumn{2}{c}{0.110454913} & \multicolumn{2}{c}{0.104895733}  & \multicolumn{2}{c}{0.292928414} & \multicolumn{2}{c}{0.0976600210} & \multicolumn{2}{c}{0.292936129} & \multicolumn{2}{c}{0.0976599874} & \multicolumn{2}{c}{0.292943845} & \multicolumn{2}{c}{0.0976599539} \\
\multicolumn{2}{c}{0.10} & \multicolumn{2}{c}{0.110533109} & \multicolumn{2}{c}{0.104801471}  & \multicolumn{2}{c}{0.285570271} & \multicolumn{2}{c}{0.0976259558} & \multicolumn{2}{c}{0.293126998} & \multicolumn{2}{c}{0.0975792557} & \multicolumn{2}{c}{0.301044527} & \multicolumn{2}{c}{0.0975471605} \\
\multicolumn{2}{c}{0.30} & \multicolumn{2}{c}{0.111157450} & \multicolumn{2}{c}{0.104008560}  & \multicolumn{2}{c}{0.272634567} & \multicolumn{2}{c}{0.0972279278} & \multicolumn{2}{c}{0.294679901} & \multicolumn{2}{c}{0.0969029582} & \multicolumn{2}{c}{0.320126453} & \multicolumn{2}{c}{0.0966913155} \\
\multicolumn{2}{c}{0.50} & \multicolumn{2}{c}{0.112380928} & \multicolumn{2}{c}{0.102183117}  & \multicolumn{2}{c}{0.261572209} & \multicolumn{2}{c}{0.0965054617} & \multicolumn{2}{c}{0.297930444} & \multicolumn{2}{c}{0.0953641940} & \multicolumn{2}{c}{0.344753175} & \multicolumn{2}{c}{0.0943945177} \\
\multicolumn{2}{c}{0.70} & \multicolumn{2}{c}{0.113979264} & \multicolumn{2}{c}{0.0986319888} & \multicolumn{2}{c}{0.251928309} & \multicolumn{2}{c}{0.0955465431} & \multicolumn{2}{c}{0.303188347} & \multicolumn{2}{c}{0.0924361432} & \multicolumn{2}{c}{0.379158524} & \multicolumn{2}{c}{0.0888481913} \\
\multicolumn{2}{c}{0.90} & \multicolumn{2}{c}{0.113848733} & \multicolumn{2}{c}{0.0915695056} & \multicolumn{2}{c}{0.243370881} & \multicolumn{2}{c}{0.0944221623} & \multicolumn{2}{c}{0.310761816} & \multicolumn{2}{c}{0.0866543529} & \multicolumn{2}{c}{0.437233799} & \multicolumn{2}{c}{0.0718481324} \\
\multicolumn{2}{c}{0.99} & \multicolumn{2}{c}{0.110441725} & \multicolumn{2}{c}{0.0895496732} & \multicolumn{2}{c}{0.239804794} & \multicolumn{2}{c}{0.0938775422} & \multicolumn{2}{c}{0.314578594} & \multicolumn{2}{c}{0.0822830637} & \multicolumn{2}{c}{0.493423266} & \multicolumn{2}{c}{0.0367119888}\\ \hline\hline
\end{tabular}}
\label{t2}
\end{table}
%%%%%\mu=0 CDM
\begin{table}[t!]
\setlength{\abovecaptionskip}{0cm} %%%tiaozhengjianju
\setlength{\belowcaptionskip}{0.3cm}
\centering
\caption{The fundamental quasinormal modes frequencies of the Kerr-like black hole for the states $l=0$ and $l=1$ in the massless scalar field with the CDM model. The corresponding calculation parameter is $M=1$, $\rho_c=2.45 \times 10^{-3}$ $M_{\bigodot}/pc^{3}$ and $R_c=5.7$ $kpc$. We have converted these main calculation parameters to the black hole units before calculating.}
\scalebox{0.7}{
\begin{tabular}{clclclclclclclclcl}
\hline\hline
\multicolumn{2}{c}{}       & \multicolumn{4}{c}{$l=0,m=0$}                                 & \multicolumn{4}{c}{$l=1,m=-1$}                                & \multicolumn{4}{c}{$l=1,m=0$}                                 & \multicolumn{4}{c}{$l=1,m=1$}                                 \\ \hline
\multicolumn{2}{c}{$a$}    & \multicolumn{2}{c}{Re}        & \multicolumn{2}{c}{-Im}       & \multicolumn{2}{c}{Re}        & \multicolumn{2}{c}{-Im}       & \multicolumn{2}{c}{Re}        & \multicolumn{2}{c}{-Im}       & \multicolumn{2}{c}{Re}        & \multicolumn{2}{c}{-Im}       \\ \hline
\multicolumn{2}{c}{0.00} & \multicolumn{2}{c}{0.110454909} & \multicolumn{2}{c}{0.104895726}  & \multicolumn{2}{c}{0.292928389} & \multicolumn{2}{c}{0.0976600127} & \multicolumn{2}{c}{0.292936105} & \multicolumn{2}{c}{0.0976599792} & \multicolumn{2}{c}{0.292943820} & \multicolumn{2}{c}{0.0976599456} \\
\multicolumn{2}{c}{0.10} & \multicolumn{2}{c}{0.110533104} & \multicolumn{2}{c}{0.104801464}  & \multicolumn{2}{c}{0.285570247} & \multicolumn{2}{c}{0.0976259476} & \multicolumn{2}{c}{0.293126973} & \multicolumn{2}{c}{0.0975792475} & \multicolumn{2}{c}{0.301044501} & \multicolumn{2}{c}{0.0975471523} \\
\multicolumn{2}{c}{0.30} & \multicolumn{2}{c}{0.111157444} & \multicolumn{2}{c}{0.104008557}  & \multicolumn{2}{c}{0.272634546} & \multicolumn{2}{c}{0.0972279197} & \multicolumn{2}{c}{0.294679876} & \multicolumn{2}{c}{0.0969029502} & \multicolumn{2}{c}{0.320126423} & \multicolumn{2}{c}{0.0966913075} \\
\multicolumn{2}{c}{0.50} & \multicolumn{2}{c}{0.112380913} & \multicolumn{2}{c}{0.102183116}  & \multicolumn{2}{c}{0.261572190} & \multicolumn{2}{c}{0.0965054539} & \multicolumn{2}{c}{0.297930419} & \multicolumn{2}{c}{0.0953641865} & \multicolumn{2}{c}{0.344753142} & \multicolumn{2}{c}{0.0943945105} \\
\multicolumn{2}{c}{0.70} & \multicolumn{2}{c}{0.113979259} & \multicolumn{2}{c}{0.0986319601} & \multicolumn{2}{c}{0.251928292} & \multicolumn{2}{c}{0.0955465356} & \multicolumn{2}{c}{0.303188322} & \multicolumn{2}{c}{0.0924361365} & \multicolumn{2}{c}{0.379158484} & \multicolumn{2}{c}{0.0888481858} \\
\multicolumn{2}{c}{0.90} & \multicolumn{2}{c}{0.113848624} & \multicolumn{2}{c}{0.0915693104} & \multicolumn{2}{c}{0.243370866} & \multicolumn{2}{c}{0.0944221542} & \multicolumn{2}{c}{0.310761791} & \multicolumn{2}{c}{0.0866543478} & \multicolumn{2}{c}{0.437233745} & \multicolumn{2}{c}{0.0718481341} \\
\multicolumn{2}{c}{0.99} & \multicolumn{2}{c}{0.110436980} & \multicolumn{2}{c}{0.0895433180} & \multicolumn{2}{c}{0.239804940} & \multicolumn{2}{c}{0.0938789531} & \multicolumn{2}{c}{0.314578766} & \multicolumn{2}{c}{0.0822830436} & \multicolumn{2}{c}{0.493423181} & \multicolumn{2}{c}{0.0367120381}\\ \hline\hline
\end{tabular}}
\label{t3}
\end{table}

\begin{figure*}[tbp]
\centering
{
\includegraphics[width=0.31\columnwidth]{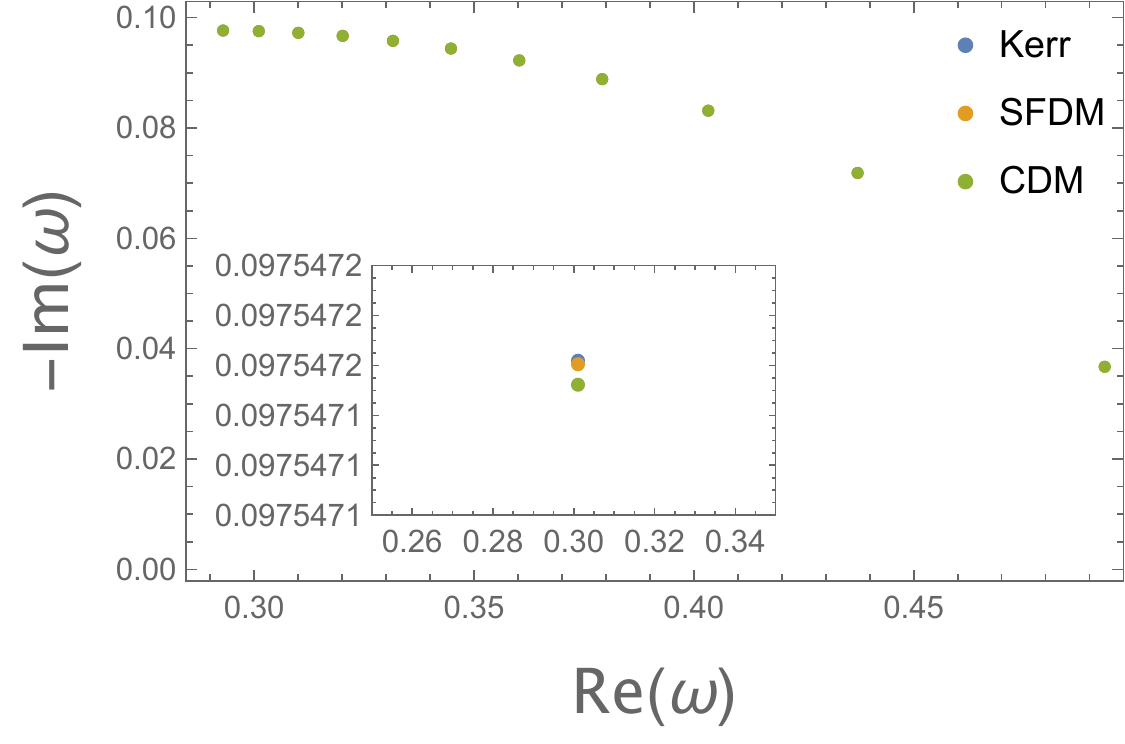}
}
{
\includegraphics[width=0.31\columnwidth]{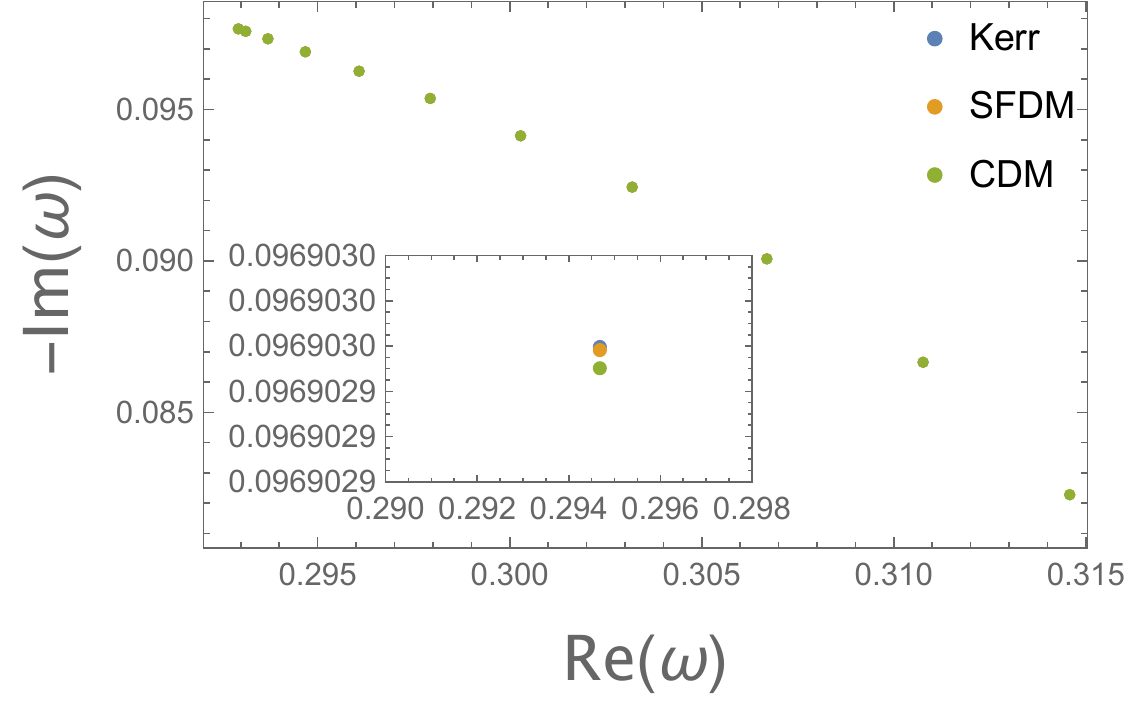}
}
{
\includegraphics[width=0.31\columnwidth]{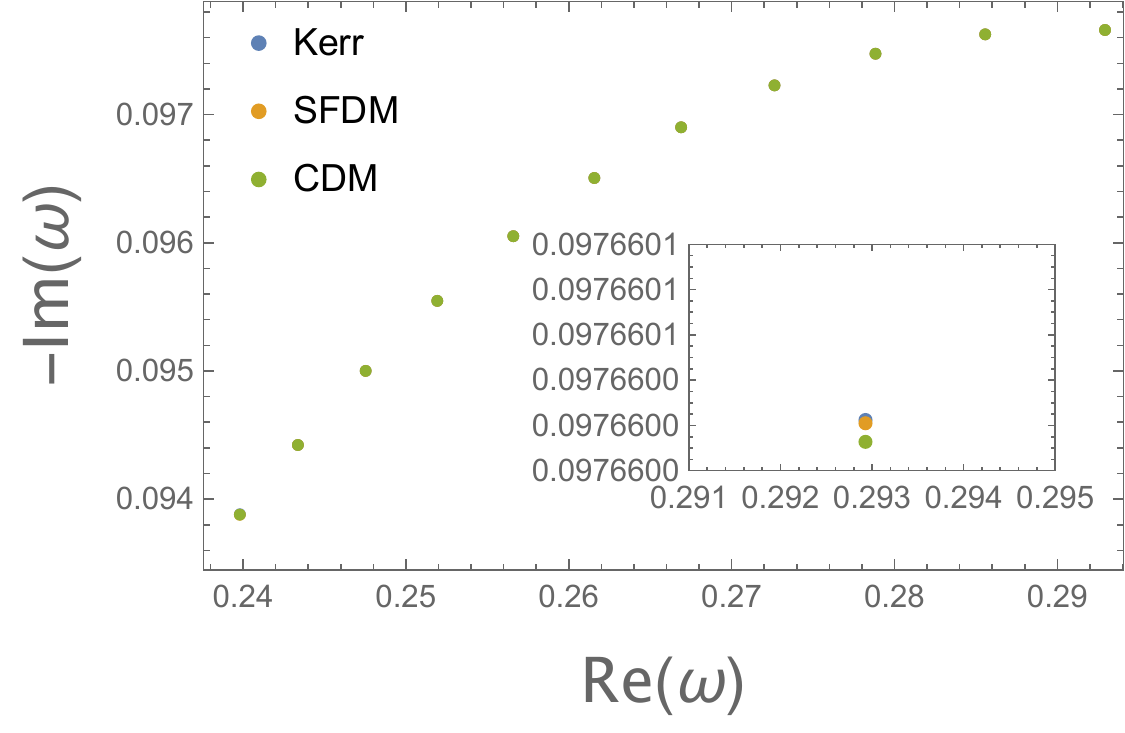}
}
\caption{The fundamental quasinormal modes of the black holes in massless scalar field for the state $l=1$ (left to right panel are the state ($l=1, m=1$), ($l=1, m=0$) and ($l=1, m=-1$). These points in each panel correspond to different rotation parameters $a$ from left to right, they are $a=0,0.1,0.2,0.3,0.4,0.5,0.6,0.7,0.8,0.9,0.99$, respectively (left and middle panels), and the right panel are $a=0.99,0.9,0.8,0.7,0.6,0.5,0.4,0.3,0.2,0.1,0$. The main calculation parameters are $M=1$, $\rho_c=2.45 \times 10^{-3}$ $M_{\bigodot}/pc^{3}$, $R_c=5.7$ $kpc$, $\rho_s=13.66 \times 10^{-3}$ $M_{\bigodot}/pc^{3}$, $R_s=2.92$ $kpc$. We have converted these main calculation parameters to the black hole units before plotting.}
\label{nf2}
\end{figure*}

\indent Now, let’s return to the discussion of QNM frequencies for the Kerr-like black holes in a dark matter halo. For a black hole, its oscillatory behavior is always related to dissipation, originating QNM. This oscillatory behavior usually appears as a pure incoming wave at the horizon and a pure outgoing wave at infinity. Therefore, the process of QNM is a stable mode. Here, we first study the oscillatory behavior of black holes in massless scalar field both in CDM and SFDM models, and compare them with the Kerr black hole. We use the continued fraction method to calculate these frequencies at the states of the angular quantum $l=0, 1$ and use the Schwarzschild fundamental modes as our initial guess. We divide these QNM frequencies into two parts, real and imaginary parts. where the real part represents the oscillation frequency of this system, and the imaginary part represents the decay rate, as a function of rotating parameter $a$. Then, we record the QNM frequencies of the black holes about these two models at the states $l=0$ and $l=1$ in Tables \ref{t2} and \ref{t3}, respectively. The main calculation parameters are $M=1$, $\rho_c=2.45 \times 10^{-3}$ $M_{\bigodot}/pc^{3}$, $R_c=5.7$ $kpc$, $\rho_s=13.66 \times 10^{-3}$ $M_{\bigodot}/pc^{3}$, $R_s=2.92$ $kpc$. In Section \ref{s2}, we have converted these dark matter parameters into black hole units, and then used them to participate in calculations. Firstly, on the whole, from Tables \ref{t1}, \ref{t2}, and \ref{t3}, we find that the QNM frequencies of black holes in a dark matter halo and kerr black holes at the same state are very close. In other words, the difference between the QNM frequencies of dark matter black holes and Kerr black holes is small. Taking the state $(l=m=1, a=0.99)$ as an example, both the real and imaginary parts of the QNM frequency of a Kerr black hole are greater than the SFDM model, and the SFDM model is greater than the CDM model. The QNM difference between the CDM model and Kerr spacetime is about $1 \times 10^{-7}$ in real part and $5 \times 10^{-8}$ in imaginary part. The difference of that between SFDM model and Kerr spacetime  is about $2 \times 10^{-8}$ in real part and $4 \times 10^{-9}$ in imaginary part. In view of the inefficient signal-to-noise ratio of the current gravitational wave detectors \cite{Cotesta:2022pci}, it may not be possible to accurately detect the frequency of dark matter and the frequency of kerr black holes. However, with the continuous update and development of detectors, future gravitational wave detectors such as LISA, Taiji \cite{Ruan:2018tsw}, Tianqin \cite{Shi:2019hqa} and DECIGO \cite{Moore:2014lga} may detect this kind of QNM signal. This may provide an effective method for detecting dark matter around black holes. Actually, from Table \ref{t2}, we found that when the mode is $l=1,m=-1$, the real and imaginary parts of the QNM frequencies in the SFDM model both decrease with the increasing of the rotation parameter $a$. However, when the modes are $l=0,m=0$, $l=1,m=0$ and $l=1,m=1$, the real part of the QNM frequencies increase with the increasing of the rotation parameter $a$, and its imaginary part decreases with the increasing of the parameter $a$. This feature indicates that the large rotation parameter of the rotating black hole has a positive contribution to the QNM frequency. On the other hand, we found that when $l=1$, the real part of the QNM frequency of a black hole in the SFDM model at the same rotation parameter $a$ increases with the increase of $m$, and its imaginary part decreases with the increasing of $m$. Similarly, when $m=0$, the real part of the QNM frequency in the SFDM model at the same rotation parameter $a$ increases with the increasing of $l$, and its imaginary part decreases with the increasing of $l$. Analyzing Table \ref{t3} as we did with Table \ref{t2}, we find that this trend of the oscillating behavior of black holes in the CDM model is much the same as that in the SFDM model. Finally, according to Tables \ref{t1}, \ref{t2} and \ref{t3}, at the same mode (when $a, l, m$ are uniquely determined), we find that the oscillation frequency and decay rate of Kerr black hole are greater than that of the SFDM model, while the SFDM model is greater than the CDM model. Besides, we also graphically present these results in Figure \ref{nf2}.
%%%%%%%%%%%%%%%%%%%%%%%%%%%%%%%%%%%%%%%%%%%%%%%%%%%%%%%%%%%%%%
%%%%%%%%%%%%%%%%%%%%%%%%%%%%%%%%%%%%%%%%%%%%%%%%%%%%%%%%%%%%%%%%%%%%%
\begin{figure*}[tbp]
\centering
{
\includegraphics[width=0.45\columnwidth]{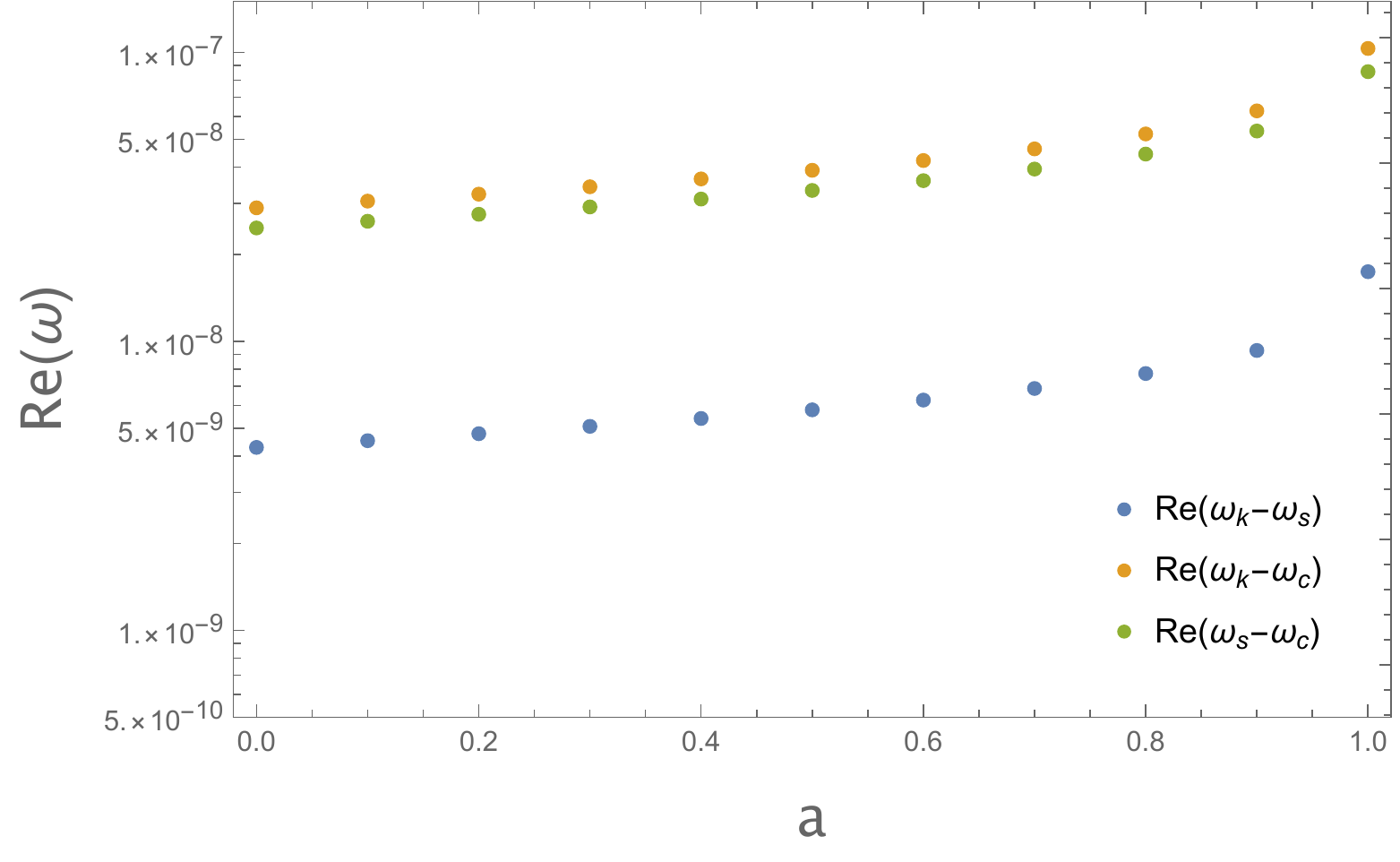}
}
{
\includegraphics[width=0.45\columnwidth]{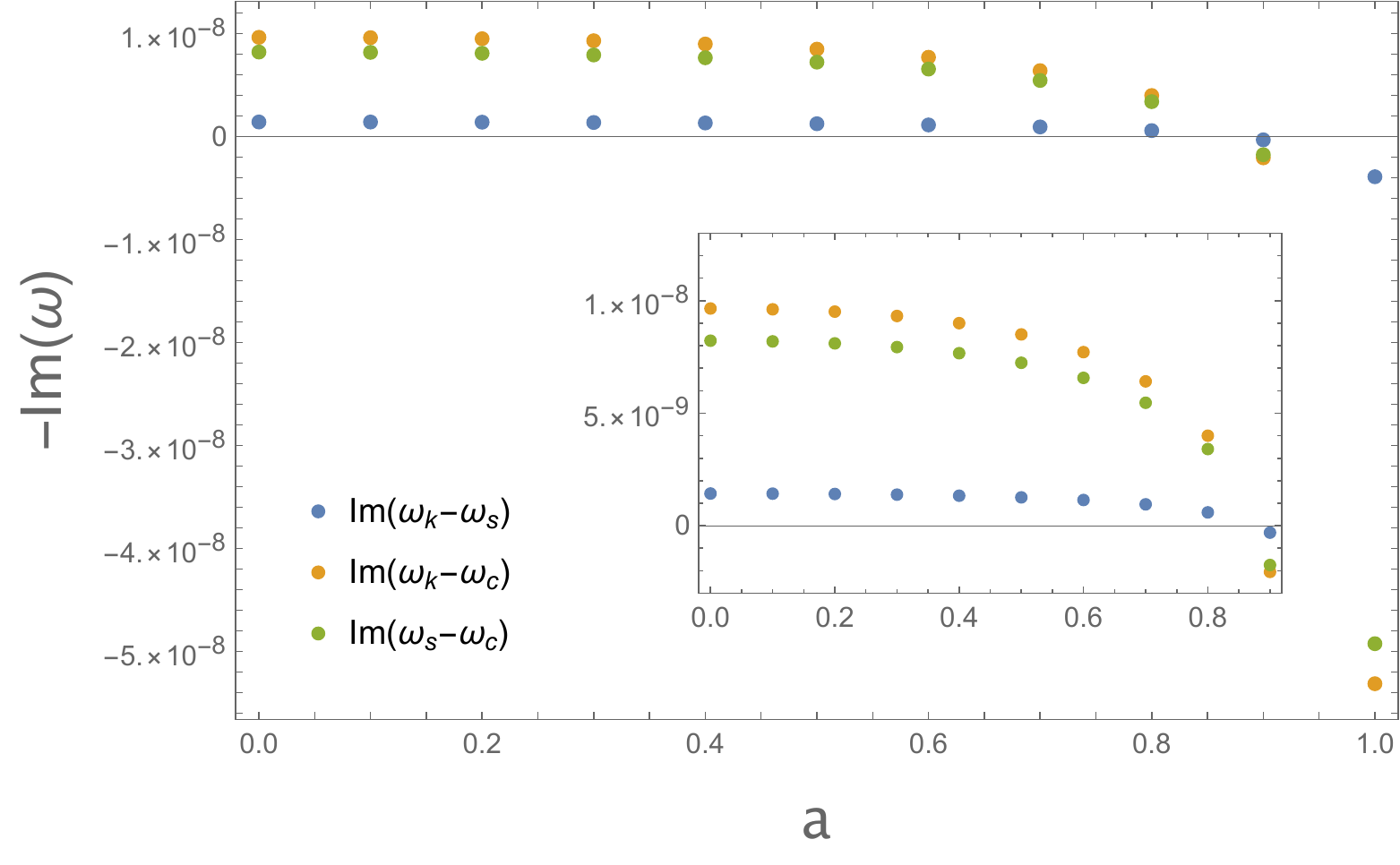}
}
\caption{The real part (left) and imaginary part (right) of the fundamental quasinormal modes of the black holes in massless scalar field for the state $l=1,m=1$. These points in each panel correspond to different rotation parameters $a$ from left to right, they are $a=0,0.1,0.2,0.3,0.4,0.5,0.6,0.7,0.8,0.9,0.99$, respectively. The main calculation parameters are $M=1$, $\rho_c=2.45 \times 10^{-3}$ $M_{\bigodot}/pc^{3}$, $R_c=5.7$ $kpc$, $\rho_s=13.66 \times 10^{-3}$ $M_{\bigodot}/pc^{3}$, $R_s=2.92$ $kpc$. We have converted these main calculation parameters to the black hole units before plotting.}
\label{nf3}
\end{figure*}
%%%%%%%%%%%%%%%%%%%%%%%%%%%%%%%%%%%%%%%%%%%%%%%%%%%%%%%%%%
We give the comparison charts of the QNM frequency of black holes in the Kerr spacetime, SFDM model and CDM model. We find that the QNM frequencies of black holes are very close to Kerr spacetime in  dark matter models. To distinguish them, we show this difference in subfigures. The oscillation behavior of the QNM frequencies in the three states are consistent, that is, the QNM frequency of the Kerr black hole is larger than that of the SFDM model, and the SFDM model is larger than the CDM model. Finally, among the states we study, we find that the state $l=m=1$ is quite special. The real and imaginary parts of the QNM frequencies vary widely at the different rotation parameter $a$. In other words, the real part of the QNM frequency increases rapidly with the increasing of the rotation parameter $a$, while its imaginary part decreases rapidly. But in other states, the change trend of QNM frequency is slowly increasing or decreasing. Based on this consideration, we analyze the QNM frequencies of the black hole at the state $l=m=1$ in SFDM/CDM models, and compare them with the Kerr black hole. We give the results of our computations in Figure \ref{nf3}. In order to find the difference between them, we combine black holes in SFDM model, CDM model and the Kerr spacetime in pairs and investigate the differences of QNM frequencies. The QNM frequencies of black hole both in SFDM/CDM models are lower than that of the Kerr black hole. In other words, when the mode can be in the same state (the parameters $a, l, m$ are determined), both the real part and the imaginary part of the QNM frequency in Kerr black hole are greater than that of the SFDM model, and the SFDM model is greater than that of the CDM model. Among them, the QNM frequency has the largest difference between the CDM model and the Kerr black hole, and the value of this difference at the same state is about $3 \times 10^{-7}$ in real part and $5 \times 10^{-8}$ in imaginary part. This difference of that could be that the impact of dark matter halo is too small. It is now generally believed that the dark matter distribution near black holes is a ``spike” structure. This ``spike” structure might amplify this difference, boosting the QNM frequency by an order of magnitude. When better-precision gravitational-wave detectors arrive, they may be able to detect this difference.

\begin{figure*}[tbp]
\centering
{
\includegraphics[width=0.45\columnwidth]{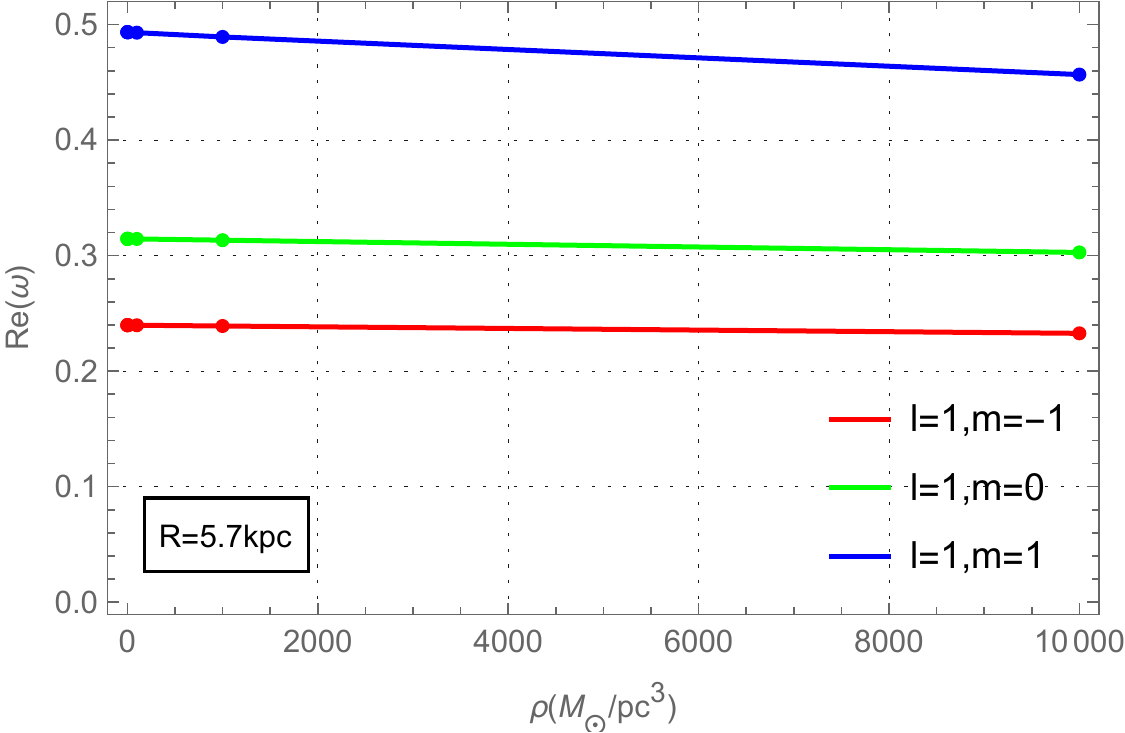}
}
{
\includegraphics[width=0.45\columnwidth]{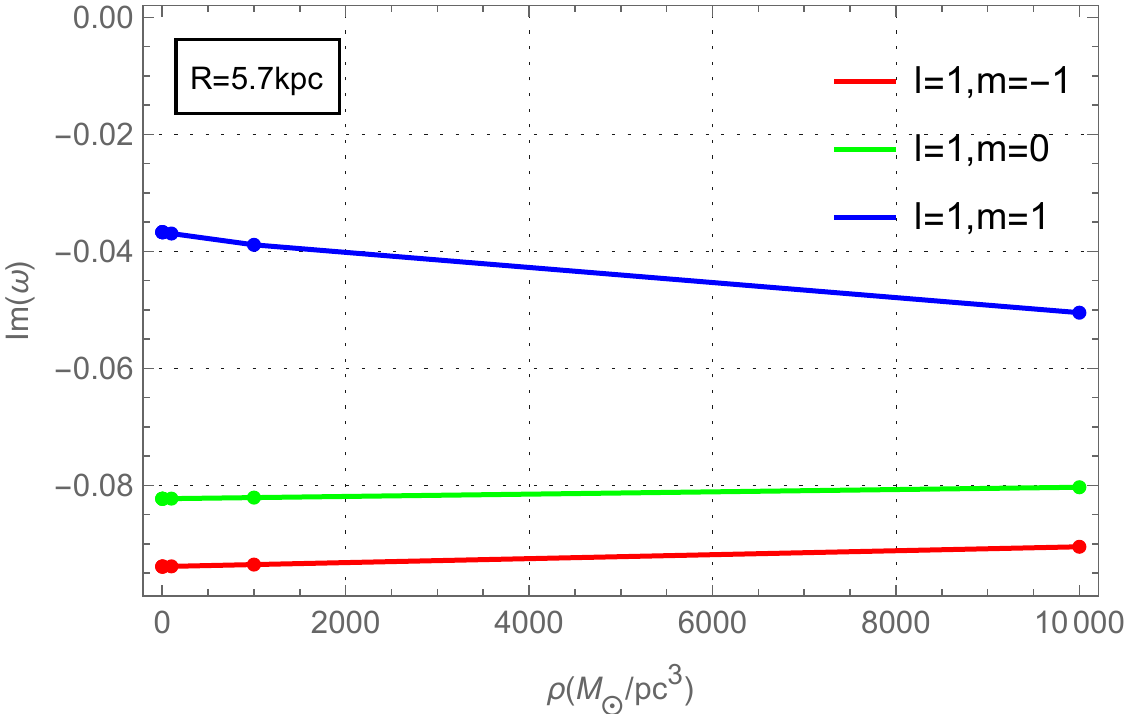}
}
{
\includegraphics[width=0.45\columnwidth]{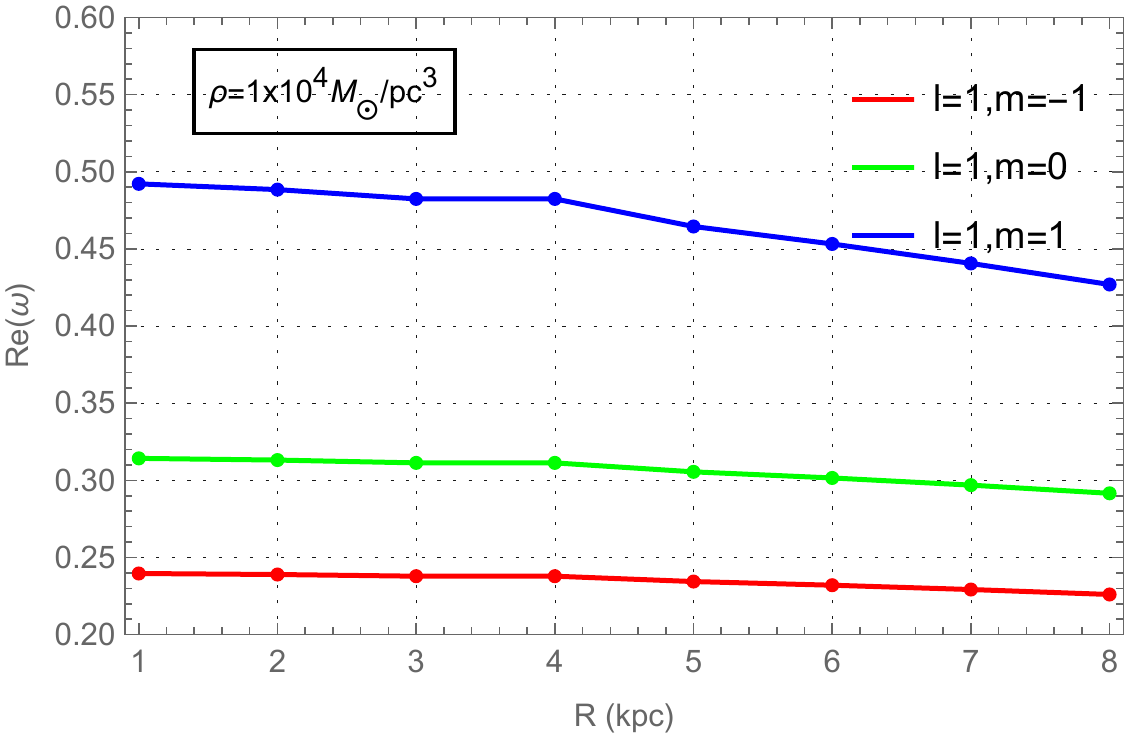}
}
{
\includegraphics[width=0.45\columnwidth]{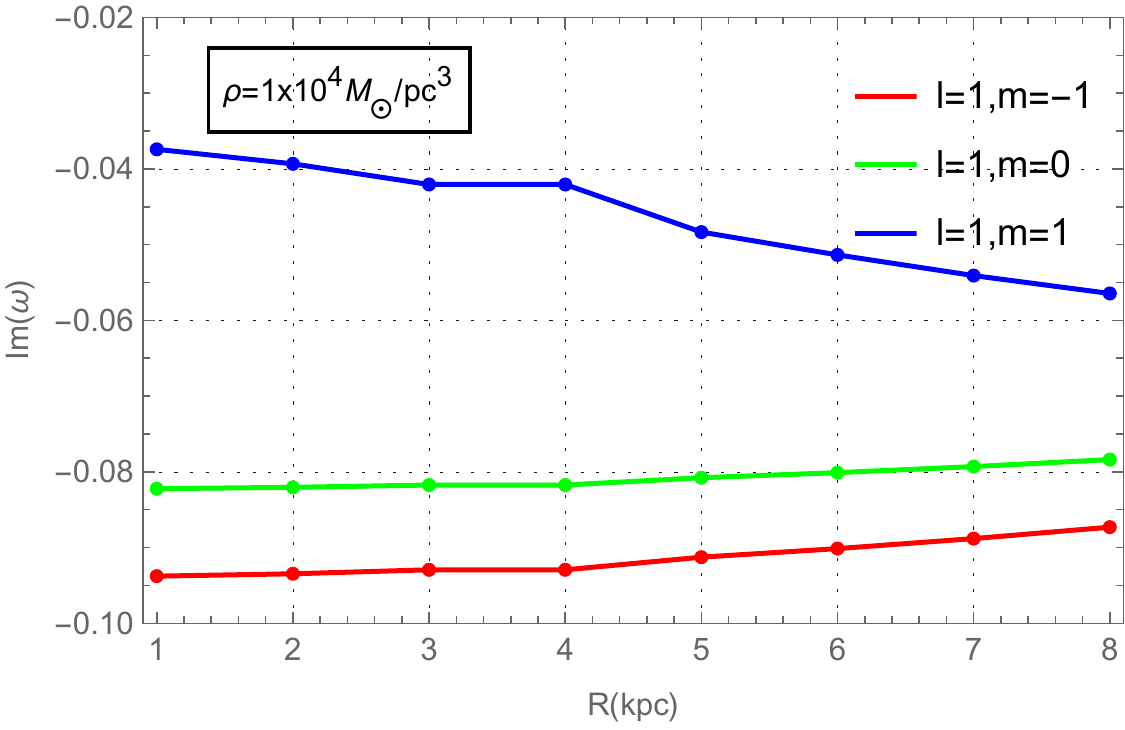}
}
\caption{The fundamental quasinormal modes of the black holes in massless scalar field $(\mu=0)$ for the state $l=1,a=0.99$ in SFDM model, as a function of $\rho$ (top) and $R$ (bottom). The top panels we fix the $R=5.7kpc$ and the bottom panels we fix the $\rho=1 \times 10^{4}$ $M_{\bigodot}/pc^{3}$. We have converted these main calculation parameters to the black hole units before plotting.}
\label{rr1}
\end{figure*}

\begin{figure*}[t!]
\centering
{
\includegraphics[width=0.45\columnwidth]{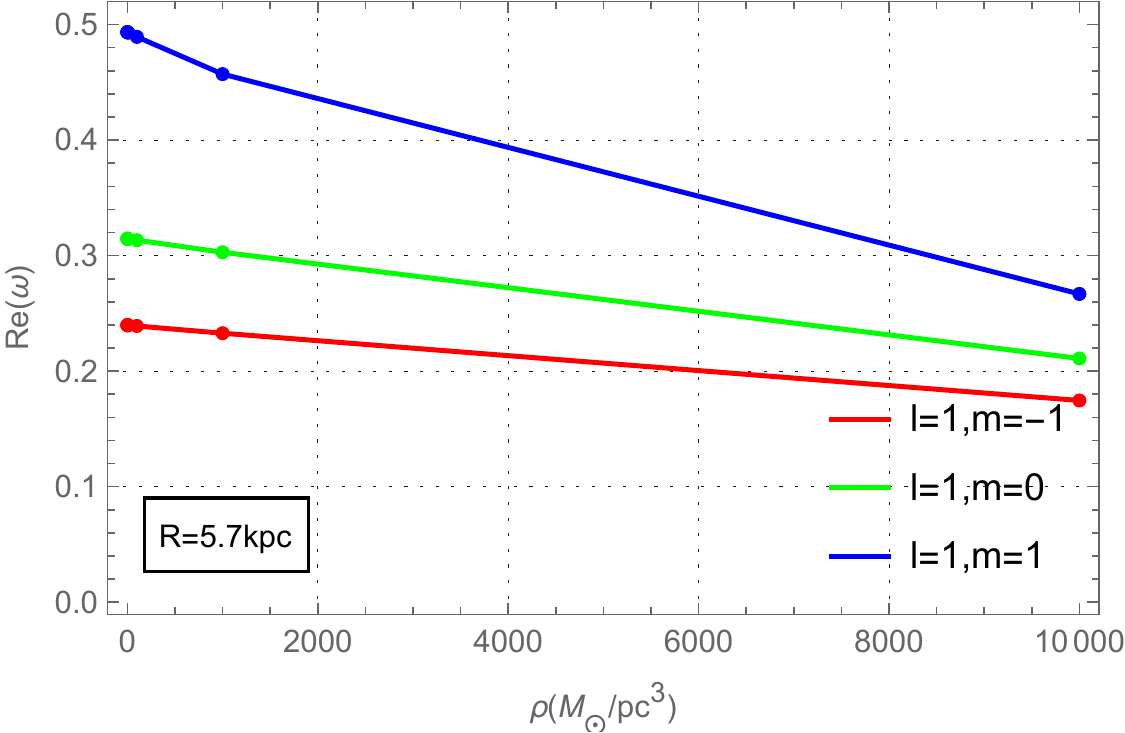}
}
{
\includegraphics[width=0.45\columnwidth]{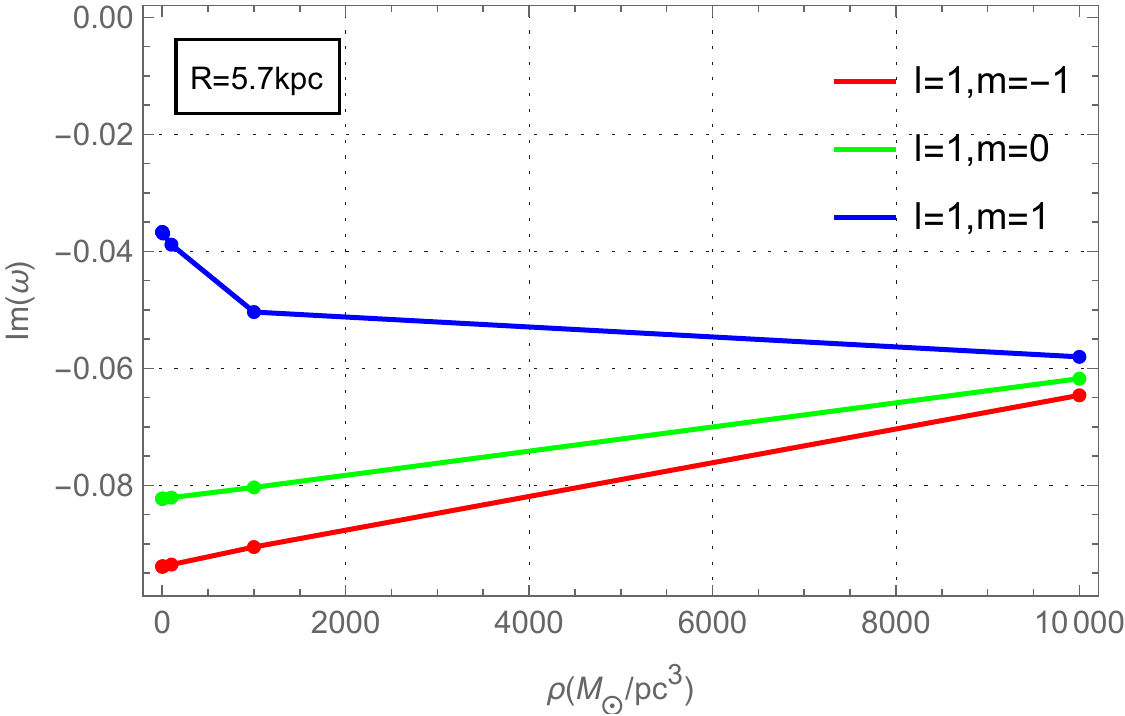}
}
{
\includegraphics[width=0.45\columnwidth]{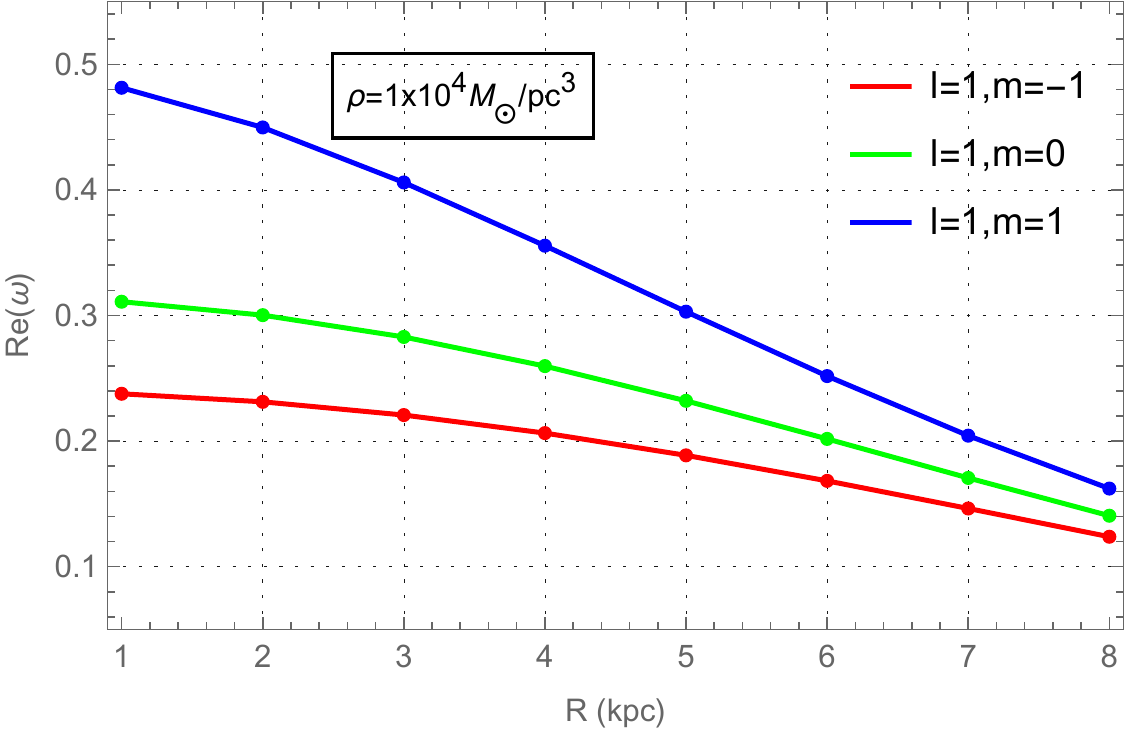}
}
{
\includegraphics[width=0.45\columnwidth]{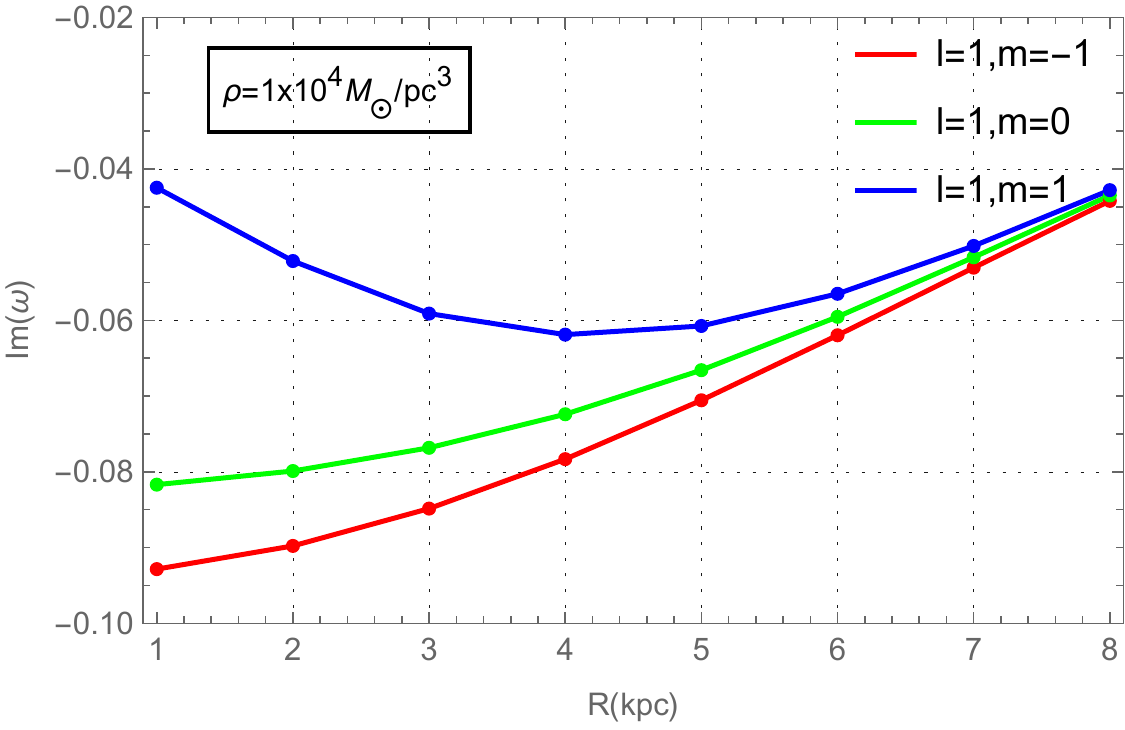}
}
\caption{The fundamental quasinormal modes of the black holes in massless scalar field $(\mu=0)$ for the state $l=1,a=0.99$ in CDM model, as a function of $\rho$ (top) and $R$ (bottom). The top panels we fix the $R=5.7kpc$ and the bottom panels we fix the $\rho=1 \times 10^{4}$ $M_{\bigodot}/pc^{3}$. We have converted these main calculation parameters to the black hole units before plotting.}
\label{rr2}
\end{figure*}

In the above we have analyzed the QNM frequencies from specific galaxy in detail. Next, we will study the impacts of dark matter parameters on the QNM frequencies both in SFDM model and CDM model. For example, in LSB galaxies, these parameters of dark matter are recorded in Table \ref{t11}. In fact, C. Zhang et al. have investigated the impacts of dark matter parameters on the QNM frequencies in non-rotating black hole geometries \cite{Zhang:2021bdr}. However, there is evidence that the impacts of dark matter on the QNM may be enhanced in rapidly rotating black holes \cite{Ferrer:2017xwm}. Therefore, here, we mainly study the QNM frequencies at the state $l=1,a=0.99$ in a massless scalar field. Firstly, for the convenience of our analysis, for the SFDM model, we fix the density parameter $\rho=1\times 10^{4}M_{\bigodot}/pc^{3}$ and the characteristic radius $R=5.7kpc$ respectively. For the CDM model, we also fixed the density parameter $\rho=1\times 10^{3}M_{\bigodot}/pc^{3}$ and the characteristic radius $R=5.7kpc$. In other words, when the density parameter is a certain value, we calculate the QNM frequencies of the black hole, as a function of the characteristic radius $R$. And vice versa. Then, we calculate the QNM frequencies of black holes in the massless scalar field at state $l=1,a=0.99$. We record our results of SFDM model and CDM model in Figures. \ref{rr1} and \ref{rr2}, respectively.
%%%%%%%%%%%%%%%%%%%%%%%%%%%%%%%%%%%%%%%%%%%%%%%%%%
%%%%%%%%%%%%%%%%%%%%%%%%%%%%%%%
%%%%%%%%%%%%%%%%%%%%%%%%%%%%%%%%%%%%%
In Figure \ref{rr1}, we find that when the characteristic radius is fixed, the real part of the QNM frequencies of the black holes in SFDM model decreases with the increasing of the density parameter for the state $l=1,m=-1,0$, while their imaginary parts increase with the increasing of the density increases. At the state $l=m=1$, the real and imaginary parts of the QNM frequencies of the black hole both decrease with the increasing of the density parameter. Similarly, when the density parameter is fixed, we investigate the impacts of the characteristic radius on the QNM frequencies of black holes. In fact, the behaviors of QNM frequencies and characteristic radius are consistent with the density parameter. This feature indicates that the dark matter around the black hole will reduce the real part of the QNM frequency of the black hole and increase the imaginary part. But from the slope of the image, the QNM frequencies seem to be more sensitive to the characteristic radius. Figure \ref{rr2} is the case of the CDM model. The behavior of the CDM model is roughly the same as that of the SFDM model. The difference is that in the state $l=m=1$, the imaginary part of QNM frequency first increases and then decreases with the characteristic radius. These results indicate that the QNM frequencies depend on the choice of dark matter parameters in these galaxies.
%%%%%%%%%%%%%%%%%%%%%%%%%%%%%%
\subsection{Quasibound states and superradiant instabilities}
In the previous subsection, we considered a massless scalar field and calculated the QNM of the black holes both in CDM and SFDM models. Next, we will continue to investigate the oscillatory behavior of black holes in these two models by considering a massive scalar field, and compare them with the Kerr black hole. The oscillation behavior at this time is called quasibound states (QBS). This state often corresponds to an unusually long-lived mode \cite{Cardoso:2011xi,Barranco:2012qs,Zouros:1979iw}. Unlike QNM, QBS behaves in a form of exponential decay as it is far away from the black hole. In other words, QBS are localized inside the potential well formed by the mass of the field. On the other hand, this oscillation frequency of the QBS usually satisfies the condition of superradiant, which has the following form
\begin{equation}
\text{Re}(\omega) < \mu  \leq m\Omega.
\label{e52}
\end{equation}
where, $\Omega$ is the angular velocity of rotation. Similar to QNM, the imaginary part of QBS is less than zero $\text{Im}(\omega) < 0$, which corresponds to a stable mode. However, for an unstable mode, it corresponds to $\text{Im}(\omega)>0$. In this case, if it satisfy the condition of superradiance, it means superradiant instability emerging. If $\text{Im}(\omega)=0$, it corresponds to the bound states of the so-called scalar clouds \cite{Hod:2012px,Siqueira:2022tbc}.

\begin{figure*}[tbp]
\centering
{
\includegraphics[width=0.31\columnwidth]{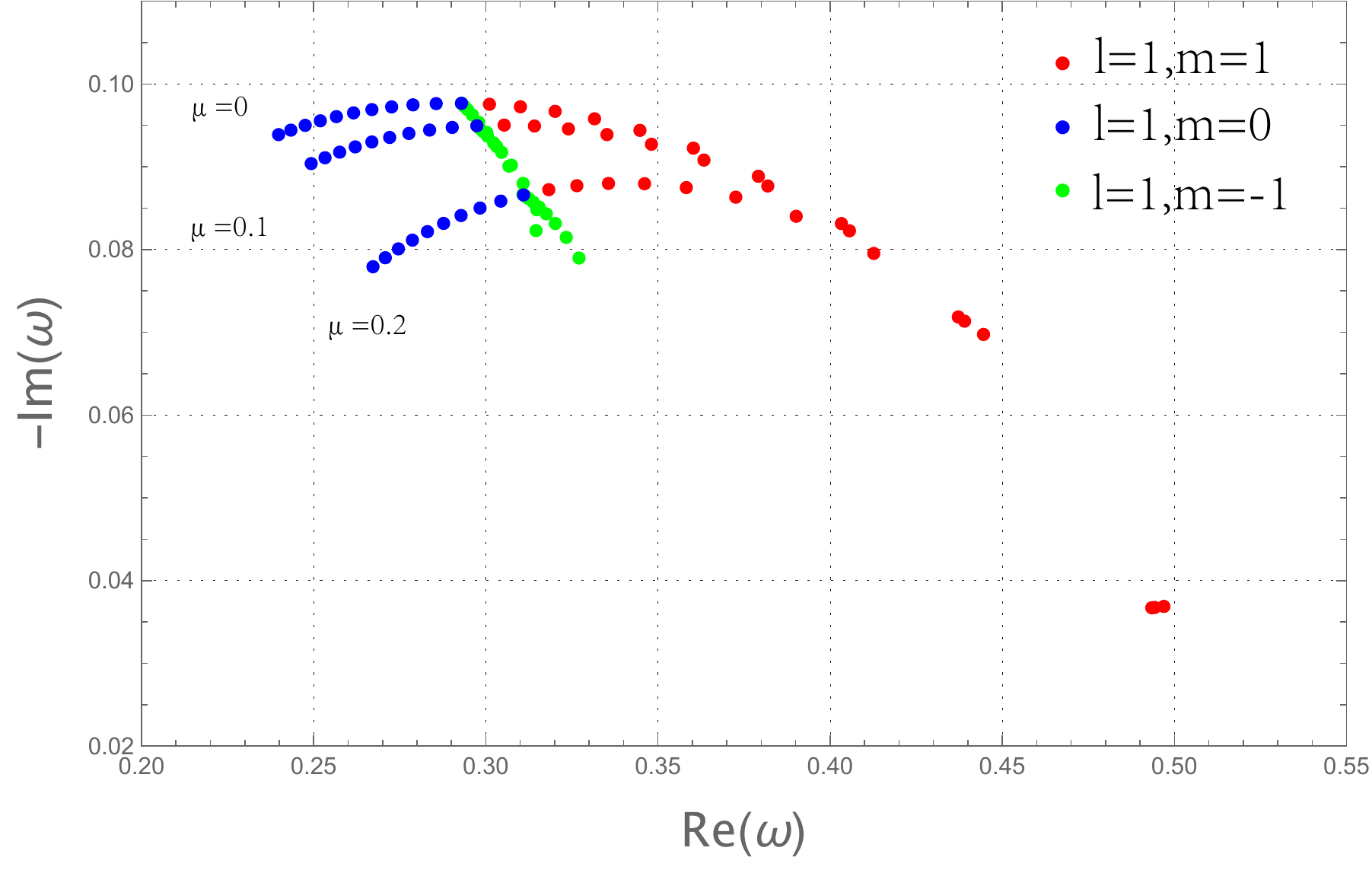}
}
{
\includegraphics[width=0.31\columnwidth]{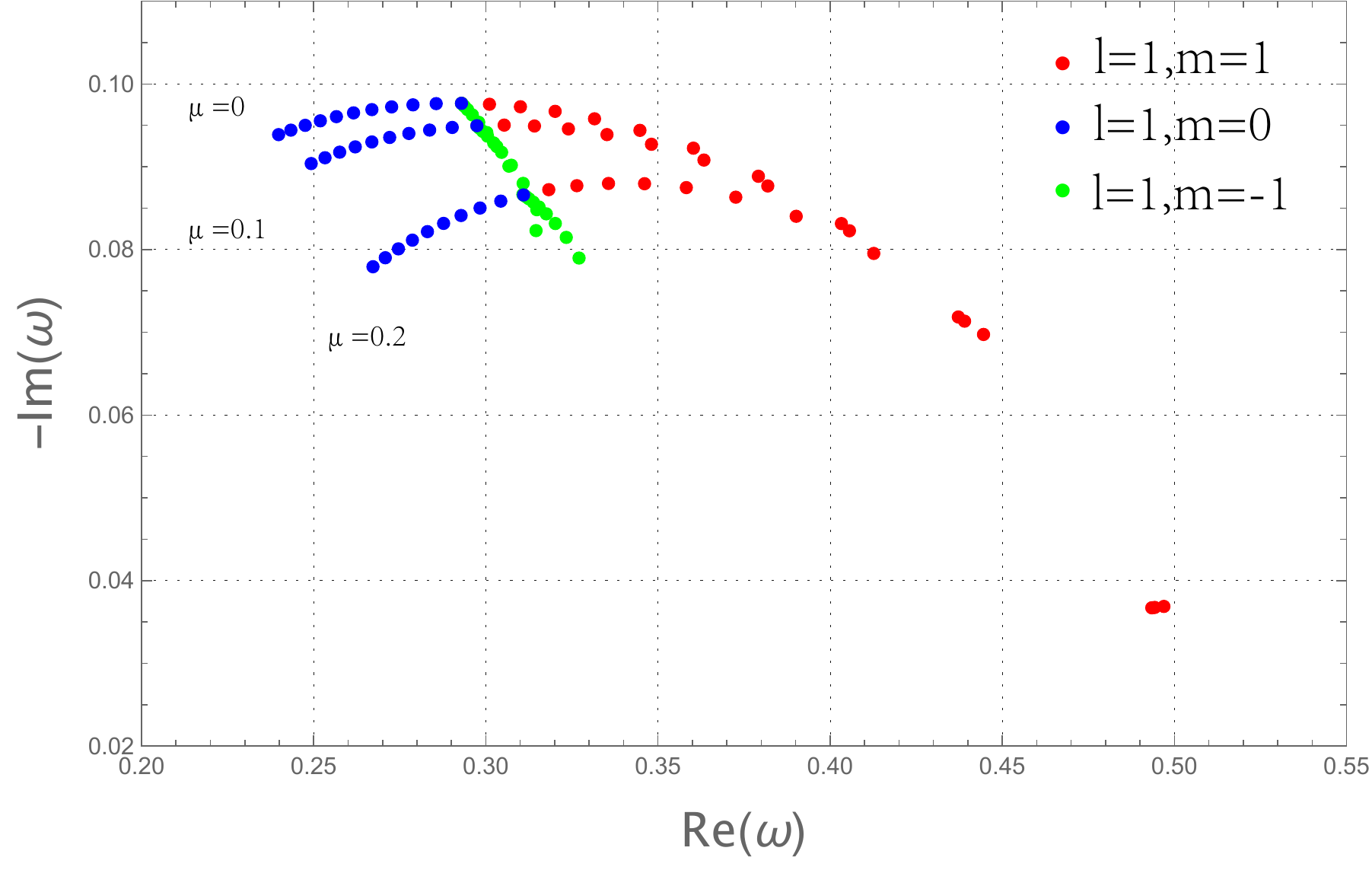}
}
{
\includegraphics[width=0.31\columnwidth]{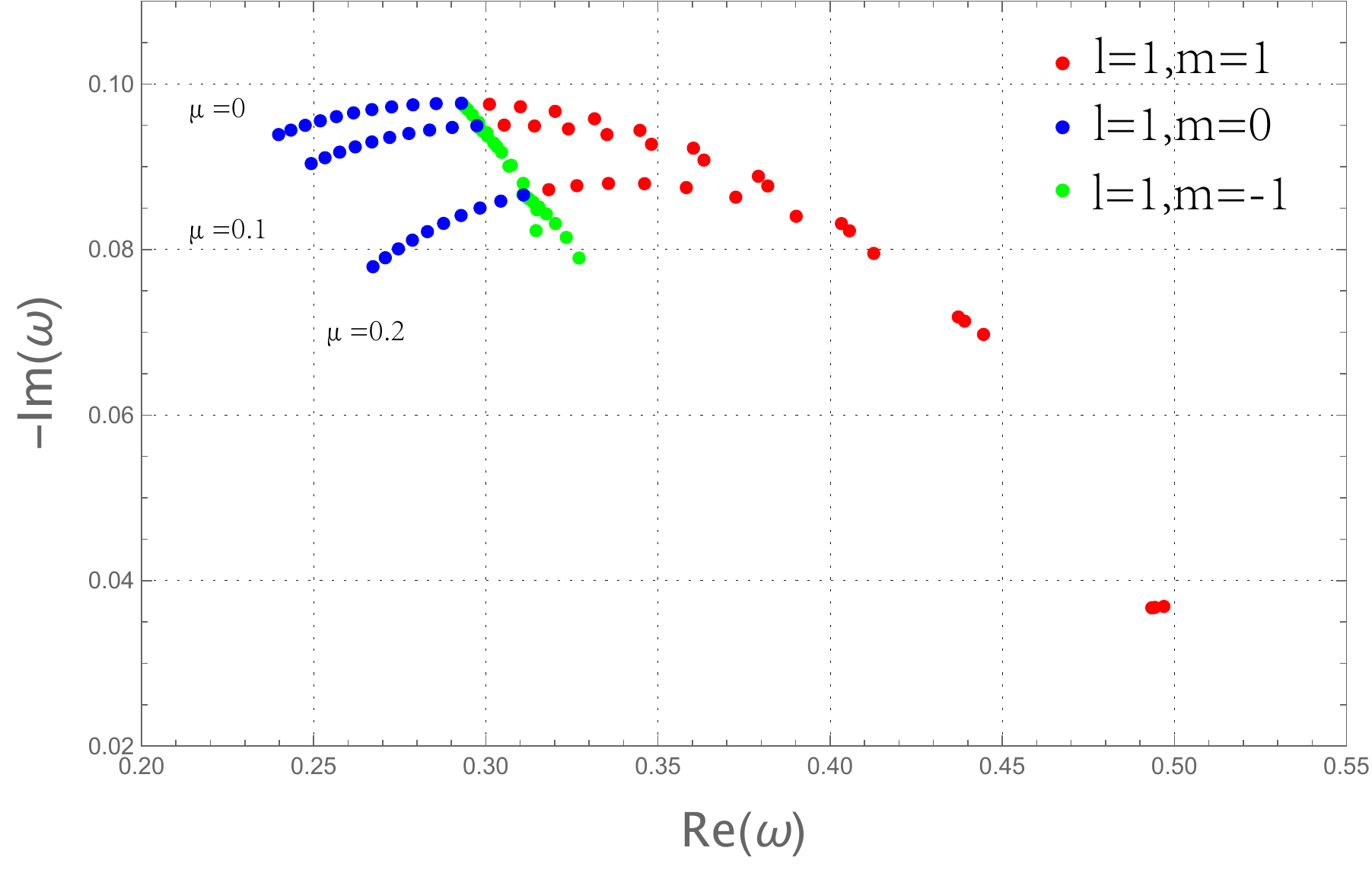}
}
\caption{The fundamental quasinormal modes of the black holes in massive scalar field for the state $l=1, m=1$ (left panel is CDM model, middle panel is SFDM model and right panel is Kerr spacetime). Three panels all reflect the stable modes. In these three pictures, three colors of red, green and blue represent three modes respectively. These points from left to right of each color are given by the increasing value of the rotation parameter $a$, corresponding to the values $a=0,0.1,0.2,0.3,0.4,0.5,0.6,0.7,0.8,0.9,0.99$. The main calculation parameters are $M=1$, $\rho_c=2.45 \times 10^{-3}$ $M_{\bigodot}/pc^{3}$, $R_c=5.7$ $kpc$, $\rho_s=13.66 \times 10^{-3}$ $M_{\bigodot}/pc^{3}$, $R_s=2.92$ $kpc$.}
\label{nf4}
\end{figure*}

\begin{figure*}[tbp]  %\subfigure[]{zitubiaohao}
\centering
{
\includegraphics[width=0.31\columnwidth]{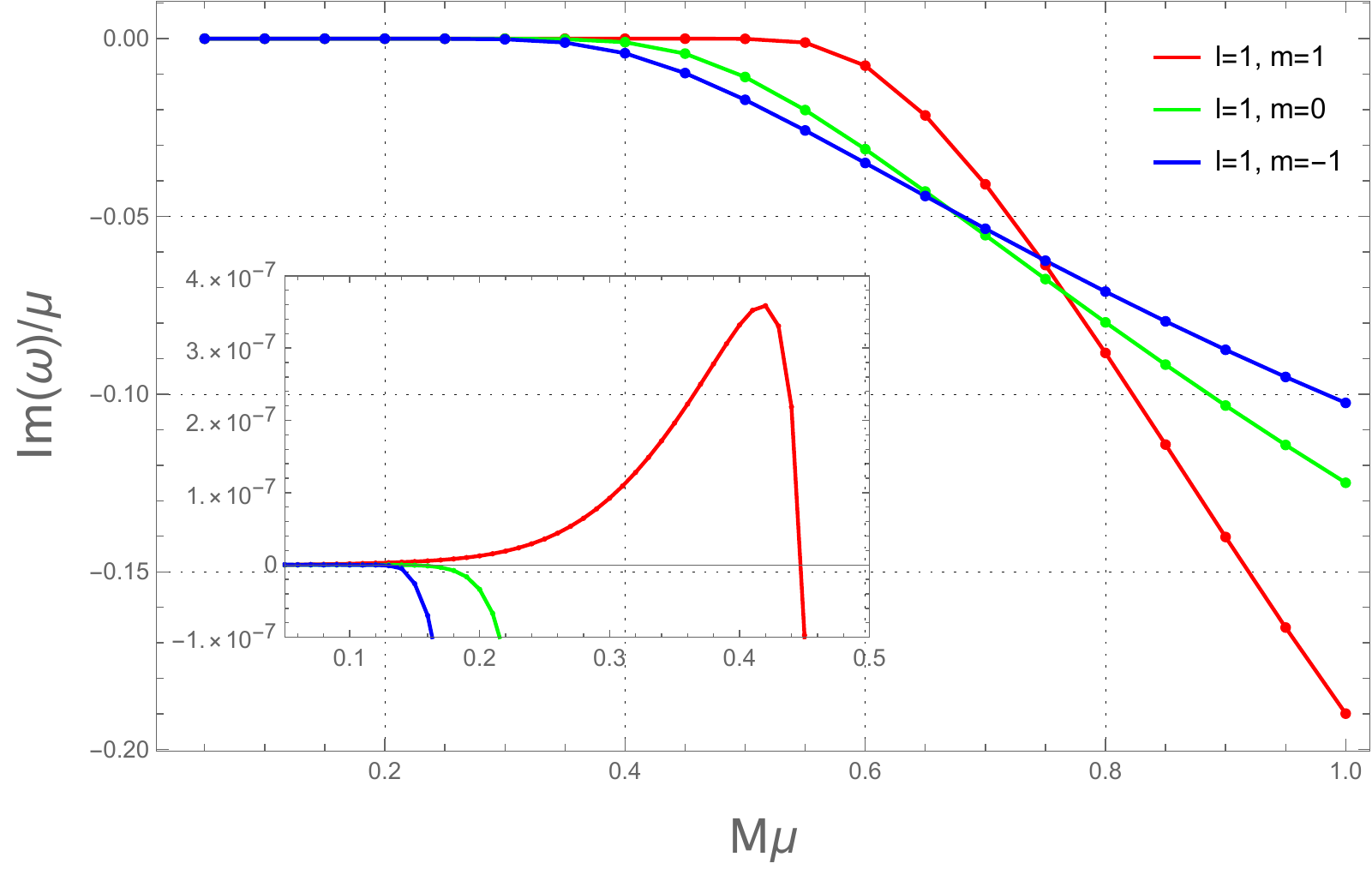}
}
{
\includegraphics[width=0.31\columnwidth]{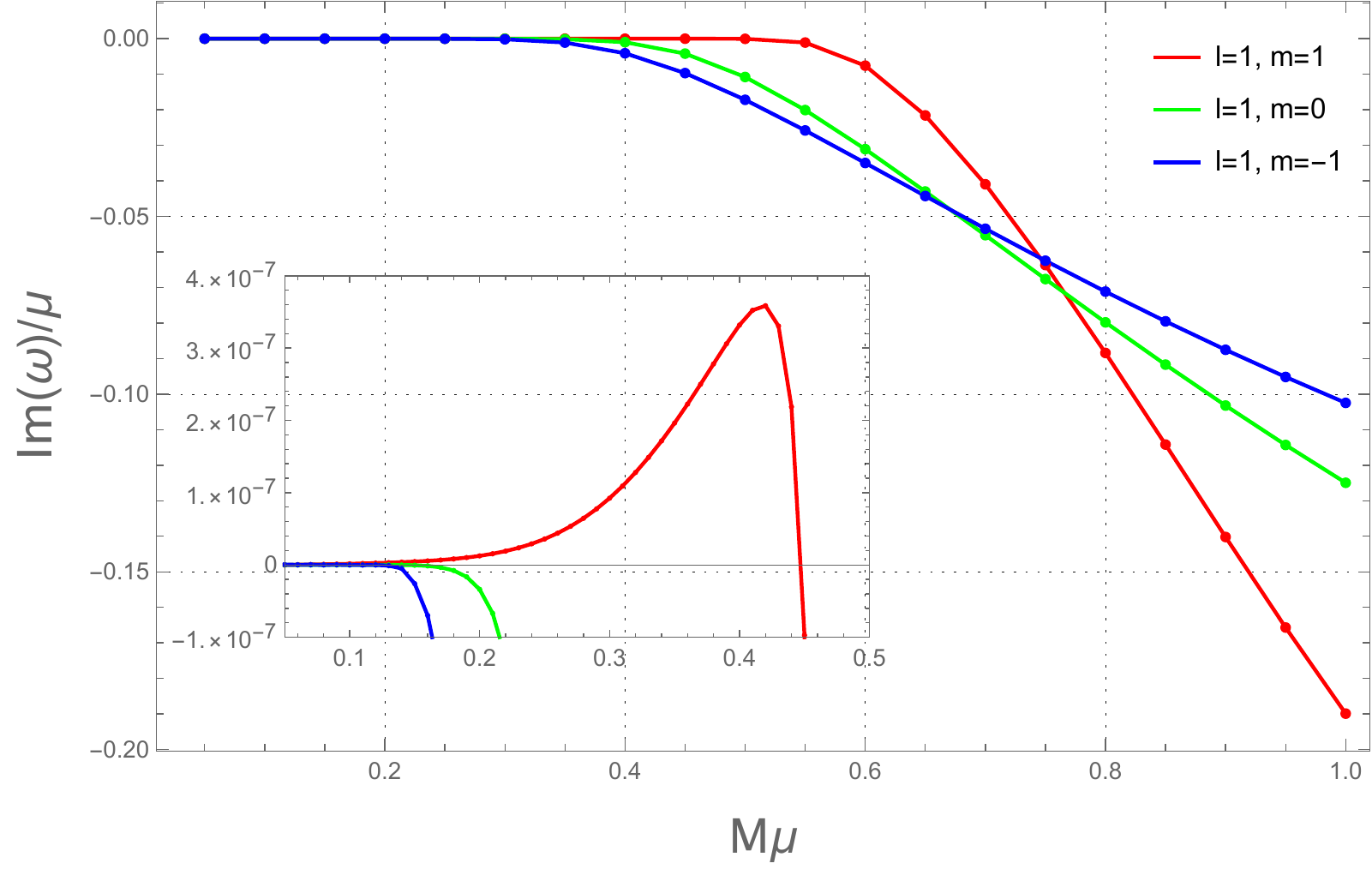}
}
{
\includegraphics[width=0.31\columnwidth]{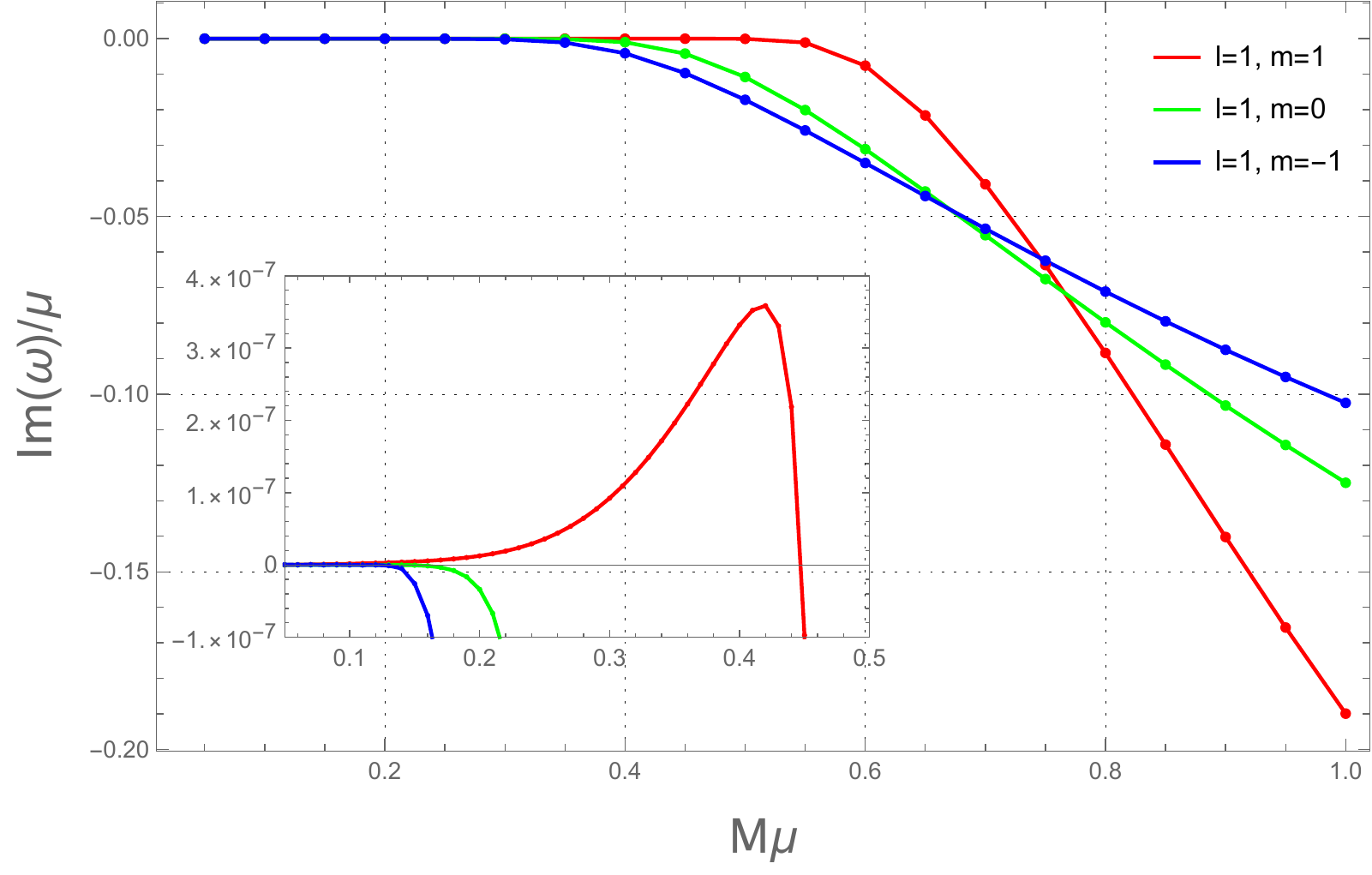}
}
\caption{The quasibound states of the black holes in massive scalar field for the state $l=1$ (left panel is CDM model, middle panel is SFDM model and the right panel is Kerr spacetime). Subfigures in three panels all reflect the superradiant instabilities. The maximum instability occurs for the state ($l=1, m=1$). The main calculation parameters are $a=0.99$, $M=1$, $\rho_c=2.45 \times 10^{-3}$ $M_{\bigodot}/pc^{3}$, $R_c=5.7$ $kpc$, $\rho_s=13.66 \times 10^{-3}$ $M_{\bigodot}/pc^{3}$, $R_s=2.92$ $kpc$. We have converted these main calculation parameters to the black hole units before plotting.}
\label{nf5}
\end{figure*}

\begin{figure*}[tbp]
\centering
{
\includegraphics[width=0.31\columnwidth]{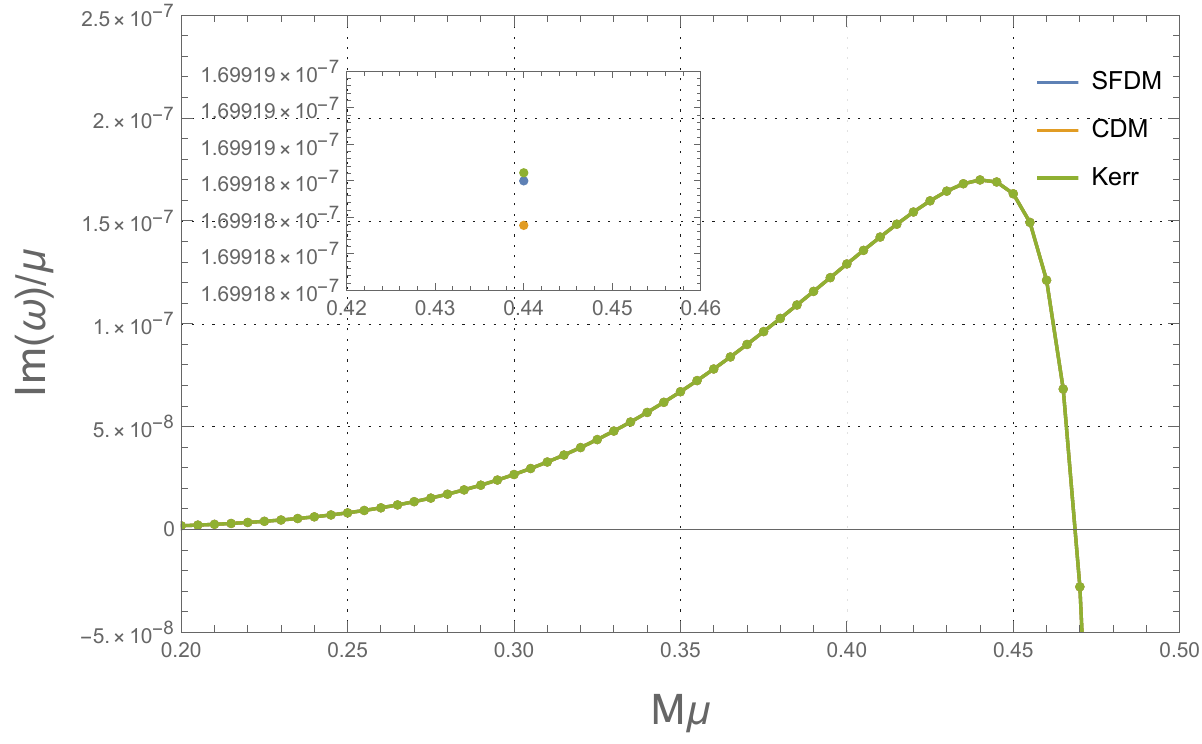}
}
{
\includegraphics[width=0.31\columnwidth]{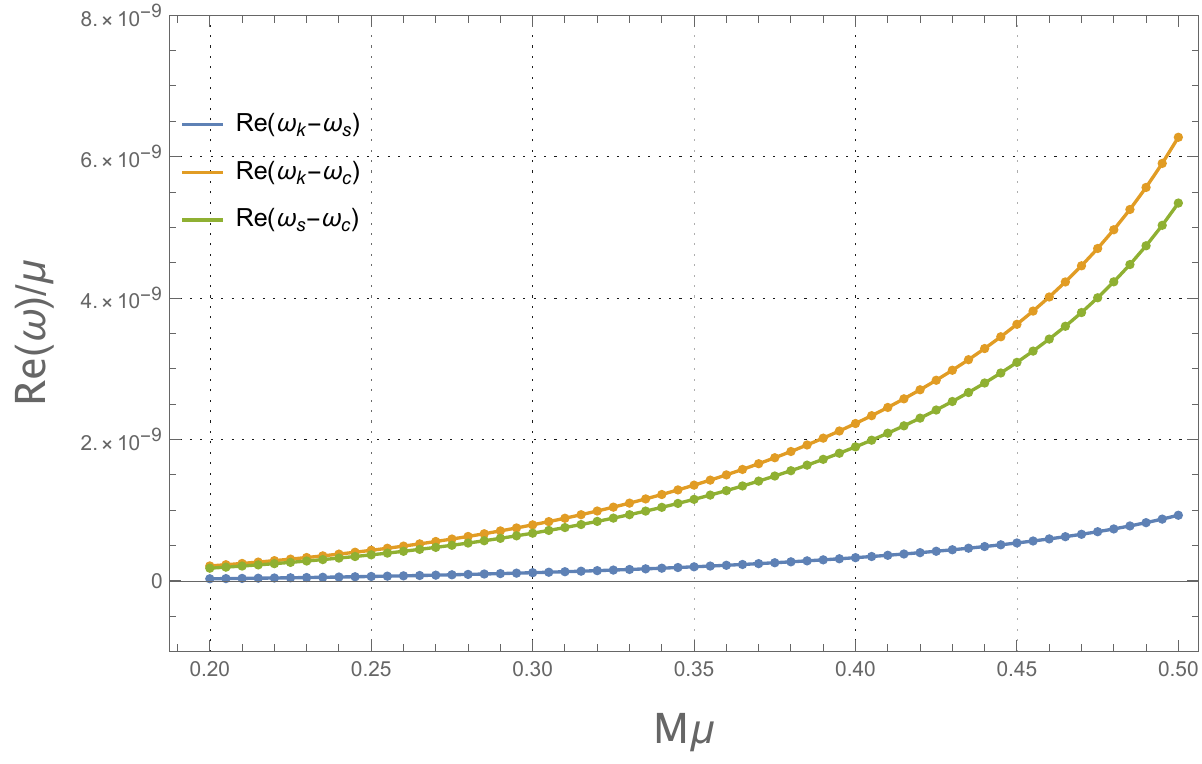}
}
{
\includegraphics[width=0.31\columnwidth]{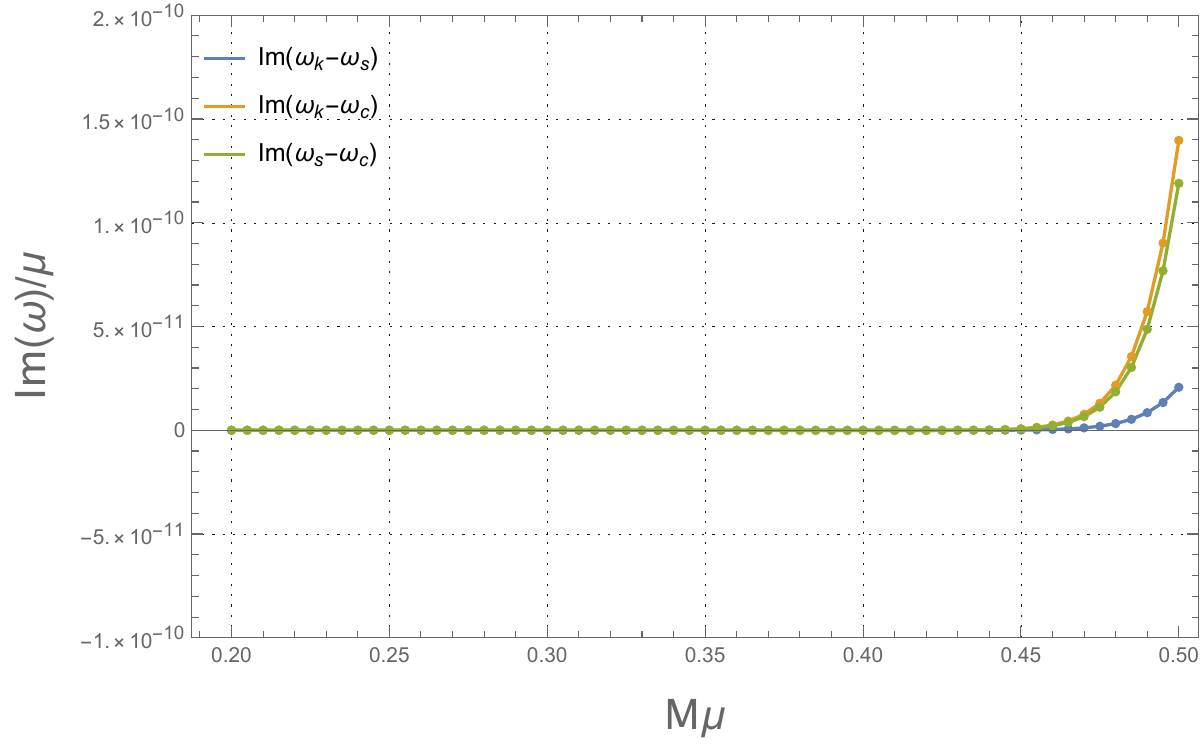}
}
\caption{The superradiant instabilities of the black holes in massive scalar field at the state ($l=1, m=1, a=0.995$) in CDM model, SFDM model and Kerr spacetime. The instability of the Kerr black hole is larger than that of the SFDM model, and the SFDM model is larger than the CDM model (left panel). After comparing Kerr black holes both with the SFDM model and CDM model, the real part of the comparing results are shown in the middle panel and the imaginary part are shown in the right panel. The main calculation parameters are $M=1$, $\rho_c=2.45 \times 10^{-3}$ $M_{\bigodot}/pc^{3}$, $R_c=5.7$ $kpc$, $\rho_s=13.66 \times 10^{-3}$ $M_{\bigodot}/pc^{3}$, $R_s=2.92$ $kpc$.}
\label{nf8}
\end{figure*}

\begin{figure*}[tbp]
\centering
{
\includegraphics[width=0.31\columnwidth]{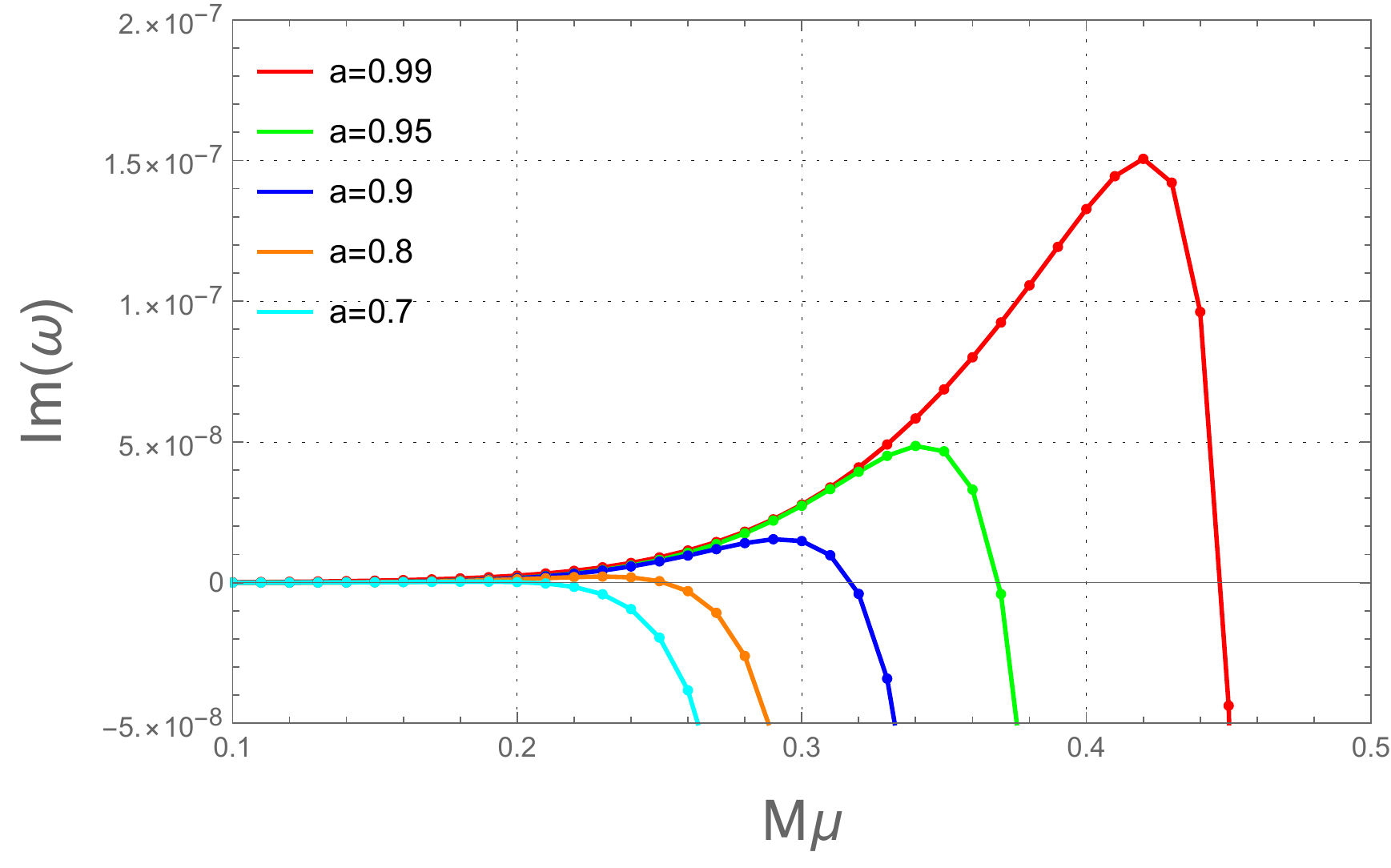}
}
{
\includegraphics[width=0.31\columnwidth]{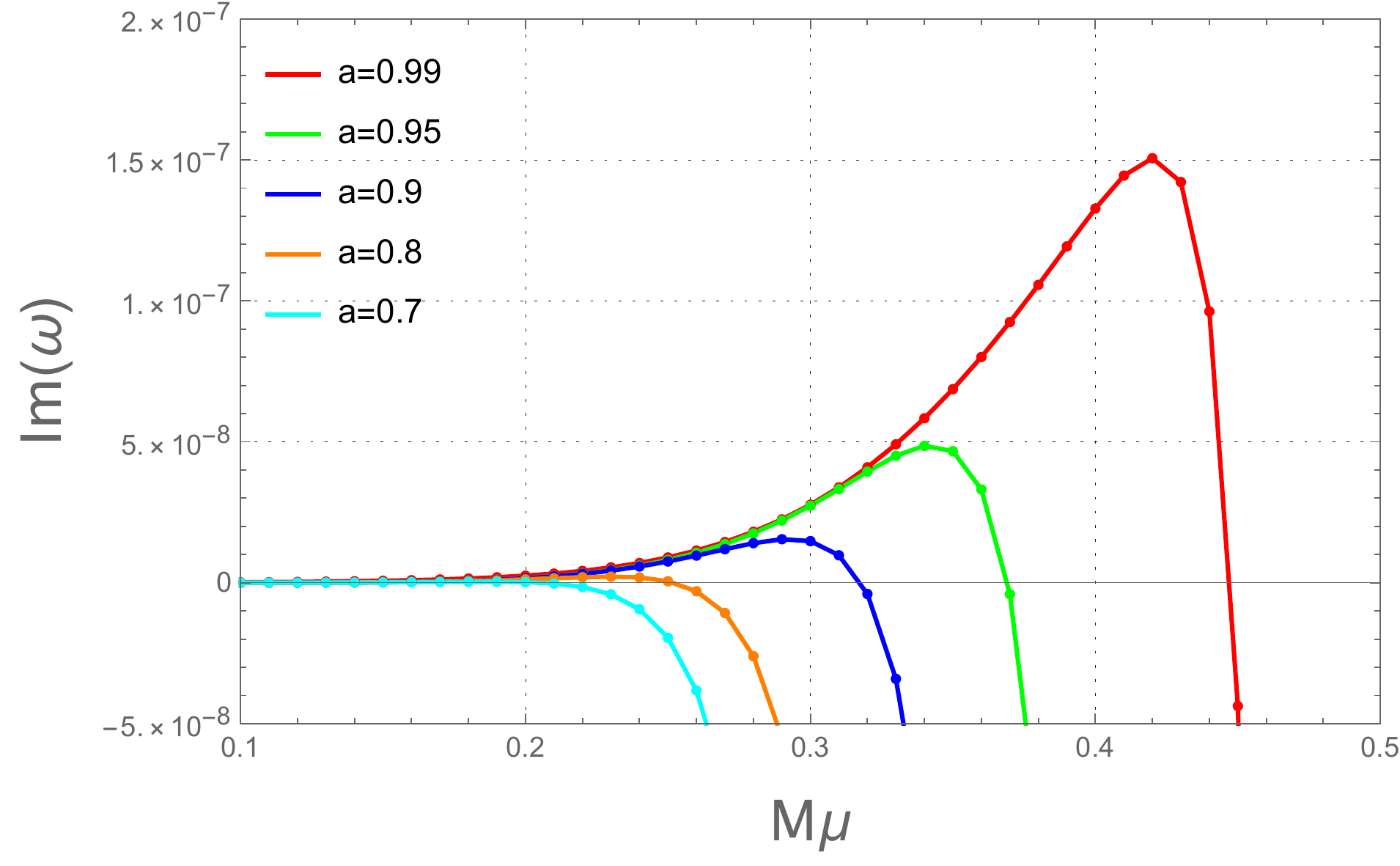}
}
{
\includegraphics[width=0.31\columnwidth]{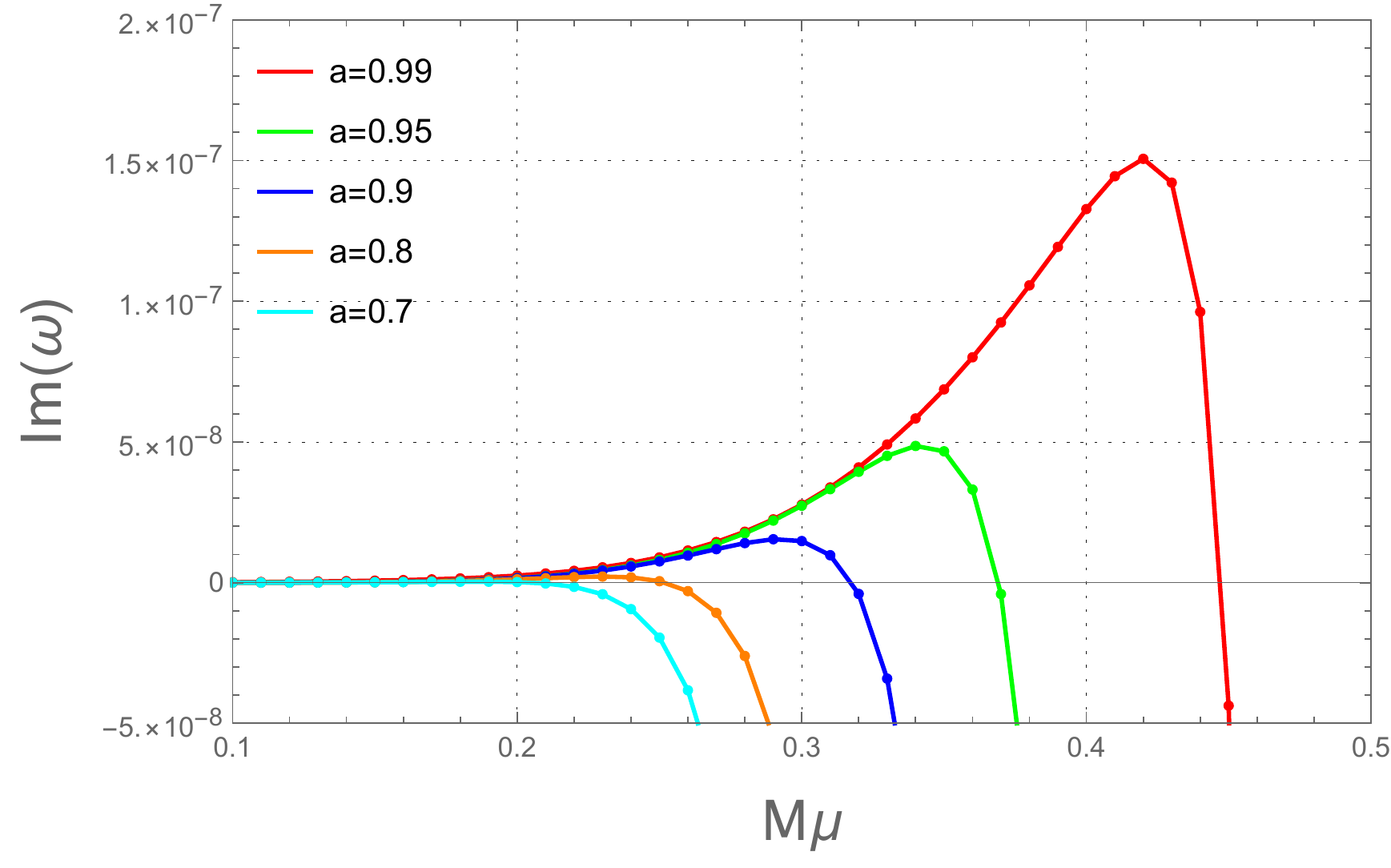}
}
{
\includegraphics[width=0.31\columnwidth]{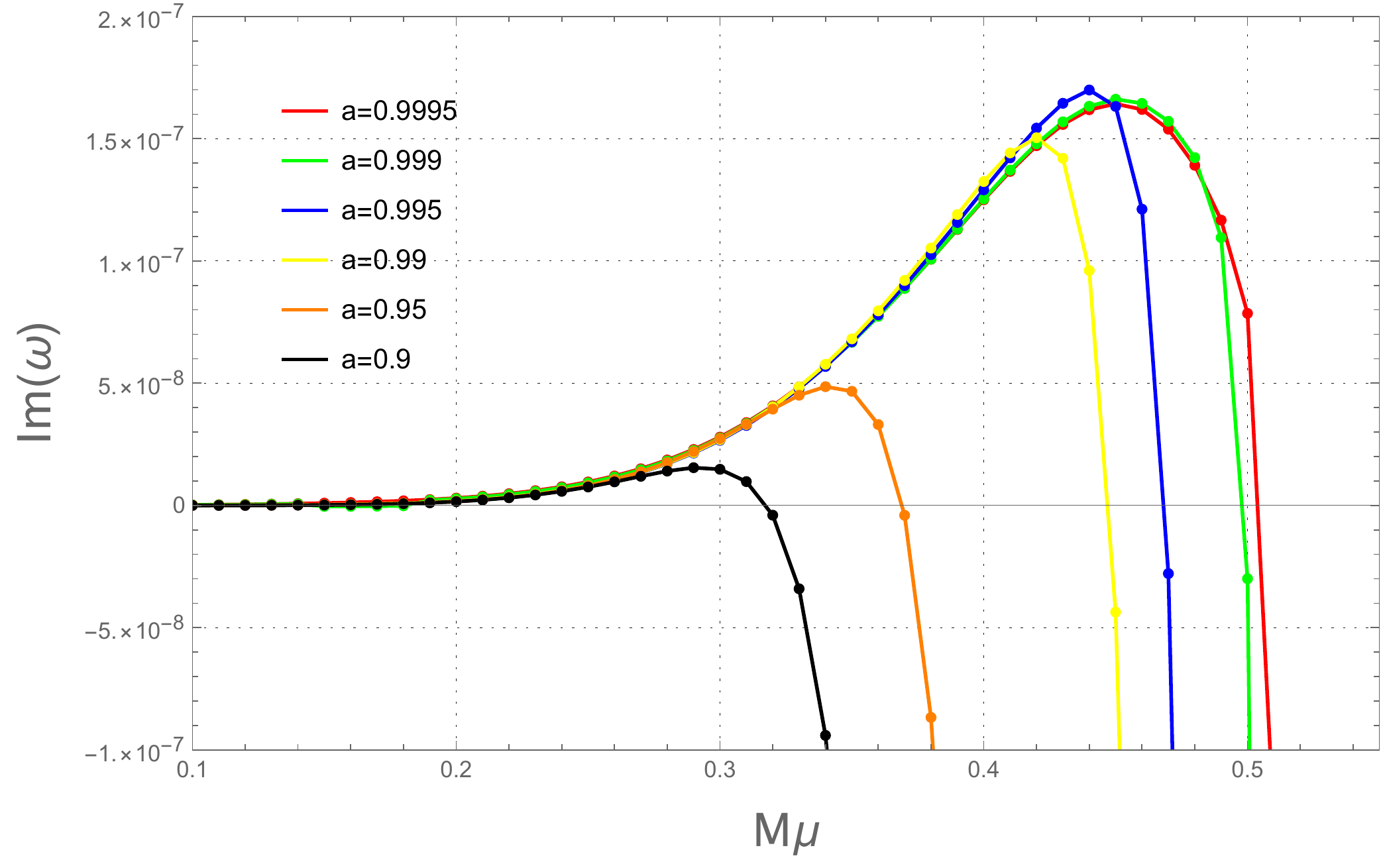}
}
{
\includegraphics[width=0.31\columnwidth]{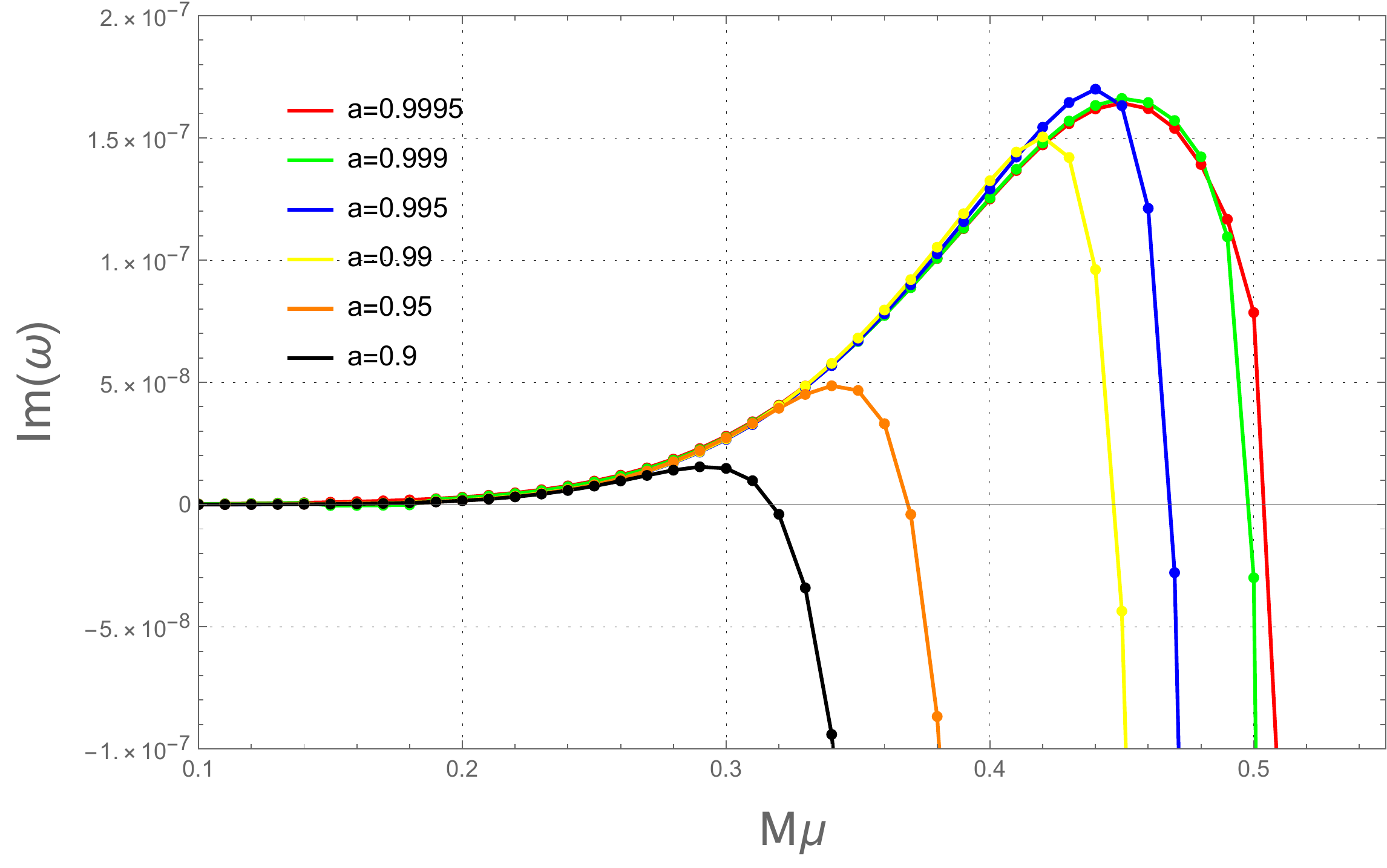}
}
{
\includegraphics[width=0.31\columnwidth]{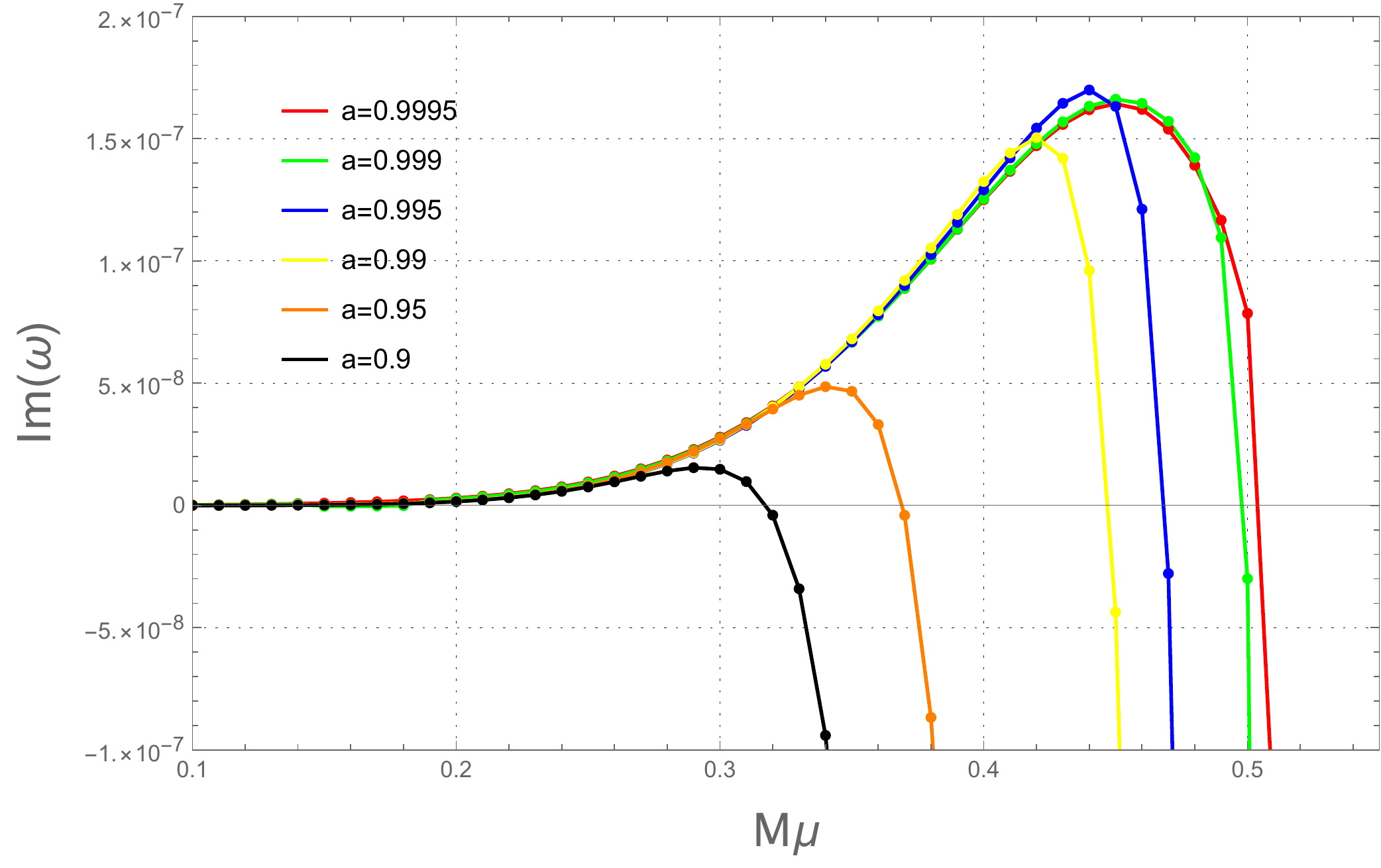}
}
\caption{The superradiant instabilities of the black holes for different rotation parameter $a$ in massive scalar field at the state ($l=1, m=1$) in CDM model (left panel), SFDM model and Kerr spacetime (right panel). In the three panels of the top, the maximum instabilities all increase with the increasing of the rotation parameter $a$. But in three panels of the bottom, the maximum instabilities first increase and then decrease with the increasing of the rotation parameter $a$. This turning point is $a=0.995$. The main calculation parameters are $M=1$, $\rho_c=2.45 \times 10^{-3}$ $M_{\bigodot}/pc^{3}$, $R_c=5.7$ $kpc$, $\rho_s=13.66 \times 10^{-3}$ $M_{\bigodot}/pc^{3}$, $R_s=2.92$ $kpc$.}
\label{nf9}
\end{figure*}

\begin{figure*}[t!]
\centering
{
\includegraphics[width=0.45\columnwidth]{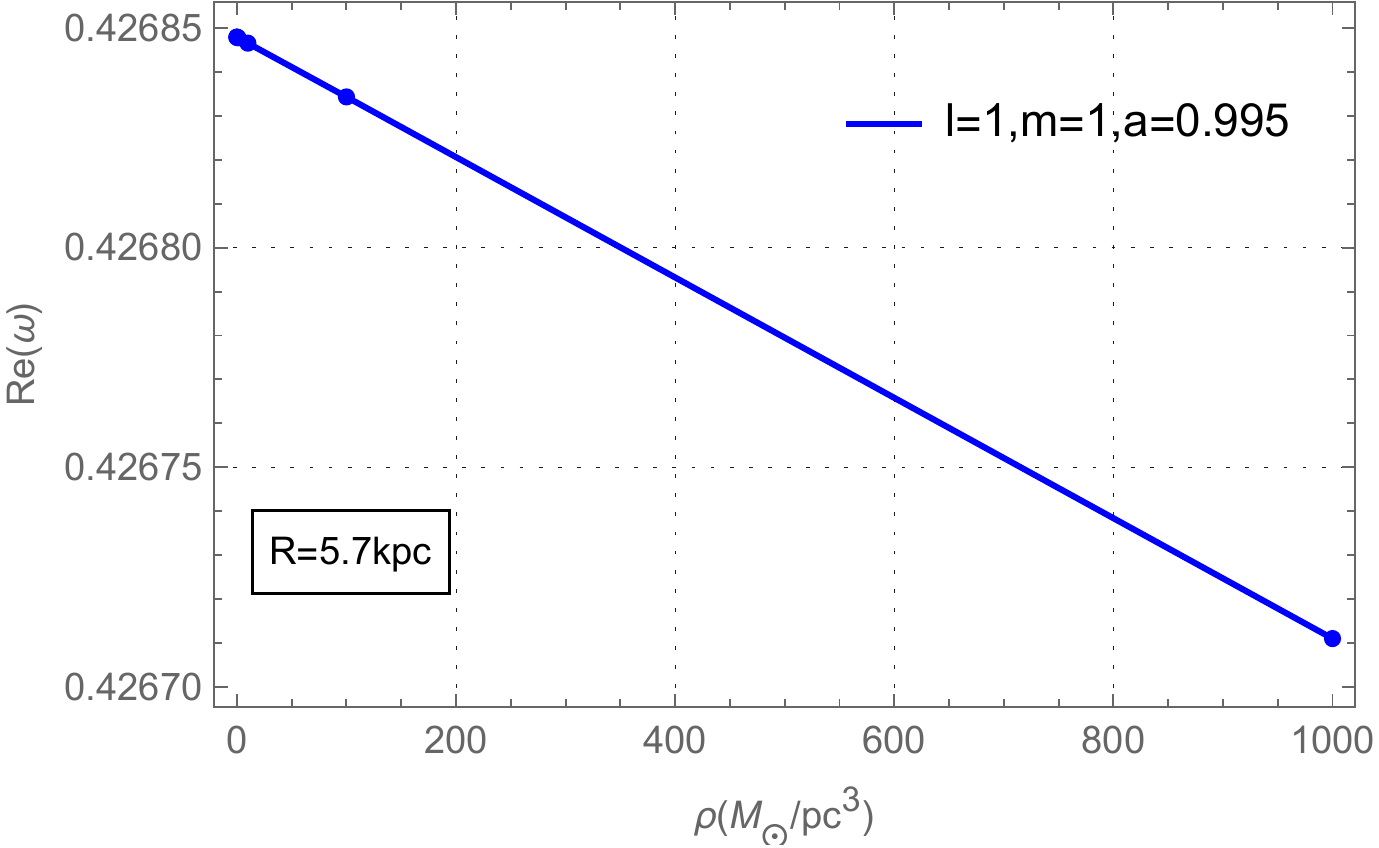}
}
{
\includegraphics[width=0.45\columnwidth]{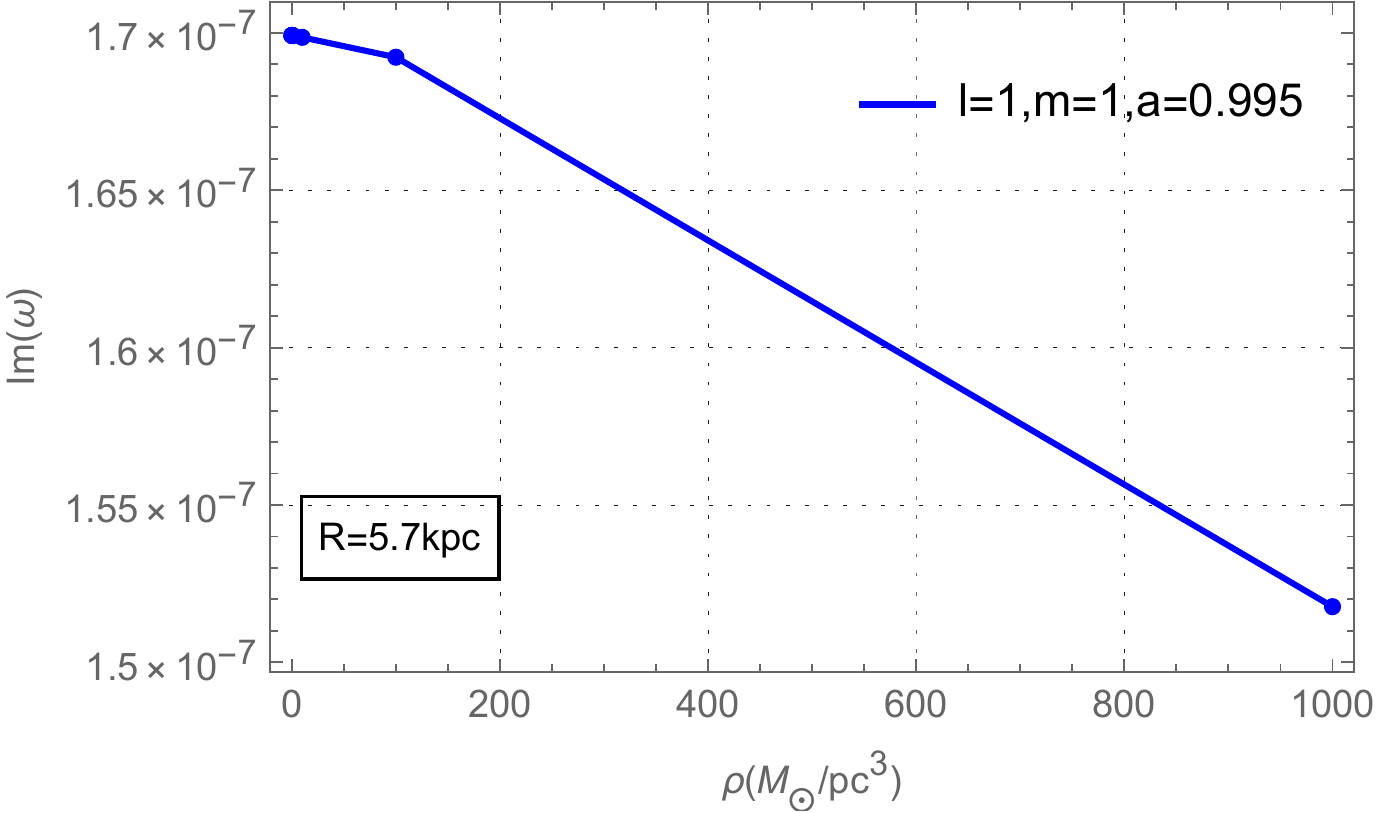}
}
{
\includegraphics[width=0.45\columnwidth]{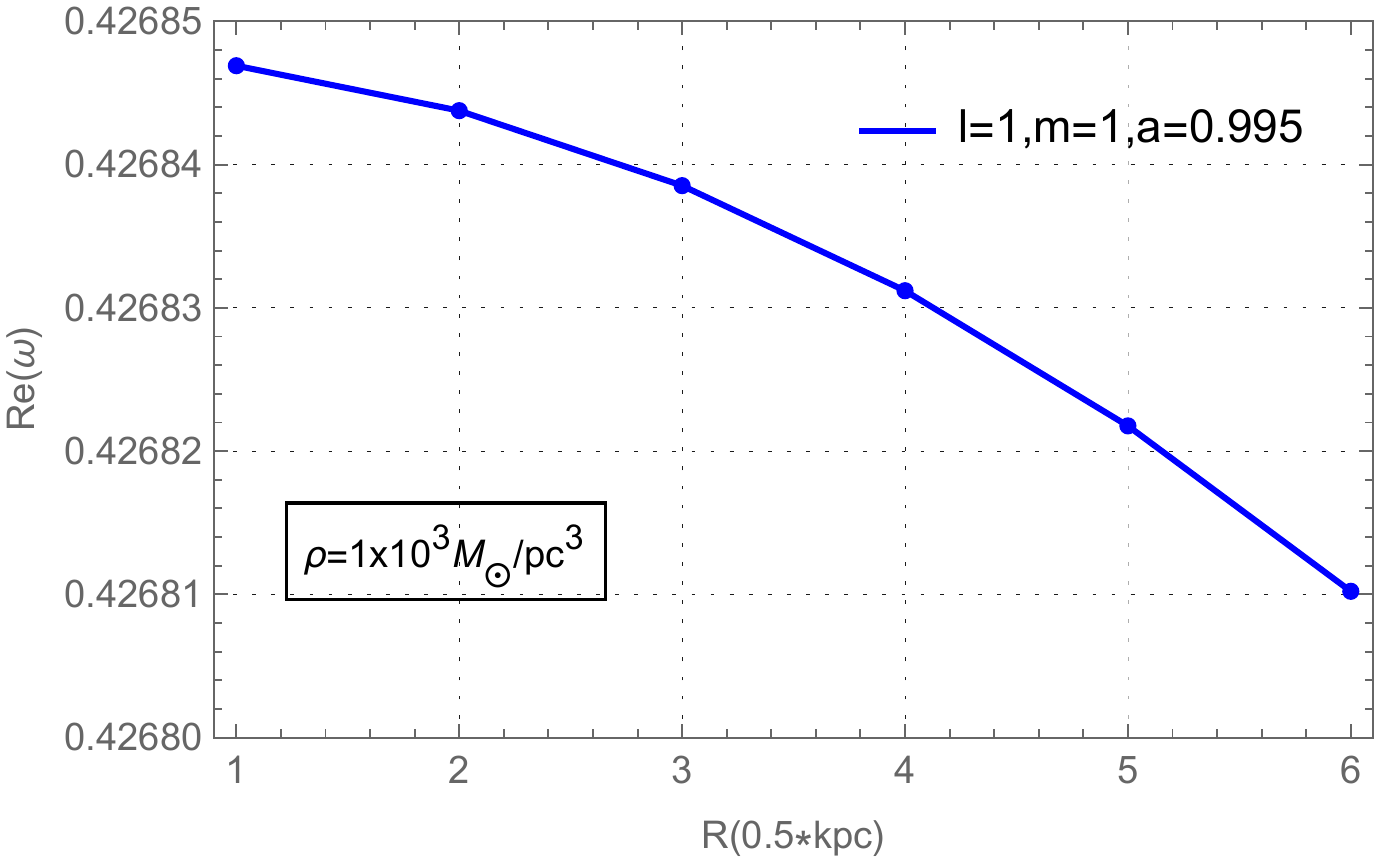}
}
{
\includegraphics[width=0.45\columnwidth]{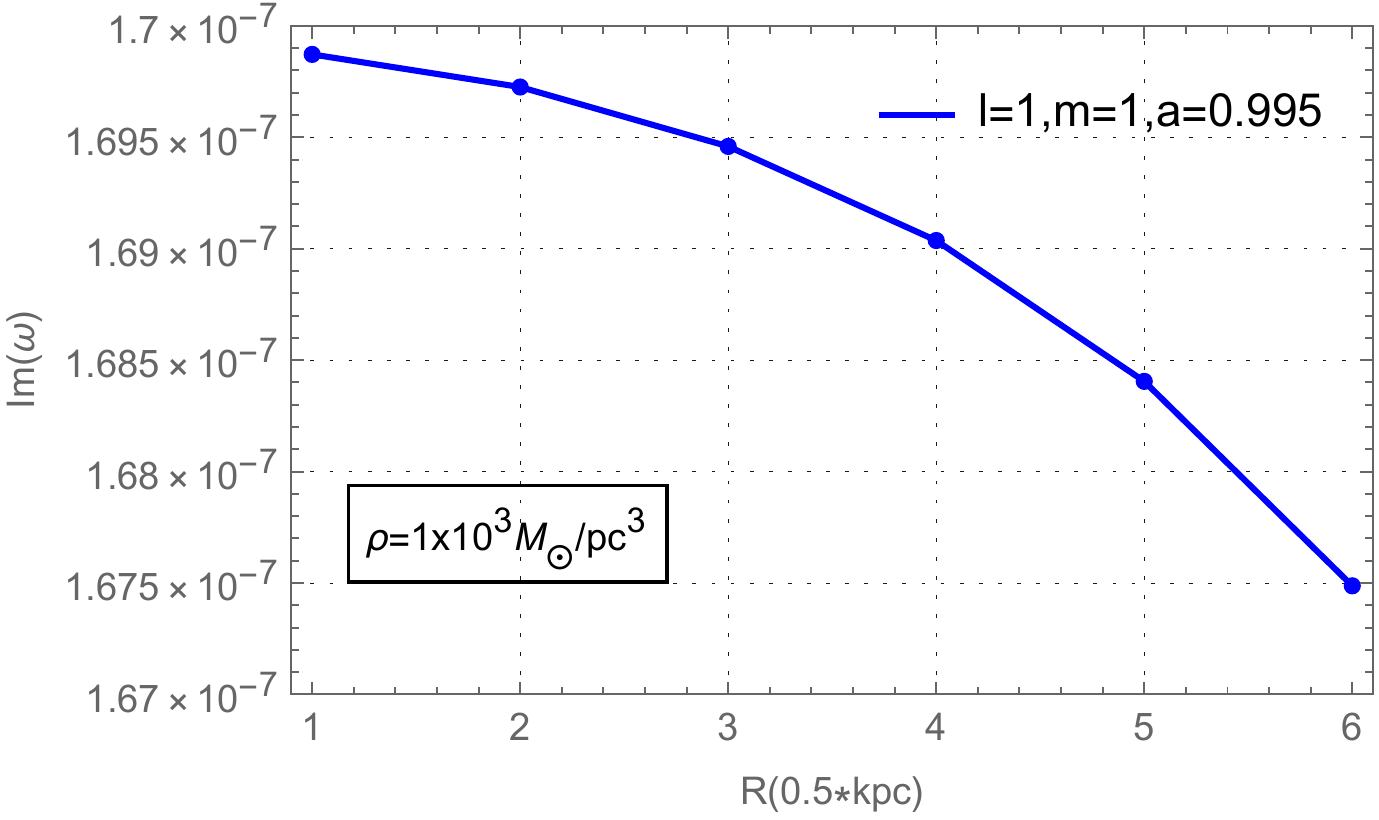}
}
\caption{The quasibound states of the black holes in massive scalar field $(\mu=0.440)$ at the state $l=1, m=1, a=0.995$ in SFDM model, as a function of $\rho$ (top) and $R$ (bottom). The top panels we fix the $R=5.7kpc$ and the bottom panels we fix the $\rho=1 \times 10^{3}$ $M_{\bigodot}/pc^{3}$. The right panels reveal the instabilities. We have converted these main calculation parameters to the black hole units before plotting.}
\label{rr3}
\end{figure*}

\begin{figure*}[t!]
\centering
{
\includegraphics[width=0.45\columnwidth]{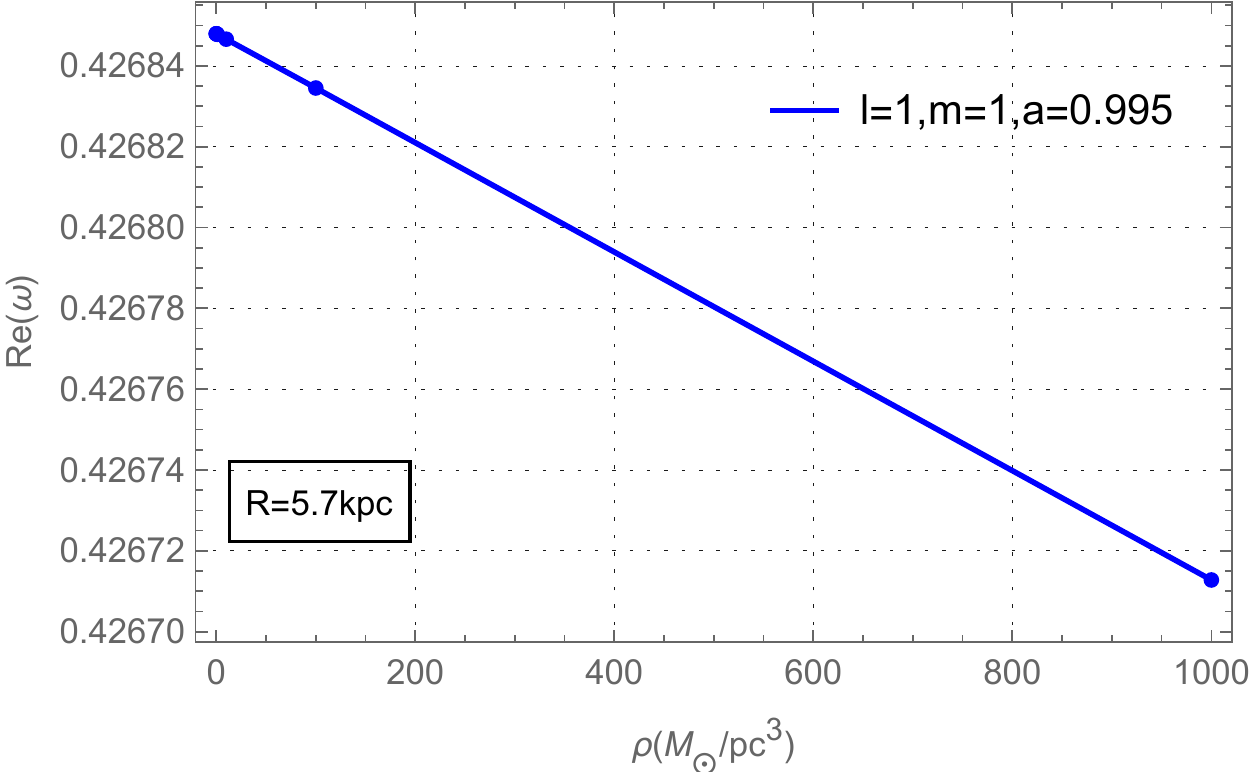}
}
{
\includegraphics[width=0.45\columnwidth]{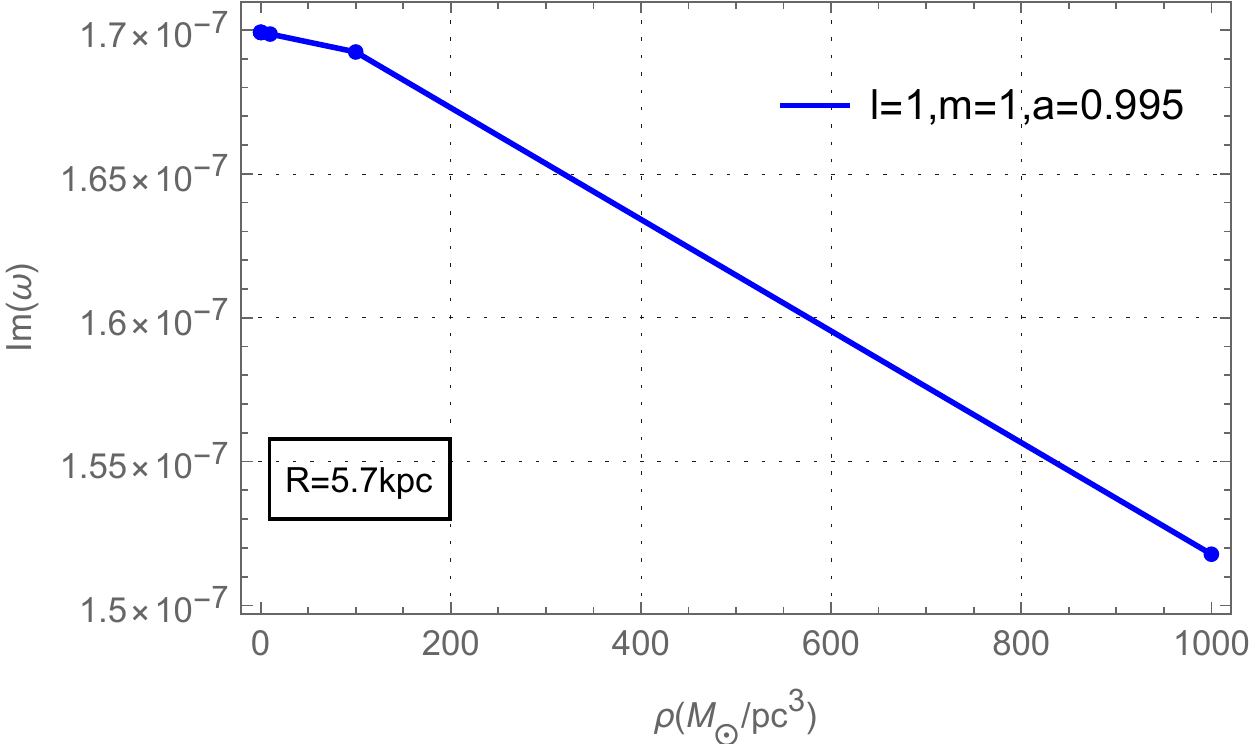}
}
{
\includegraphics[width=0.45\columnwidth]{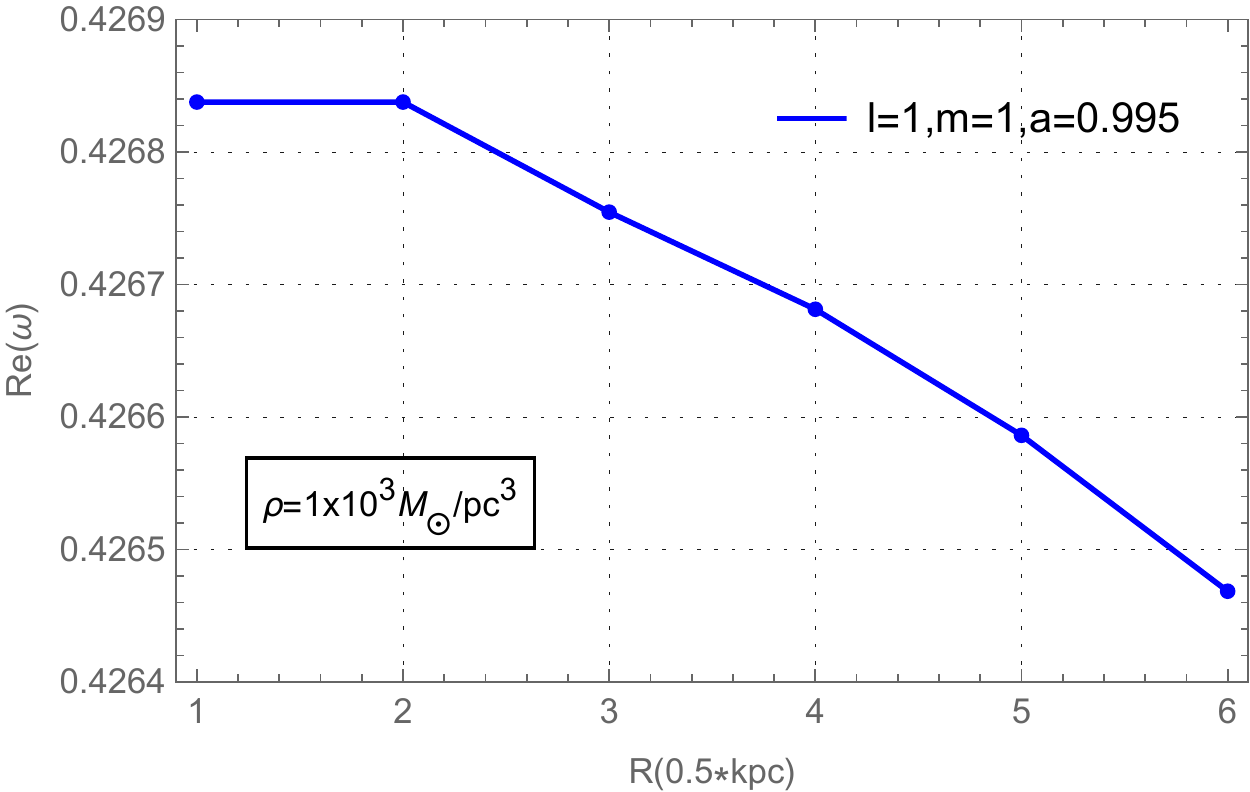}
}
{
\includegraphics[width=0.45\columnwidth]{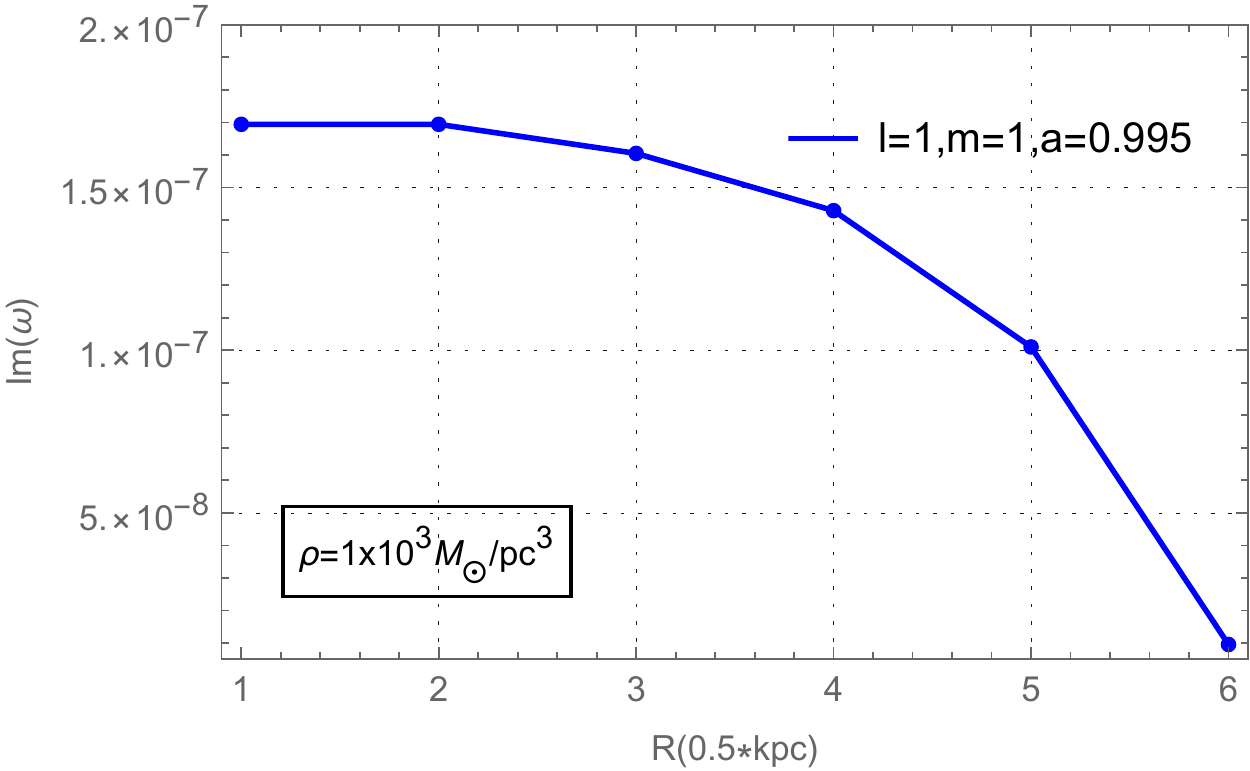}
}
\caption{The quasibound states of the black holes in massive scalar field $(\mu=0.440)$ at the state $l=1, m=1, a=0.995$ in CDM model, as a function of $\rho$ (top) and $R$ (bottom). The top panels we fix the $R=5.7kpc$ and the bottom panels we fix the $\rho=1 \times 10^{3}$ $M_{\bigodot}/pc^{3}$. The right panels reveal the instabilities. We have converted these main calculation parameters to the black hole units before plotting.}
\label{rr4}
\end{figure*}

\begin{figure*}[t!]
\centering
{
\includegraphics[width=0.45\columnwidth]{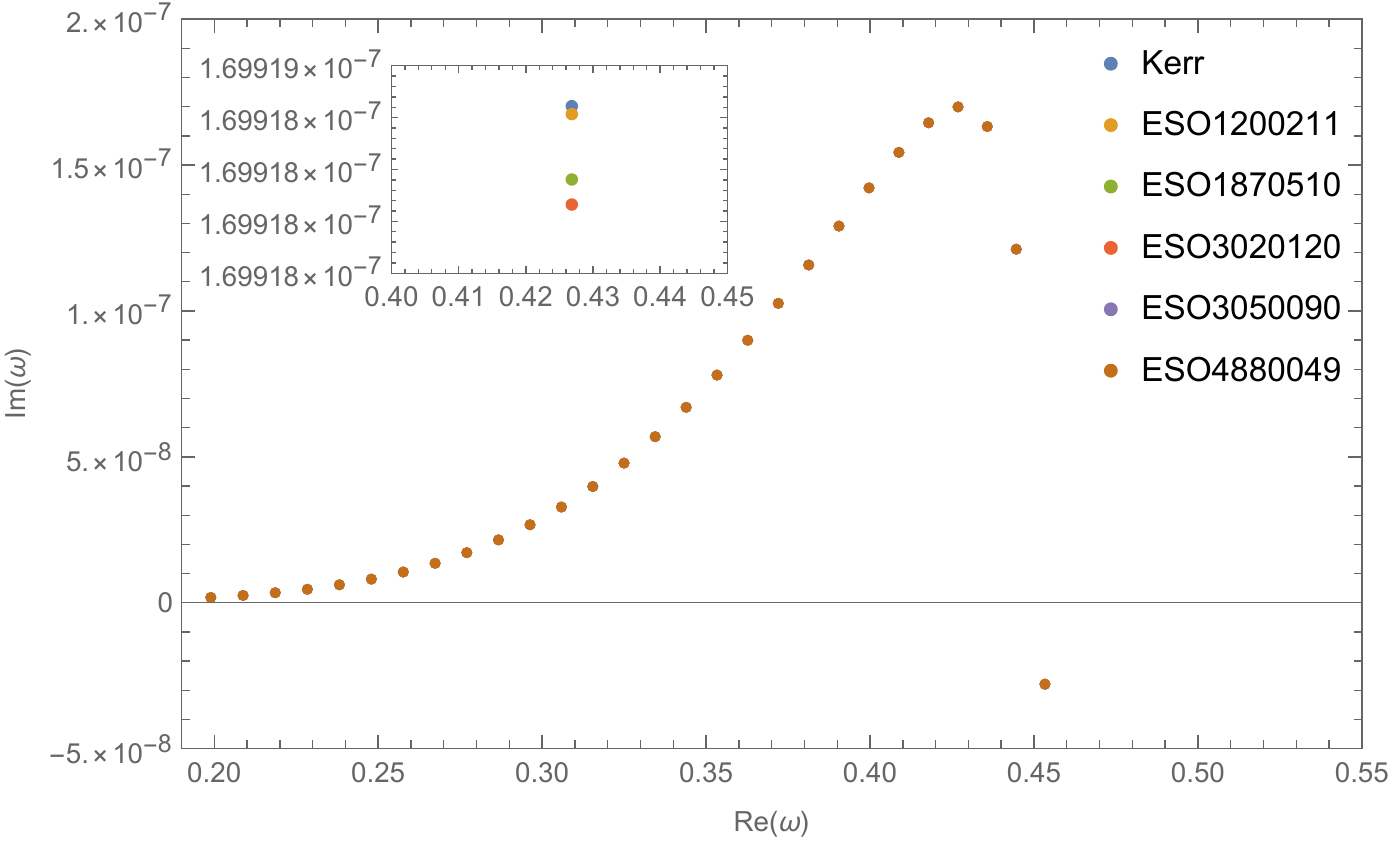}
}
{
\includegraphics[width=0.45\columnwidth]{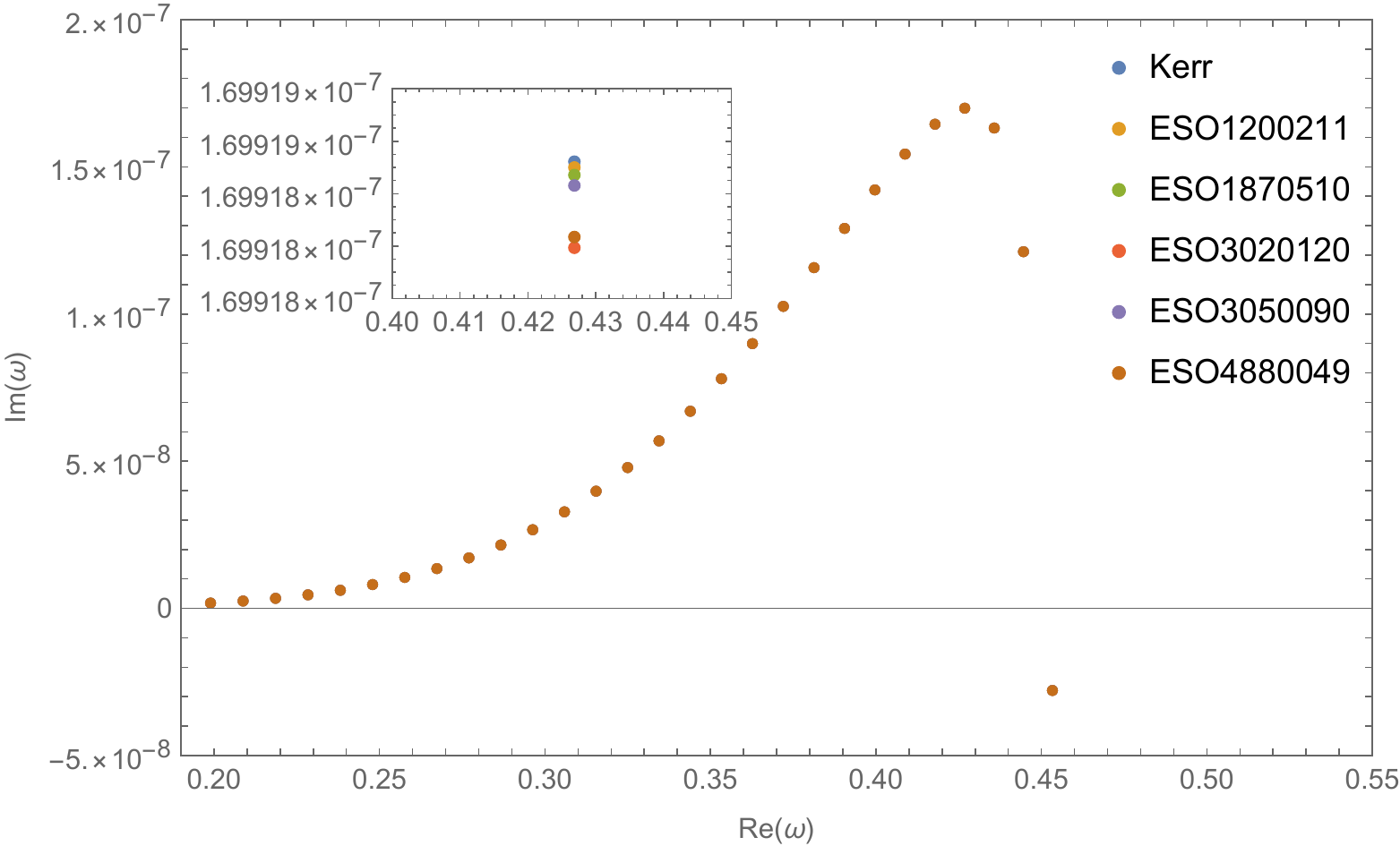}
}
\caption{The quasibound states of the black holes from different galaxies in massive scalar field at the state ($l=1, m=1, a=0.995$) in the CDM model (left panel) and SFDM model (right panel). These two panels show superradiant instabilities. The maximum instability of Kerr black holes is larger than that of other galaxies. The points in each panel correspond to different mass, from left to right are the $\mu=0.2$ to $\mu=0.55$ , with a step-size of $0.01$. The main dark matter parameters are from the data in Table \ref{t11}.}
\label{nf10}
\end{figure*}

\indent Just like the previous calculation and analysis of QNM frequencies, we firstly study the QNM frequencies in CDM model, SFDM model and Kerr spacetime at the state $l=1$ in the massive scalar field, and compare them with the massless scalar field. Here, we also use the Schwarzschild fundamental modes as our initial guess. We present our results in Figure \ref{nf4}. Compared with the massless scalar field, the imaginary part of QNM frequency at the same state decreases rapidly with the increasing of mass $\mu$ in CDM model, SFDM model and Kerr spacetime, while their real parts increase with the increasing of mass $\mu$. This seems to be predictable, as the mass increases to a certain critical value, there will be superradiant instability, and the state $l=m=1$ may be the fastest and most important one. Actually, in Figure \ref{nf4}, we find that these QNM frequencies are still the stable modes at these states.
%%%%%%%%%%%%%%%%%%%%%%%%%%%%%%%%%%%%%%%%%%%%%%%%%%%%%%%%%
%%%%%%%%%%%%%%%%%%%%%%%%%%%%%%%%%%%%%%%%%%%%%%%%%%
In order to further confirm whether these two dark matter models have superradiant instability, we use the bound state spectrum of the Schwarzschild black hole as our initial guess. At the same time, we rewrite the oscillation frequency and decay rate as $\text{Re}(\omega)/\mu$ and $\text{Im}(\omega)/\mu$, respectively, as a function of the mass parameter $M\mu$. Then, we investigate all the state for the angular quantum $l=1$ in the case of the nearly extremal parameter ($a=0.99$) in CDM model, SFDM model and Kerr spacetime. We show our calculation results in Figure \ref{nf5}.
%%%%%%%%%%%%%%%%%%%%%%%%%%%%%%%%%%%%%%%%%%%%%%%%%%%%%%%%%
%%%%%%%%%%%%%%%%%%%%%%%%%%%%%%%%%%%%%%%%%%%%%%%%
From Figure \ref{nf5}, we found that the imaginary part of the QBS in these three panels are negative numbers at the range of large mass, which indicates that QBS is a stable mode in the large mass. However, at the range of the low mass, the value of the decay rate $\text{lm}(\omega)/\mu$ is very close to $0$ (may be greater than $0$ or less than $0$), and the former case may lead to the occurrence of superradiant instability. So, in the subfigures of each panel, we zoomed in the QBS frequencies in the low mass range. These subfigures all demonstrate superradiant instabilities. Among the three state $(l=1,m=1), (l=1,m=1)$ and $(l=1,m=1)$, the maximum instability of the black hole appears at the state $l=1, m=1$. Up to here, our results show that there are instabilities both in SFDM/CDM models. This appears to be due to bound state spectra. This instability seems to be sensitive to the initial guess value. Actually, this bound state spectrum plays an important role in analyzing QBS.
%%%%%%%%%%%%%%%%%%%%%%%%%%%%%%%%%%%%%%%%%%%%%%%%%%%%%%%%%%%%%%%%
\begin{figure*}[tbp]
\centering
{
\includegraphics[width=0.31\columnwidth]{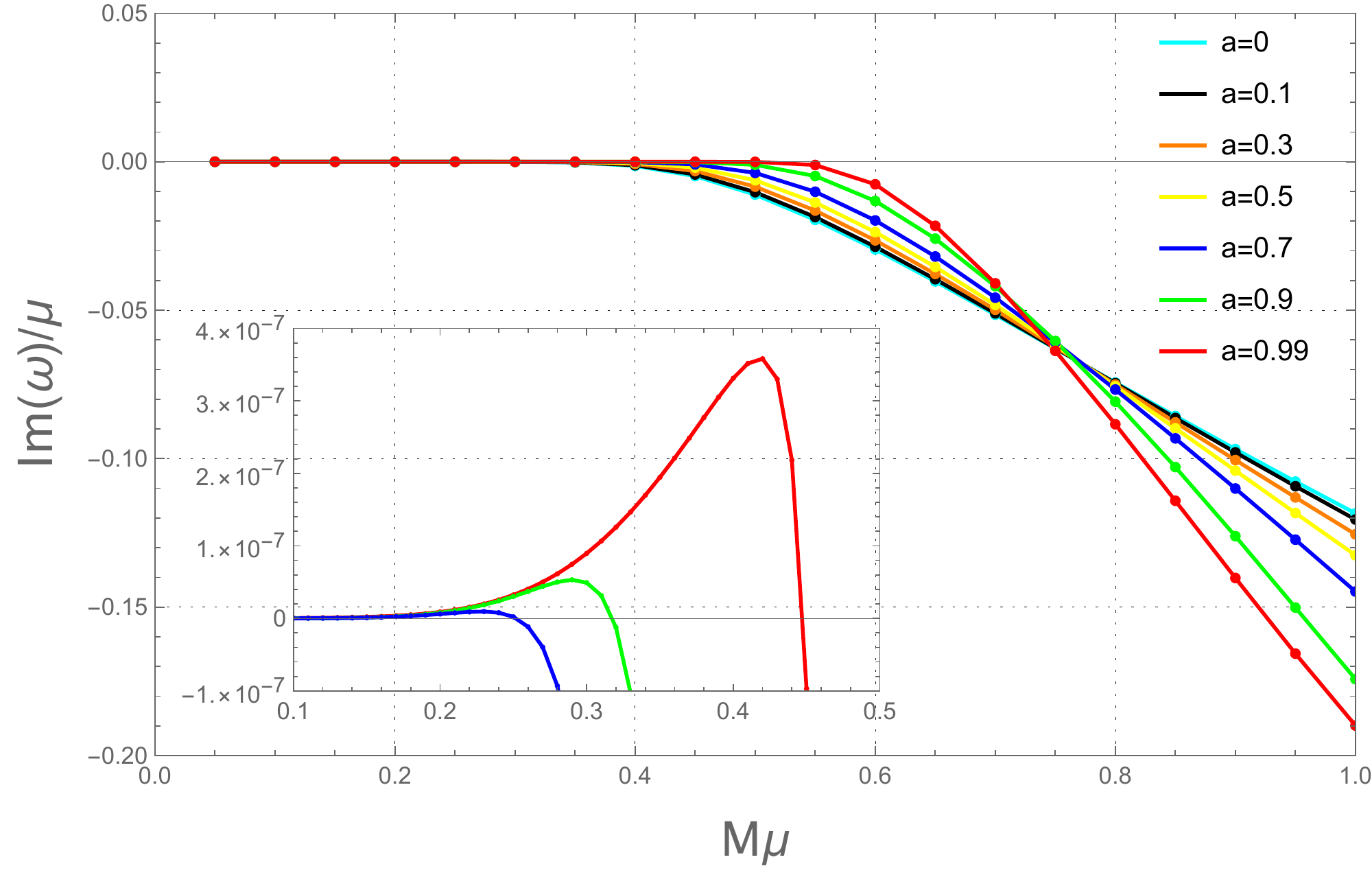}
}
{
\includegraphics[width=0.31\columnwidth]{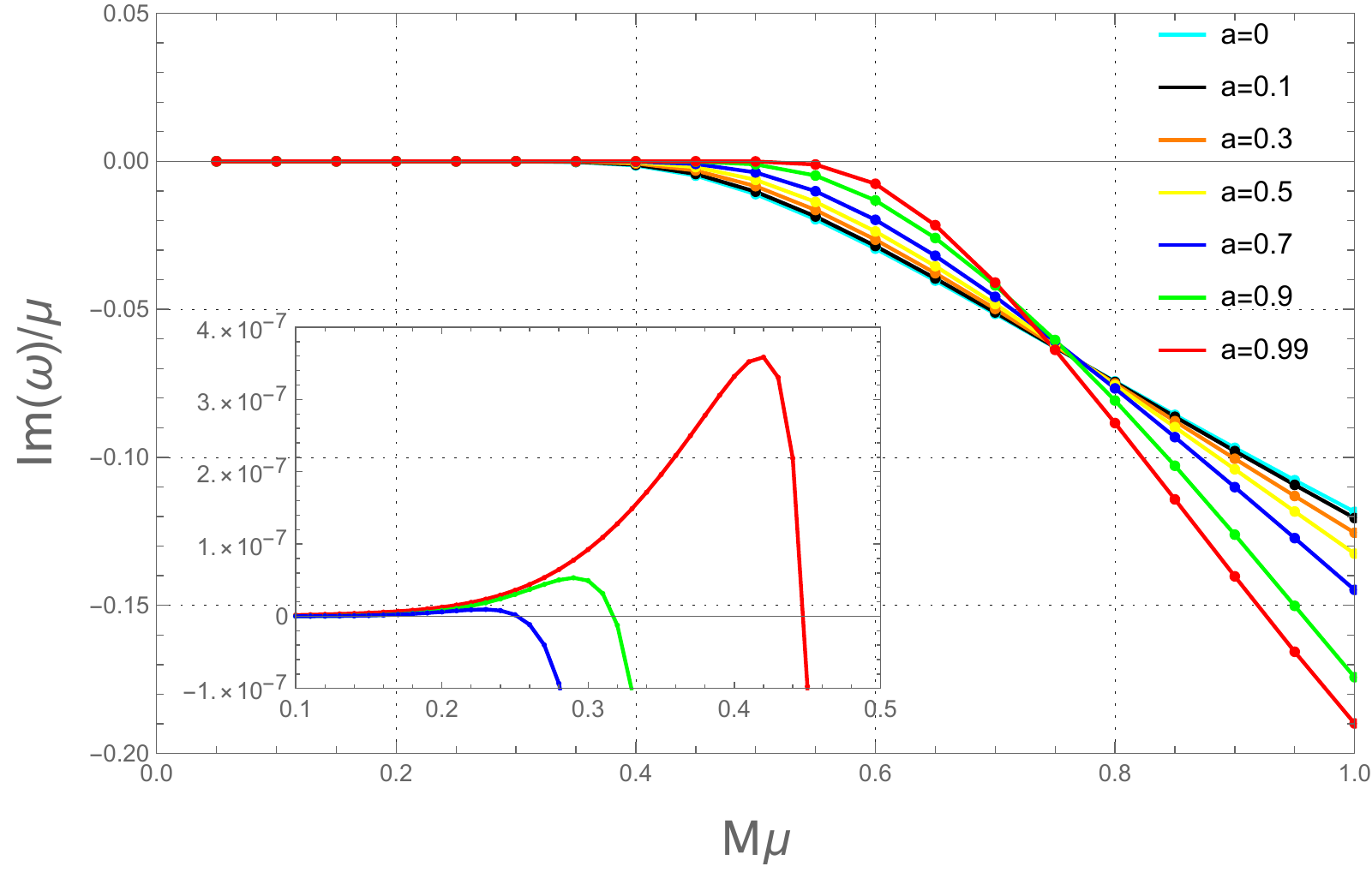}
}
{
\includegraphics[width=0.31\columnwidth]{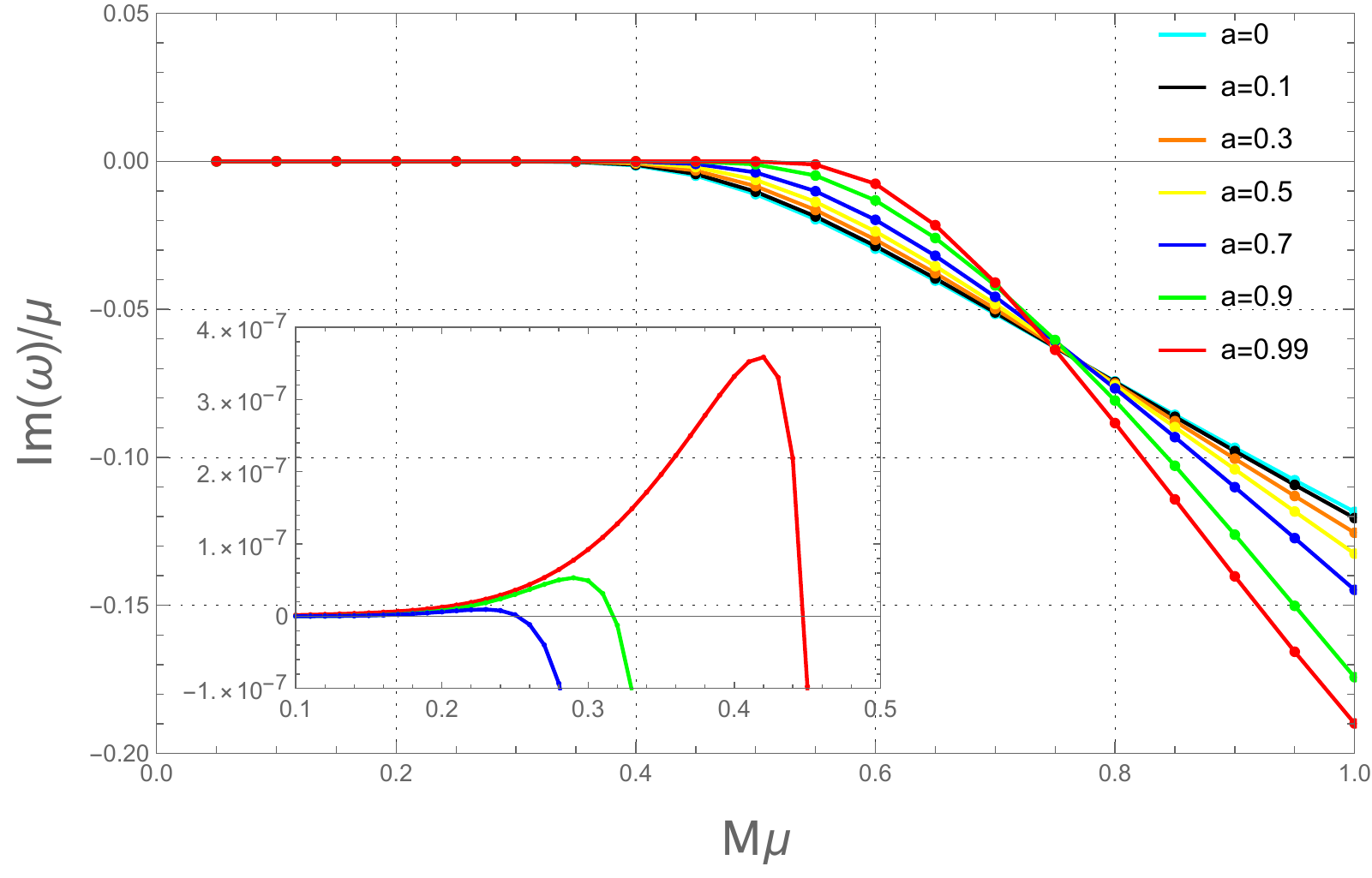}
}
{
\includegraphics[width=0.31\columnwidth]{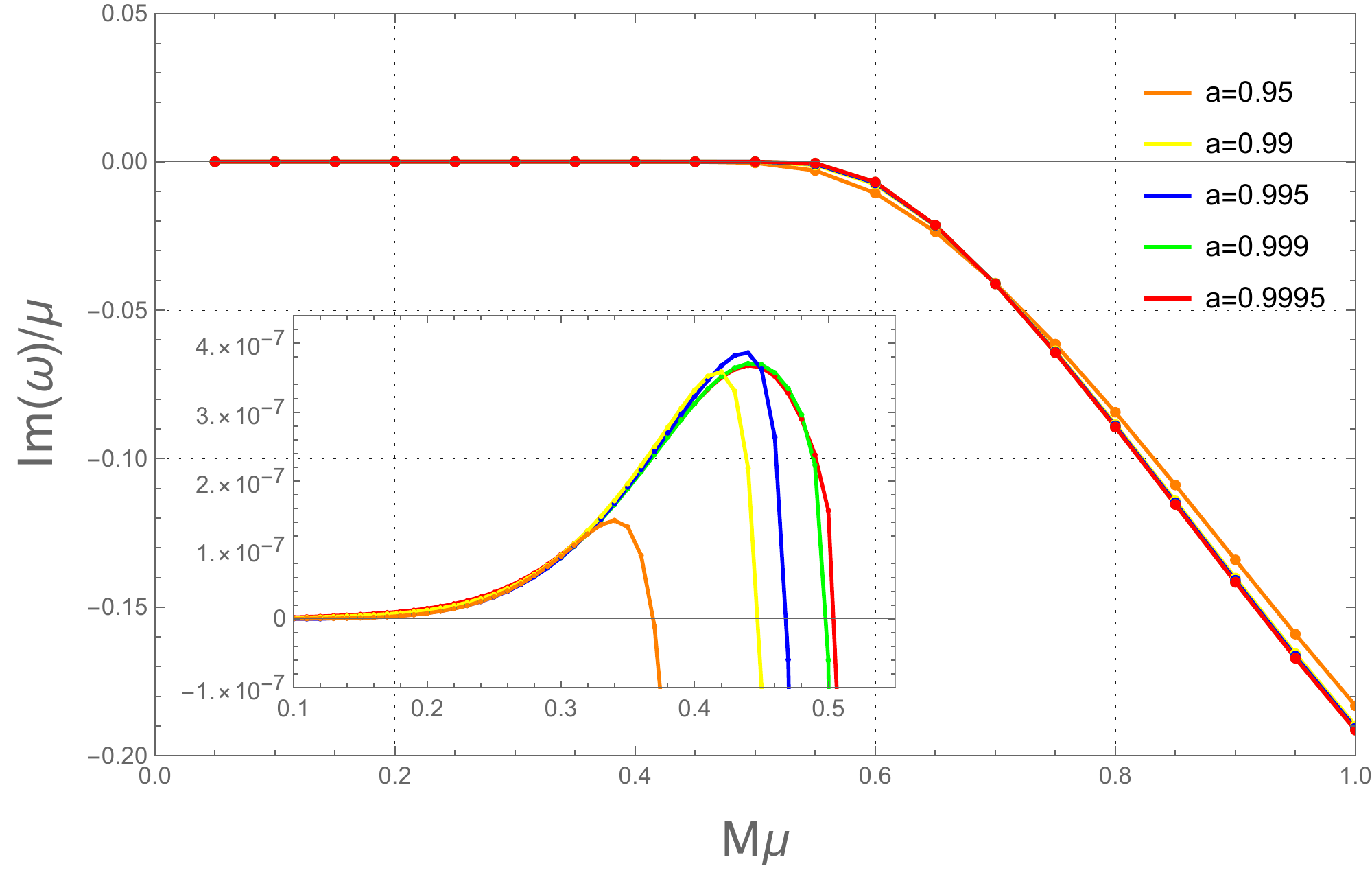}
}
{
\includegraphics[width=0.31\columnwidth]{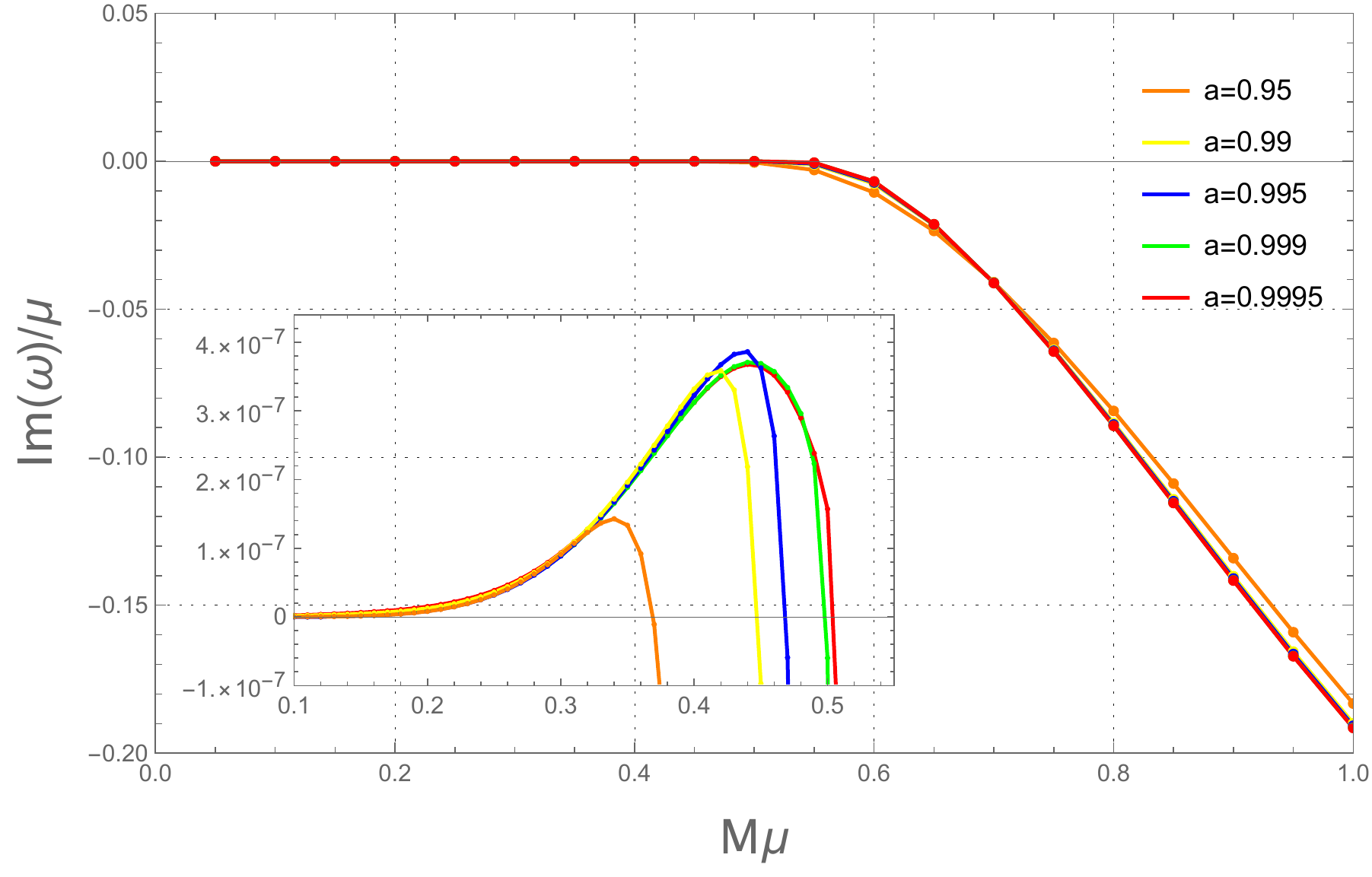}
}
{
\includegraphics[width=0.31\columnwidth]{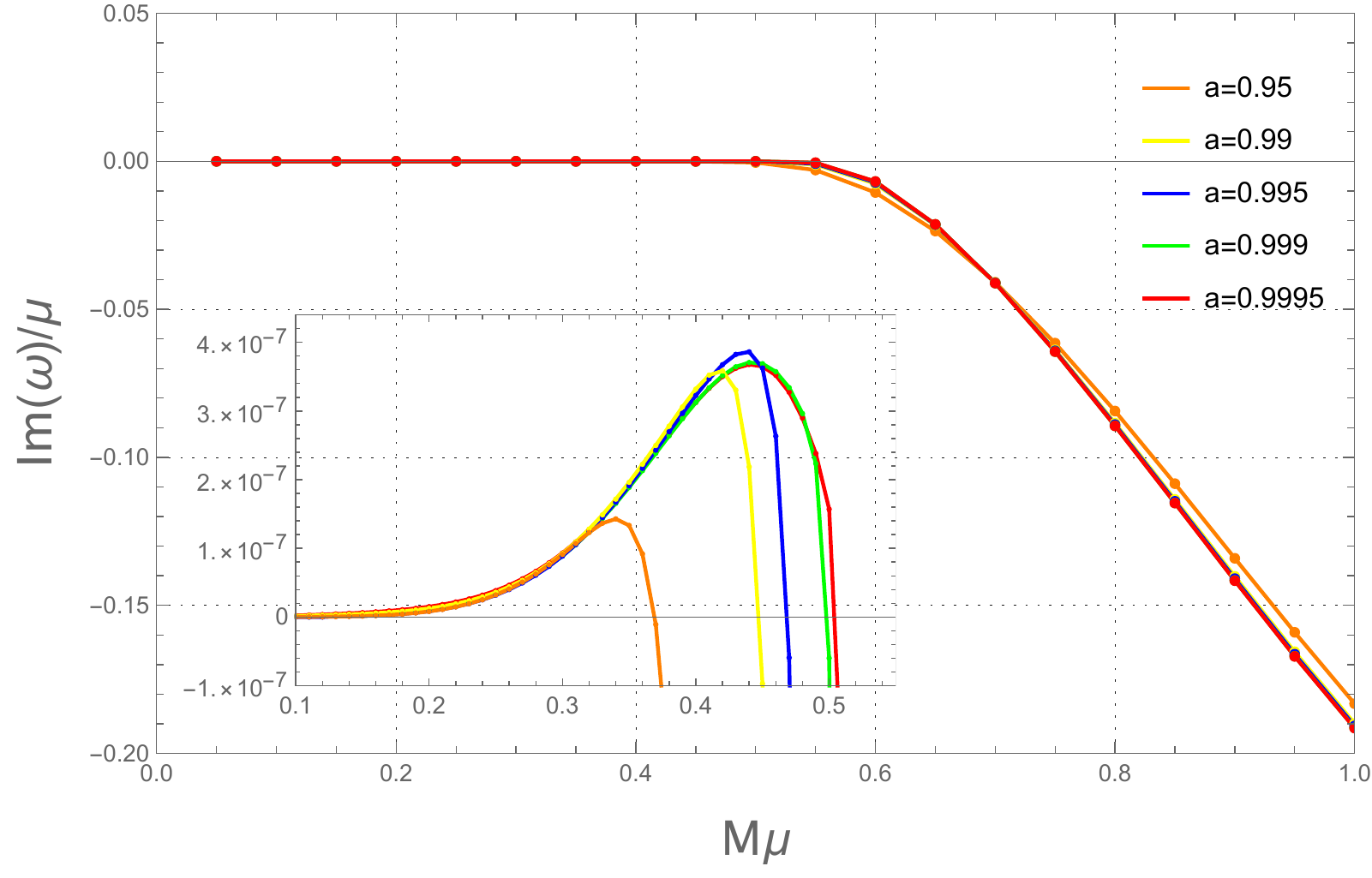}
}
\caption{The quasibound states of the black holes in massive scalar field for the state $l=1, m=1$ (left panel is CDM model, middle panel is SFDM model and the right panel is Kerr spacetime). Subfigures in three panels all reflect the superradiant instabilities, and the maximum instabilities all occur approximately for the state $a=0.995$. The main calculation parameters are $M=1$, $\rho_c=2.45 \times 10^{-3}$ $M_{\bigodot}/pc^{3}$, $R_c=5.7$ $kpc$, $\rho_s=13.66 \times 10^{-3}$ $M_{\bigodot}/pc^{3}$, $R_s=2.92$ $kpc$. We have converted these main calculation parameters to the black hole units before plotting.}
\label{nf6}
\end{figure*}
%%%%%%%%%%%%%%%%%%%%%%%%%%%%%%%%%%%%%%%%%%%%%%%%%%%%%
In the above, our results show the existence of superradiant instabilities both in the SFDM model and the CDM model. Next, we will qualitatively analyze and demonstrate in which state the maximum instability occurs. Firstly, based on the instabilities at the state $l=1$, what we need in the following is that determining the rotation parameter $a$ in the maximum instabilities. By the fixed rotation parameter $a$, we show the instabilities as the function of mass $M\mu$ in Figure \ref{nf6}. The rotation parameter $a$ is $0\leq a \leq 0.9995 (a_{max} \approx 1)$. From these subfigures, for each rotation parameter $a$, there is always a maximum value of the instability. For these subfigures in the top three panels, the maximum value of these maximum instabilities increases with the increasing of the rotation parameter $a$. But in three panels of the bottom, the maximum value of instabilities first increases and then decreases with the increasing of the rotation parameter $a$. This turning point is rotation parameter $a= 0.995$. These results indicate that there is a maximum value of instability when the rotation parameter $0 \leq a \leq 0.9995 (a_{max} \approx 1)$, i.e., this turning point $a= 0.995$. Therefore, at the state $l=1$, the maximum instability occurs approximately for the state $l=1, m=1, a=0.995$. Secondly, we need to determine what the maximum instabilities occur in at the states with different $l$.
%%%%%%%%%%%%%%%%%%%%%%%%%%%%%%%%%%%%%%%%%%%%%%
\begin{figure*}[tbp]
\centering
{
\includegraphics[width=0.31\columnwidth]{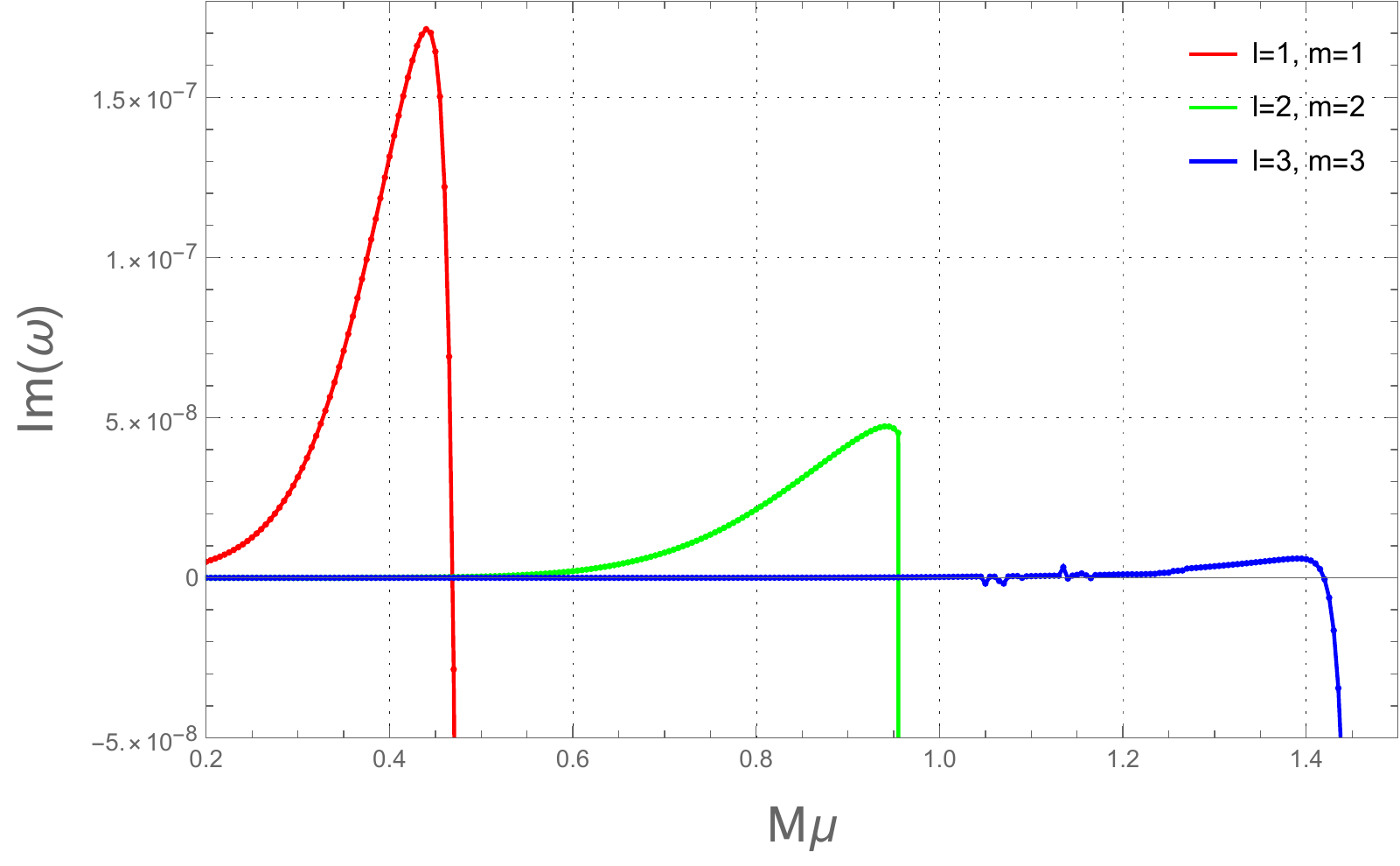}
}
{
\includegraphics[width=0.31\columnwidth]{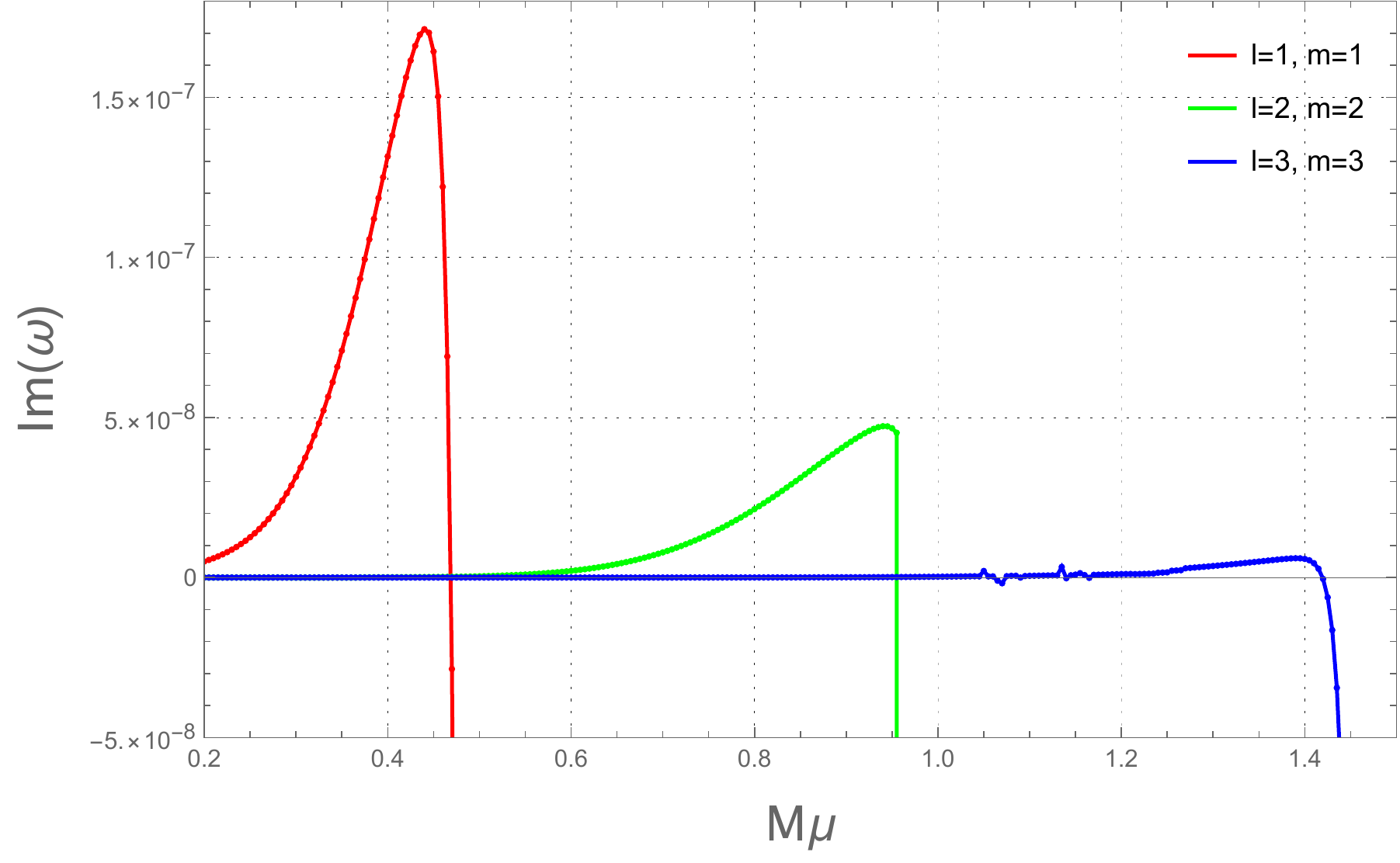}
}
{
\includegraphics[width=0.31\columnwidth]{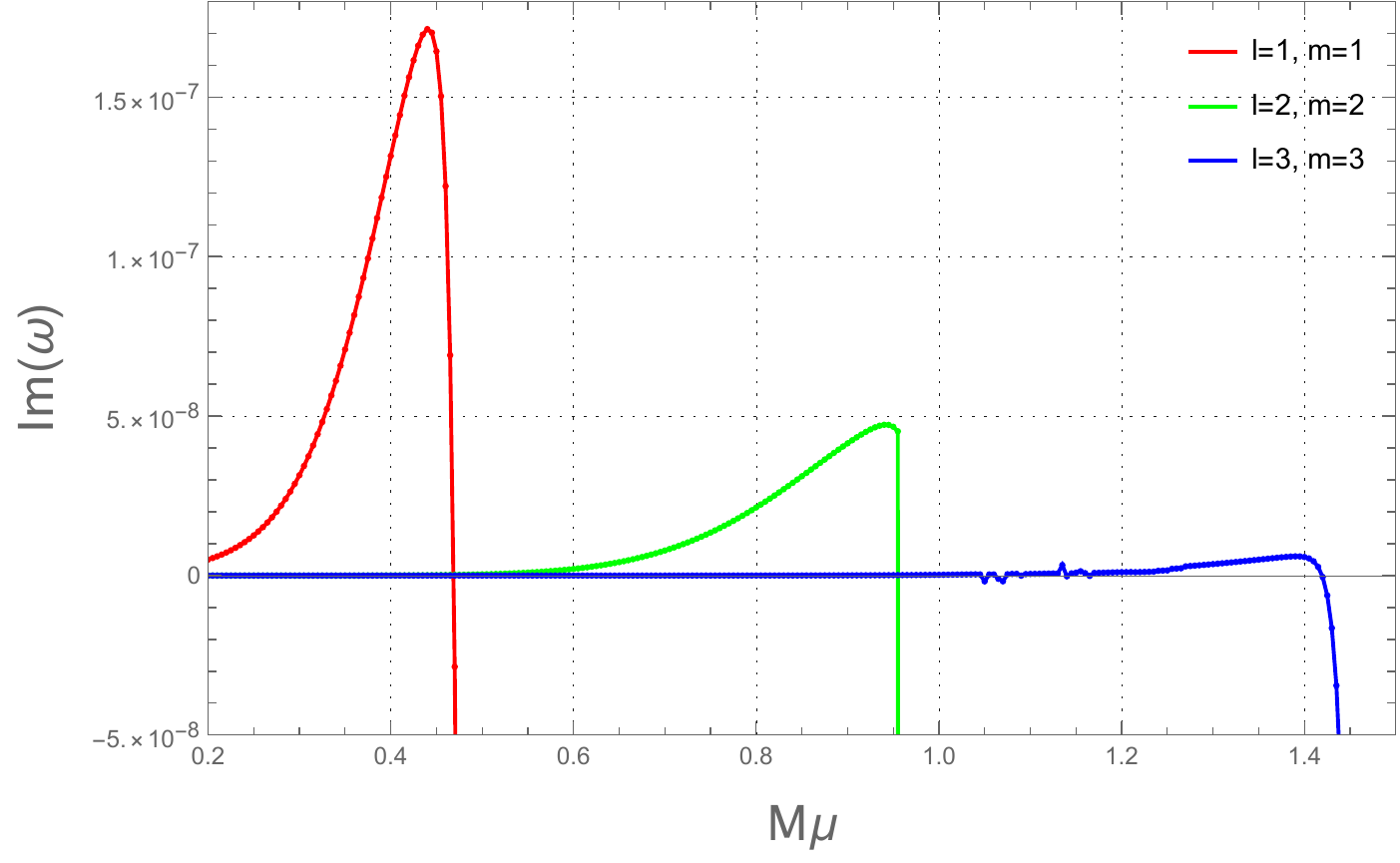}
}
\caption{The quasibound states of the black holes in massive scalar field at the state $l=m=1, 2, 3, a=0.995$ (left panel is CDM model, middle panel is SFDM model and the right panel is Kerr spacetime). Three panels all reflect the superradiant instabilities, and the maximum instabilities all occur for the state $l=m=1$. The main calculation parameters are $M=1$, $\rho_c=2.45 \times 10^{-3}$ $M_{\bigodot}/pc^{3}$, $R_c=5.7$ $kpc$, $\rho_s=13.66 \times 10^{-3}$ $M_{\bigodot}/pc^{3}$, $R_s=2.92$ $kpc$. We have converted these main calculation parameters to the black hole units before plotting.}
\label{nf7}
\end{figure*}
%%%%%%%%%%%%%%%%%%%%%%%%%%%%%%%%%%%%%%%%%
In Figure \ref{nf7}, we show the instabilities at the states $l=m=1, 2, 3$. The maximum instability decreases both with the increasing of $l, m$. So, the maximum instability occur for the state $l=m=1$. Finally, based on the above discussions, we prove that in CDM model, SFDM model and Kerr spacetime, the maximum instability occurs approximately for the state $l=1, m=1, a=0.995$. This result was as previously expected. In order to further study and analyze the difference in the maximum instability between dark matter black holes and Kerr black holes, in Figure \ref{nf8}, we show the instabilities at the state $l=1, m=1, a=0.995$ in the CDM model, the SFDM model and the Kerr spacetime. The maximum instabilities of the Kerr black hole is larger than that of the SFDM model, and the SFDM model is larger than the CDM model. After comparing Kerr black holes both with the SFDM model and CDM model, the difference of the maximum instability between the black hole in CDM model and the Kerr black hole is approximately $1.4\times10^{-10}$ with $M\mu=0.5$. The difference of the maximum instability between Kerr black hole and black hole in the SFDM model is approximately $2.0\times10^{-11}$ with $M\mu=0.5$. These differences increase with the increasing of mass. This result directly reflects the impacts of the mass on the superradiant mechanism. With the increasing of the mass, that is, when the mass satisfies $\text{Re}(\omega)<\mu$, this mass can work as a mirror \cite{Detweiler:1980uk,Zouros:1979iw,Furuhashi:2004jk}. This causes the wave to bounce back and forth between the black hole and the mirror, and it can amplify itself each time. This is the so-called the black hole bomb.
%%%%%%%%%%%%%%%%%%%%%%%%%%%%%%%%%%%%%%%%
%%%%%%%%%%%%%%%%%%%%%%%%%%%%%%%%
According to the above analysis, we have proved that the maximum instability of QBS occurs approximately for the state $l=1, m=1, a=0.995$ in the SFDM model, CDM model and Kerr spacetime. Next, we will quantitatively analyze the maximum instability at the state $l=1, m=1, a=0.995$. Here, we show that how to find these maximum instabilities of the state $l=1, m=1, a=0.995$ in the CDM model, SFDM model and Kerr spacetime. First of all, through the results in Figure \ref{nf6}, we can roughly determine a range of mass corresponding to the maximum instability, and form an array of the mass $M\mu$ and decay rate $\text{Im}(\omega)$. Secondly, we reduce the mass stepwise and use the continued fraction method to determine the frequency of each step. where the initial guess of frequency is the frequency obtained in the previous step. Finally, we stop iterating as soon as a local maximum occurs among a series of frequencies. Following this procedure strictly, we obtain the maximum instability of black holes for the state $l=1, m=1,a=0.995$ in the CDM model, SFDM model and Kerr spacetime.  We show these results in Figure \ref{nf9}.
From these figures, the maximum instabilities increase with the increasing of the rotation parameter $a$ in the three panels of the top. But in three panels of the bottom, the maximum instabilities first increase and then decrease with the increasing of the rotation parameter $a$. This turning point is $a = 0.995$. Finally, by the rotation parameter $a$ at this state, we quickly obtain the value of mass $\mu$. Since these results for black holes in CDM/SFDM models are very close to Kerr black hole, we give their numerical results of the maximum instabilities growth rate ($\tau^{-1}=M\text{Im}(\omega)$) in Tables \ref{t4}, \ref{t5} and \ref{t6}.
%%%%%%%%%%%%%%%%%%%%%%%%%%%%%%%%%%%%%%%%   kerr
\begin{table}[t!]
\setlength{\abovecaptionskip}{0cm}
\setlength{\belowcaptionskip}{0.3cm}
\centering
\caption{The maximum instability growth rate with different rotation parameter $a$ at the state $l=m=1$ in Kerr black hole. And the maximum instability growth rate ($\tau^{-1}=M\text{Im}(\omega)$) and the corresponding mass $\mu$ are given. The $\Delta$ is the relative error relative to the Table III in Ref. \cite{Dolan:2007mj}.}
\scalebox{0.68}{
\begin{tabular}{ccccccc}
\hline \hline
$a$    & 0.70 & 0.80  & 0.90 & 0.95 & 0.99 &0.995 \\ \hline
$\mu$  & 0.185    &    0.230   &  0.0.295   & 0.345    &  0.420  &0.440 \\
$\tau^{-1}$ & $3.31676821\times10^{-10}$    & $2.16321174\times10^{-9}$    &  $1.54706240\times10^{-8}$   &  $4.85965449\times10^{-8}$       & $1.50601550\times10^{-7}$     &$1.69918521\times10^{-7}$\\
$\Delta$ & $0.3974\%$ &$0.1487\%$& $0.1895\%$& $0.4169\%$ & $0.4010\%$ &/\\
\hline \hline
\end{tabular}}
\label{t4}
\end{table}
%%%%%%%%%%%%%%%%%%%%%%%%%%%%%%%%%%%%%%%%%%%%%%%%%%%%SFDM
\begin{table}[t!]
\setlength{\abovecaptionskip}{0cm}
\setlength{\belowcaptionskip}{0.3cm}
\centering
\caption{The maximum instability growth rate with different rotation parameter $a$ at the state $l=m=1$ in SFDM model. The maximum instability growth rate ($\tau^{-1}=M\text{Im}(\omega)$) and the mass $\mu$ are given. The main calculation parameters are $M=1$, $\rho_s=13.66 \times 10^{-3}$ $M_{\bigodot}/pc^{3}$, $R_s=2.92$ $kpc$.}
\scalebox{0.68}{
\begin{tabular}{ccccccc}
\hline \hline
$a$    & 0.70 & 0.80 & 0.90 & 0.95 & 0.99 &0.995 \\ \hline
$\mu$  & 0.185    &    0.230 &  0.295   &  0.345   & 0.420        &0.440\\
$\tau^{-1}$ & $3.31676758\times10^{-10}$    & $2.16321137\times10^{-9}$    &  $1.54706203\times10^{-8}$   &  $4.85965313\times10^{-8}$   & $1.50601515\times10^{-7}$        &$1.69918500\times10^{-7}$\\ \hline \hline
\end{tabular}}
\label{t5}
\end{table}
%%%%%\mu/=0 SFDM l1m0%%%%%%%%%%%%%%%%%%%%%%%CDM
\begin{table}[t!]
\setlength{\abovecaptionskip}{0cm}
\setlength{\belowcaptionskip}{0.3cm}
\centering
\caption{The maximum instability growth rate with different rotation parameter $a$ at the state $l=m=1$ in CDM model. The maximum instability growth rate ($\tau^{-1}=M\text{Im}(\omega)$) and the corresponding mass $\mu$ are given. The main calculation parameters are $M=1$, $\rho_c=2.45 \times 10^{-3}$ $M_{\bigodot}/pc^{3}$, $R_c=5.7$ $kpc$.}
\scalebox{0.68}{
\begin{tabular}{ccccccc}
\hline \hline
$a$    & 0.70 & 0.80 & 0.90 & 0.95 & 0.99 & 0.995  \\ \hline
$\mu$  &   0.185  &  0.230  &   0.295  & 0.345    &  0.420       &0.440\\
$\tau^{-1}$ & $3.31676487\times10^{-10}$    & $2.16320929\times10^{-9}$    &  $1.54705991\times10^{-8}$   & $4.85964529\times10^{-8}$    & $1.50601311\times10^{-7}$       & $1.69918377\times10^{-7}$\\ \hline \hline
\end{tabular}}
\label{t6}
\end{table}
We list the maximum instability growth rate of CDM model, SFDM model and Kerr spacetime with different rotation parameters $a$ at the state $l=1, m=1$ in Tables  \ref{t4}, \ref{t5} and \ref{t6}, respectively. Besides, in Table \ref{t4}, we also give the relative error between the maximum instability and the test value from Ref. \cite{Dolan:2007mj} in the last row. These results demonstrate that our approach is reliable. From the data from Tables \ref{t4}, \ref{t5} and \ref{t6}, both in CDM and SFDM models, the maximum instability increases with the increasing of the rotation parameter $a$. For CDM model, the maximum instability occurs approximately for $M\mu \approx 0.440 $ and the maximum instability growth rate is approximately $\tau^{-1} =M\text{Im}(\omega) \approx 1.69918377 \times 10^{-7} (GM/c^3)^{-1}$. For SFDM model, the maximum instability occurs approximately for $M\mu \approx 0.440 $ and the maximum instability growth rate is approximately $\tau^{-1} =M\text{Im}(\omega) \approx 1.69918500 \times 10^{-7} (GM/c^3)^{-1}$. The difference between them is approximately $1.2\times10^{-13} (GM/c^3)^{-1}$. Compared the maximum instabilities of black holes in CDM/SFDM models with Kerr black hole, the values of the maximum instability of Kerr-like black holes in a dark matter halo are all smaller than that of Kerr black hole. The value of the maximum instability of the Kerr black hole is larger than that of the SFDM model, and the SFDM model is larger than that of the CDM model. The maximum instability difference of black hole between CDM model and Kerr black hole is approximately $1.4\times10^{-13}$. For SFDM model, the difference is approximately $2.0 \times10^{-14}$. These values are an upper bound on the growth rate of instabilities of black holes in massive scalar field in Table \ref{t4}, \ref{t5} and \ref{t6}. These results show that, in the three cases considered, the growth rate of the Kerr black hole instability is larger than that of the SFDM model, which is larger than that of the CDM model.\\
\indent Finally, we also study the impacts of dark matter parameters on the QBS frequencies, just like the previous analysis of how the dark matter parameters affect the QNM frequency. Here, we mainly consider to test the QBS frequencies at the state $l=m=1,a=0.995$ in a massive field ($\mu \neq 0$). Similarly, we need a set of fixed dark matter parameters for analysis. For the SFDM model, we fix the density parameter $\rho=10^{4}M_{\bigodot}/pc^{3}$ and the characteristic radius $R=5.7kpc$ respectively. For the CDM model, we also fixed the density parameter $\rho=10^{3}M_{\bigodot}/pc^{3}$ and the characteristic radius $R=5.7kpc$. Then, we calculate the QBS frequencies of black holes in the massive scalar field $(\mu=0.440)$ at state $l=m=1,a=0.995$. We record our results of SFDM model and CDM model in Figures. \ref{rr3} and \ref{rr4}, respectively. In Figure \ref{rr3}, we find that when the characteristic radius is fixed, the real and imaginary parts of the QBS frequencies of the black hole both decrease with the increasing of the density parameter. Similarly, when the density parameter is fixed, we investigate the impacts of the characteristic radius on the QBS frequencies of black holes. Actually, the behaviors of QBS frequencies and characteristic radius are consistent with the density parameter. These features indicates that the dark matter around the black hole will reduce the real and imaginary parts of the QBS frequencies. Figure \ref{rr4} is the case of the CDM model. The behavior of the CDM model is roughly the same as that of the SFDM model.
%%%%%%%%%%%%%%%%%%%%%%%%%%%%%%%%%%
%%%%%%%%%%%%%%%%%%%%%%%%%%%%%%%
%%%%%%%%%%%%%%%%%%%%%%%%%%%%%%%%%%%%%%%%
In addition, we have also calculated the superradiant instabilities of the black holes in the massive scalar field at the state $l=m=1,a=0.995$ in the galaxies numbered ESO from the Table \ref{t11}. We show the results in Figure \ref{nf10}. We find that the real part of the QBS frequencies of these galaxies increases with the increasing of mass, while the imaginary part increases first and then decreases.
%%%%%%%%%%%%%%%%%%%%%%%%%%%%%%%%%%%%%%compare
%%%%%%%%%%%%%%%%%%%%%%%%%%%%%%%%%%%%%%%%%%%%%%
Our results show that these dark matter parameter configurations can all lead to superradiant instabilities both in SFDM and CDM models. Among them, the maximum instabilities both in SFDM model and CDM model are lower than that of Kerr black holes.  In other words, for nearly extremal black holes, the distribution of dark matter reduces the imaginary part of the QBS frequency compared to Kerr black holes.

\newpage
\section{Conclusions and discussions}\label{s6}
In this paper, we mainly use the scalar field perturbation to study QNM/QBS of the Kerr-like black holes in a dark matter halo, and compare these results with Kerr black hole. The method we used for calculating the frequencies of QNM/QBS is the continued fraction method. In addition, we also study the impacts of dark matter parameters on the QNM/QBS of black holes at the specific circumstances. Our main conclusions are as follows:

(1) In the massless scalar field, the real and imaginary parts of QNM frequencies of the black hole at the state $l=1$ both increase with the increasing of $m$ in CDM model, SFDM model and Kerr spacetime. For the same state, QNM frequency of kerr black hole is greater than that of SFDM model, and SFDM model is greater than that of CDM model. For the state $l=m=1,a=0.99$, the QNM difference between the CDM model and Kerr spacetime is approximately $1 \times 10^{-7}$ in real part and $5 \times 10^{-8}$ in imaginary part. The QNM difference of that between SFDM model and Kerr spacetime is approximately $2 \times 10^{-8}$ in real part and $4 \times 10^{-9}$ in imaginary part.

(2) In the massive scalar field, by testing the states of QBS frequencies with different $l,m,a$, we confirm the existence of the superradiant instabilities when the black holes both in CDM and SFDM models. Besides, we prove that the maximum instabilities of black hole in CDM model and SFDM model occur approximately for the state $l=1,m=1, a=0.995$.

(3) In black hole units, for CDM model in ESO$1200211$, the superradiant instability of black hole occurs approximately for the mass parameter $M\mu \lesssim 0.55$. Its maximum instability occurs approximately for $M\mu \approx 0.44 $ and the maximum instability growth rate is approximately $\tau^{-1} \approx 1.69918377 \times 10^{-7} (GM/c^3)^{-1}$. For SFDM model in ESO$1200211$, the superradiant instability of black hole occurs approximately for mass parameter $M\mu \lesssim 0.55$. Its maximum instability occurs approximately for $M\mu \approx 0.44 $ and the maximum instability growth rate is approximately $\tau^{-1} \approx 1.69918500 \times 10^{-7} (GM/c^3)^{-1}$. These values are the upper bounds on the instability growth rate of black holes in the massive scalar field. The growth rate of the maximum instability of black hole in SFDM model is greater than that of CDM model. The difference between them is approximately $1.2\times10^{-13} (GM/c^3)^{-1}$.

(4) Compared the maximum instabilities of black holes in CDM/SFDM models with Kerr black hole, the values of the maximum instability of Kerr-like black holes in a dark matter halo are all smaller than that of Kerr black hole. The maximum instability difference of black hole between CDM model and Kerr black hole is approximately $1.4\times10^{-13}$. For SFDM model, the difference of that is approximately $2.0\times10^{-14}$. In the future, these differences may be detected by the gravitational wave detection, which may provide an effective method for detecting the existence of dark matter.

(5) The impacts of dark matter parameters on the QNM/QBS of black holes at the specific circumstances are studied. The dark matter parameters affecting QNM/QBS are density parameter $\rho$ and characteristic radius $R$. The method of the research is the control variable method. For QNM frequencies both in SFDM model and CDM model, at the state $l=1,m=-1,0$, the real part of QNM frequencies decrease with the increasing of dark matter parameter, and their imaginary parts increase with the increasing of dark matter parameter. At the state $l=m=1$, the real and imaginary parts of the QNM frequencies decrease with the increasing of the dark matter parameter. For QBS frequencies both in SFDM model and CDM model, at the state $l=1,m=1,a=0.995$, the real and imaginary parts of the QBS frequencies decrease with the increasing of the dark matter parameter.

At last, it is also worth mentioning that we are very interested in the exploration of the relevant physical processes of black holes around the dark matter. Based on this premise, we choose to study the QNM/QBS of a rotating black hole in a dark matter halo. In fact, the distribution of dark matter around a black hole is a ``spike” structure, and this structure is closer to the real situation. The QNM/QBS of a special black hole are the characteristic ``sound” which can provide us with a new method to identify black holes in the universe. On the other hand, in some recent studies, we found some discussion about echoes\cite{Cardoso:2017cqb,Churilova:2019cyt,Dey:2020wzm,Ikeda:2021uvc,Chowdhury:2020rfj,Churilova:2021tgn}. These studies show that the echoes appear after the quasinormal mode. Echoes are one of the important means currently used to test gravitational waves. Taking the above two points into consideration, our next plans are to study the physics of rotating black holes associated with echoes. About the study of echoes, we have already made some preliminary attempts\cite{Yang:2021cvh,Yang:2022ryf,Wu:2022eiv}. To sum up, we also hope that our work can form a complete research system in the direction of interaction between dark matter and black holes. In the future, these studies may be used for gravitational wave detection of supermassive black holes, and may provide an effective method for detecting the existence of dark matter.

\acknowledgments
We would like to acknowledge the anonymous referee for a constructive report, which significantly improved this paper. We are grateful to Prof. M. Richartz and Prof. A. Zhidenko for the useful discussions. This research was funded by the National Natural Science Foundation of China (Grant No.12265007) and the Natural Science Special Research Foundation of Guizhou University (Grant No.X2020068).

%\newpage

\bibliographystyle{EPJC}
\bibliography{EPJCexample}
\end{document}